\documentclass[superscriptaddress,prab,reprint,floatfix,nofootinbib,longbibliography]{revtex4-1}

\usepackage[T1]{fontenc}
\usepackage[utf8]{inputenc}

\usepackage{amssymb,amsmath}
\usepackage{graphicx}
\usepackage{subfigure}
  \usepackage{color}
\usepackage[dvipsnames]{xcolor}
\usepackage[colorlinks]{hyperref}
 \hypersetup{
     colorlinks=true, 
     linktoc=all,     
     linkcolor=blue,  
     citecolor=magenta,
     urlcolor=blue	
 }
\usepackage{dcolumn}
\usepackage{bm}
\usepackage{interval}
\usepackage{rotating}
\usepackage{multirow}
\usepackage[separate-uncertainty = true,multi-part-units=single,multi-part-units=brackets]{siunitx}
\usepackage{url}
\usepackage[ruled,vlined]{algorithm2e}
\usepackage{cancel}
\usepackage{tikz}
\usetikzlibrary{arrows,decorations.markings}
\usepackage{tikz-3dplot}
\usetikzlibrary{math} 
\usetikzlibrary{calc}
\usepgflibrary{arrows}
\tikzset{
    partial ellipse/.style args={#1:#2:#3}{
        insert path={+ (#1:#3) arc (#1:#2:#3)}
    }
}

\newcommand{\dd}{\text{d}}
\def\centerarc[#1](#2)(#3:#4:#5)
    { \draw[#1] ($(#2)+({#5*cos(#3)},{#5*sin(#3)})$) arc (#3:#4:#5); }

\begin{document} 

\preprint{Draft for PR AB \today}

\title{First detection of collective oscillations of a stored deuteron beam \\with an amplitude close to the quantum limit}

\author{J. Slim}
\affiliation{III. Physikalisches Institut B, RWTH Aachen University, 52056 Aachen, Germany}

\author{N.N. Nikolaev}
\affiliation{L.D. Landau Institute for Theoretical Physics, 142432 Chernogolovka, Russia}
\affiliation{Moscow Institute for Physics and Technology, 141700 Dolgoprudny, Russia}

\author{F. Rathmann}
\affiliation{Institut f\"ur Kernphysik, Forschungszentrum J\"ulich, 52425 J\"ulich, Germany}

\author{A. Wirzba}
\affiliation{Institut f\"ur Kernphysik, Forschungszentrum J\"ulich,  52425 J\"ulich, Germany}
\affiliation{Institute for Advanced Simulation, Forschungszentrum J\"ulich, 52425 J\"ulich, Germany}

\author{A. Nass}
\affiliation{Institut f\"ur Kernphysik, Forschungszentrum J\"ulich, 52425 J\"ulich, Germany}

\author{V. Hejny}
\affiliation{Institut f\"ur Kernphysik, Forschungszentrum J\"ulich, 52425 J\"ulich, Germany}

\author{J. Pretz}
\affiliation{Institut f\"ur Kernphysik, Forschungszentrum J\"ulich, 52425 J\"ulich, Germany}

\author{H.\,Soltner}
\affiliation{Zentralinstitut f\"ur Engineering, Elektronik und Analytik, Forschungszentrum J\"ulich, 52425 J\"ulich, Germany}

\author{F.\,Abusaif}
\affiliation{Institut f\"ur Kernphysik, Forschungszentrum J\"ulich, 52425 J\"ulich, Germany}

\author{A.\,Aggarwal}
\affiliation{Marian Smoluchowski Institute of Physics, Jagiellonian University, 30348 Cracow, Poland}

\author{A. Aksentev}
\affiliation{Institute for Nuclear Research, Russian Academy of Sciences, 117312 Moscow, Russia}

\author{A. Andres}
\affiliation{III. Physikalisches Institut B, RWTH Aachen University, 52056 Aachen, Germany}

\author{L. Barion}
\affiliation{University of Ferrara and Istituto Nazionale di Fisica Nucleare, 44100 Ferrara, Italy}

\author{G. Ciullo}
\affiliation{University of Ferrara and Istituto Nazionale di Fisica Nucleare, 44100 Ferrara, Italy}

\author{S.\,Dymov}
\affiliation{University of Ferrara and Istituto Nazionale di Fisica Nucleare, 44100 Ferrara, Italy}
\affiliation{Laboratory of Nuclear Problems, Joint Institute for Nuclear Research, 141980 Dubna, Russia}

\author{R.\,Gebel}
\affiliation{Institut f\"ur Kernphysik, Forschungszentrum J\"ulich, 52425 J\"ulich, Germany}

\author{M.\,Gaisser}
\affiliation{III. Physikalisches Institut B, RWTH Aachen University, 52056 Aachen, Germany}

\author{K.\,Grigoryev}
\affiliation{Institut f\"ur Kernphysik, Forschungszentrum J\"ulich, 52425 J\"ulich, Germany}

\author{D.\,Grzonka}
\affiliation{Institut f\"ur Kernphysik, Forschungszentrum J\"ulich, 52425 J\"ulich, Germany}

\author{O.\,Javakhishvili}
\affiliation{Department of Electrical and Computer Engineering, Agricultural University of Georgia, 0159 Tbilisi, Georgia}

\author{A.\,Kacharava}
\affiliation{Institut f\"ur Kernphysik, Forschungszentrum J\"ulich, 52425 J\"ulich, Germany}

\author{V.\,Kamerdzhiev}
\affiliation{Institut f\"ur Kernphysik, Forschungszentrum J\"ulich, 52425 J\"ulich, Germany}

\author{S.\,Karanth}
\affiliation{Marian Smoluchowski Institute of Physics, Jagiellonian University, 30348 Cracow, Poland}

\author{I.\,Keshelashvili}
\affiliation{Institut f\"ur Kernphysik, Forschungszentrum J\"ulich, 52425 J\"ulich, Germany}

\author{A.\,Lehrach}
\affiliation{Institut f\"ur Kernphysik, Forschungszentrum J\"ulich, 52425 J\"ulich, Germany}

\author{P. Lenisa}
\affiliation{University of Ferrara and Istituto Nazionale di Fisica Nucleare, 44100 Ferrara, Italy}

\author{N. Lomidze}
\affiliation{High Energy Physics Institute, Tbilisi State University, 0186 Tbilisi, Georgia}

\author{B.\,Lorentz}
\affiliation{GSI Helmholtzzentrum für Schwerionenforschung, 64291 Darmstadt, Germany}

\author{A. Magiera}
\affiliation{Marian Smoluchowski Institute of Physics, Jagiellonian University, 30348 Cracow, Poland}

\author{D.\,Mchedlishvili}
\affiliation{High Energy Physics Institute, Tbilisi State University, 0186 Tbilisi, Georgia}

\author{F. Müller}
\affiliation{III. Physikalisches Institut B, RWTH Aachen University, 52056 Aachen, Germany}

\author{A. Pesce}
\affiliation{Institut f\"ur Kernphysik, Forschungszentrum J\"ulich, 52425 J\"ulich, Germany}

\author{V. Poncza}
\affiliation{Institut f\"ur Kernphysik, Forschungszentrum J\"ulich, 52425 J\"ulich, Germany}

\author{D. Prasuhn}
\affiliation{Institut f\"ur Kernphysik, Forschungszentrum J\"ulich, 52425 J\"ulich, Germany}

\author{A.\,Saleev}
\affiliation{University of Ferrara and Istituto Nazionale di Fisica Nucleare, 44100 Ferrara, Italy}

\author{V. Shmakova}
\affiliation{Institut f\"ur Kernphysik, Forschungszentrum J\"ulich, 52425 J\"ulich, Germany}
\affiliation{Laboratory of Nuclear Problems, Joint Institute for Nuclear Research, 141980 Dubna, Russia}

\author{H. Ströher}
\affiliation{Institut f\"ur Kernphysik, Forschungszentrum J\"ulich, 52425 J\"ulich, Germany}

\author{M.\,Tabidze}
\affiliation{High Energy Physics Institute, Tbilisi State University, 0186 Tbilisi, Georgia}

\author{G.\,Tagliente}
\affiliation{Istituto Nazionale di Fisica Nucleare sez.\ Bari, 70125 Bari, Italy}

\author{Y. Valdau}
\affiliation{Institut f\"ur Kernphysik, Forschungszentrum J\"ulich, 52425 J\"ulich, Germany}

\author{T. Wagner}
\affiliation{Institut f\"ur Kernphysik, Forschungszentrum J\"ulich, 52425 J\"ulich, Germany}

\author{C. Weidemann}
\affiliation{Institut f\"ur Kernphysik, Forschungszentrum J\"ulich, 52425 J\"ulich, Germany}

\author{A. Wro\'{n}ska}
\affiliation{Marian Smoluchowski Institute of Physics, Jagiellonian University, 30348 Cracow, Poland}

\author{M. \.{Z}urek}
\affiliation{Lawrence Berkeley National Laboratory, Berkeley, California 94720, USA}




\begin{abstract}
We investigated coherent betatron oscillations of a deuteron beam in the storage ring COSY, excited by a detuned radio-frequency Wien filter. The beam oscillations were detected by conventional beam position monitors. 
With the currently available apparatus, we show that oscillation amplitudes down to \SI{1}{\micro \meter} can be detected. The interpretation of the response of the stored beam to the detuned radio-frequency Wien filter is based on simulations of the beam evolution in the lattice of the ring and realistic time-dependent 3D field maps of the Wien filter. 
Future measurements of the electric dipole moment of protons will, however, require control of the relative position of counter-propagating beams in the sub-picometer range. Since here the stored beam can be considered as a rarefied gas of uncorrelated particles, we moreover demonstrate that the amplitudes of the zero-point (ground state) betatron oscillations of individual particles are only a factor of about 10 larger than  the Heisenberg uncertainty limit. As a consequence of this, we conclude that quantum mechanics does not preclude the control of the beam centroids to sub-picometer accuracy. The smallest Lorentz force exerted on a single particle that we have been able to determine is \SI{10}{aN}. 
\end{abstract}

\maketitle 

\section{Introduction}

The approach to the quantum ground state, the observation of quantum effects in macroscopic systems, and the possibility to detect displacements of macroscopic bodies on the nanometer scale, are the subject of intense theoretical and experimental efforts\,\cite{Schreppler1486, Abbott:2009zz, Murch2008, Biercuk2010, Rugar2004}. A notable example is the detection of gravitational waves using an interferometric detector with mirrors in the kilogram range\,\cite{Abbott:2016izl}. In all-electric proton storage rings, coherent beam displacements down to the picometer range that are  caused by Earth's gravity pull are in principle accessible using the  spin rotations of the proton as a detector\,\cite{CPEDM}. We also mention here the ongoing discussions of the possibility to detect gravitational waves via perturbations of the beam orbit in high-energy storage rings\,\cite{ariesWS:2021}. Here we report the first detection of collective oscillations of  an intense beam of  deuterons in a storage ring with an amplitude close to the quantum limit.  The present study is part of an international effort to prepare for the search for the permanent electric dipole moment (EDM) of charged particles. The focus of these studies has been on systematic effects, e.g., imperfection magnetic fields in storage rings\,\cite{PhysRevAccelBeams.20.072801}, and orbit improvements in a  machine using beam-based alignment\,\cite{Wagner:2020akw}, thereby advancing the high-precision frontier in spin dynamics in storage rings. A comprehensive description of this activity and of the proposed stepwise approach leading to a dedicated proton EDM storage ring is presented in Ref.\,\cite{CPEDM}.

Experiments searching for electric dipole moments of charged particles using storage rings are at the forefront of the incessant quest to find new physics beyond the Standard Model of particle physics. These investigations bear the potential to shed light on the origin of the anomalously large matter-antimatter asymmetry in the Universe\,\cite{POSPELOV2005119}, for which the combined predictions of the Standard Models of particle physics and of cosmology fall short of the experimentally observed asymmetry by about seven to eight orders of magnitude\,\cite{Bernreuther:2002uj}. 

The signal for an EDM is the spin precession in electric fields, where it should be noted that the spins of charged particles can be subjected to large electric fields only in storage rings. The need to eliminate the overwhelmingly stronger spin rotations driven by the magnetic moment in magnetic fields brings to the front an all-electric, so-called frozen spin proton storage ring\,\cite{doi:10.1063/1.4967465,CPEDM}. An important advantage of such a machine is the ability to simultaneously store two counter-propagating proton beams. The concurrent measurement of the EDM-driven spin rotations of the counter-propagating beams would allow to cancel major systematic effects. To this end, to reach an ambitious sensitivity to the proton EDM of $d_p \approx \SI{e-29}{e.cm}$, it is imperative to control the relative vertical displacement of the centers of gravity of the two beams to an accuracy of about $\SI{5}{pm}$\,\cite{CPEDM}. One may wonder whether such an enormously demanding accuracy is not prohibited by the Heisenberg uncertainty principle. Towards an ultimate precision search for EDMs of charged particles, this particular aspect of the systematics of such measurements had not been investigated so far, and our experiment constitutes the first step in this direction.

\begin{figure*}[tb]
\centering
\subfigure[CAD drawing of the design of the RF Wien filter. 1: RF feed, 2: beam pipe, 3: inner mounting cylinder, 4: inner support structure, 5: lower electrode, 6: insulator, 7: RF connector, and 8: vacuum vessel.]{\includegraphics[height=0.35\textheight]{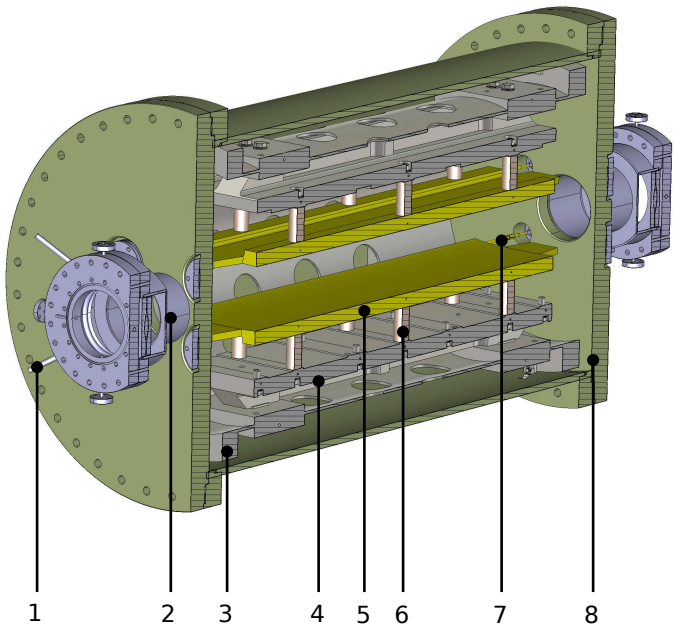}} \hspace{0.4cm}
\subfigure[Photograph with a view along the beam axis showing the gold-plated copper electrodes, which have a length of \SI{808.8}{mm}.]
{\includegraphics[height=0.35\textheight]{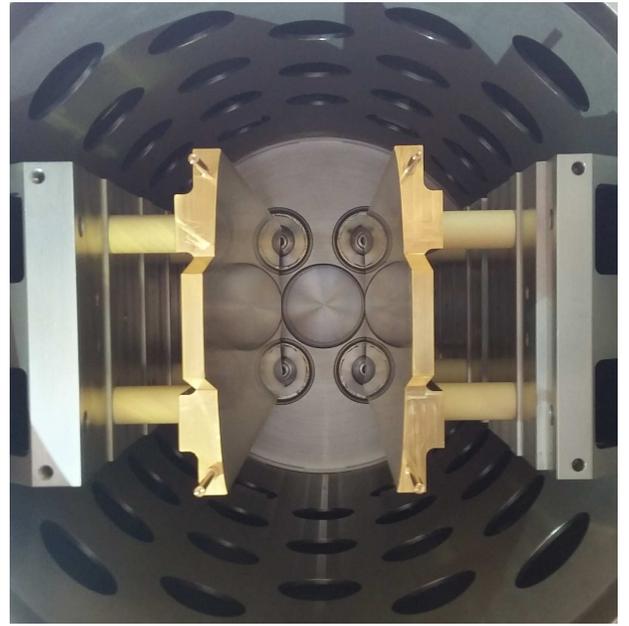}}
\caption{The waveguide RF Wien filter is mounted inside a cylindrical vessel. The effective length of the device amounts to $\ell =\SI{1.16}{m}$.The technical details are described in Refs.\,\cite{slim201752, Slim2016116}.}
\label{fig:wfcad}
\end{figure*}

Here we report on the  measurement of the amplitude of collectively excited vertical oscillations of a deuteron beam orbiting in the magnetic storage ring COSY at a momentum of about $\SI{970}{MeV/c}$\,\cite{Maier19971}. The data were taken in 2018 in the course of a dedicated experiment in the framework of systematic beam and spin dynamics studies for the deuteron EDM experiment (so-called \textit{precursor} experiment), presently carried out by the JEDI collaboration at COSY\,\cite{Rathmann:2013rqa,Morse:2013hoa,PhysRevAccelBeams.23.024601}. One of the central devices in the precursor experiment is the radio-frequency (RF) Wien filter (WF), shown in Fig.\,\ref{fig:wfcad}, which was designed to provide a cancellation of the electric and magnetic forces acting on the particle. In this operation mode, the Wien filter affects only the particle spins, but does not perturb the beam  orbit\,\cite{Slim2016116,slim201752,Slim:748558}. A slightly detuned Wien filter, however, exerts a non-vanishing Lorentz force on the orbiting beam particles.  It is shown  collective beam oscillation excited by the WF with amplitudes down to $\SI{1}{\micro m}$ can be detected with the currently available equipment. Our approach to measuring ultra-small displacements complements other measurements of ultra-small forces using different techniques\,\cite{Schreppler1486, Abbott:2009zz, Murch2008, Biercuk2010, Rugar2004}. 

In our experiment, the measurement cycles were much shorter than the intrabeam interaction time, and the beam attenuation rate was negligibly weak (see discussion in\,\cite{PAX}), therefore we treat the beam as a rarefied gas of uncorrelated particles. Individual particles undergo stable betatron oscillations around the equilibrium orbit in the horizontal and vertical planes, driven by focusing magnetic fields. Apart from their conventional individual betatron motions, all the particles in a bunch participate in one and the same collective and coherent oscillation that is driven by the Wien filter. Therefore, the upper bound of the amplitude of the collective oscillation of the entire beam corresponds to the upper bound of the oscillation amplitude of a single particle. In our approach, access to ultrasmall oscillation amplitudes results from the fact that the measured signal corresponds to a collective response of the electric charge of about $N = \num{e9}$ deuterons in the bunch.

The beam tracking simulations were carried out to predict the response of the stored beam to the detuned radio-frequency Wien filter. The simulations use the elements of the ring lattice and realistic time-dependent 3D field maps of the Wien filter. These field maps describe the spatial variation of complex electric and magnetic fields, including the fringe field areas. Furthermore, the tolerances of the elements of the circuit driving the Wien filter are also taken into consideration.


As a reference value for the Heisenberg uncertainty relation, we take an estimate of the amplitude of the single-particle zero-point betatron oscillation amplitude  $Q$. Then, our result for the smallest measured amplitude of the Wien filter-driven single-particle oscillation is only about a factor of ten larger than the quantum limit of Heisenberg’s uncertainty relation for vertical single-particle betatron oscillations. The smallest detected oscillation amplitude is by three orders of magnitude smaller than the beam size. 

We demonstrate for the first time that the accuracy with which periodic  beam oscillation amplitudes can be measured is vastly higher compared to that of static beam displacements generated by steerers using the same BPM. In a broad context, any new precision tool is of interest per se, and the latter point has important implications, for instance, for all-electric  frozen-spin EDM storage rings. Here, one aims to control the interfering radial magnetic fields by measuring the vertical spacing of the counter-propagating beams. Our result complements the potential of using beam oscillations to measure this distance, as discussed in Ref.\,\cite{Haciomeroglu:2018son}.
		

One must distinguish the RF-driven collective oscillations above the quantum limit $Q$ from the quantum uncertainty of the center of mass of the bunch circulating in a static ring. Specifically, for a rarefied-gas of $N$ uncorrelated particles, the quantum limit of the centroid of the bunch, detected by the BPMs, amounts to $Q/\sqrt{N}$.

The paper is organized as follows. In Sec.\,\ref{sec:pom}, the measurement principle is introduced, followed by a description of the  operation of the radio-frequency Wien filter in Sec.\,\ref{sec:WF}. The method to determine the beam oscillations is discussed in Sec.\,\ref{sec:measurements}, and the evolution of the beam to the combined effect of the ring lattice and the Wien filter fields are presented in Sec.\,\ref{app:sec:pce}. The time-dependent  field maps of the Wien filter are discussed in Sec.\ref{sec:time-dependent-field-maps}, and the evaluation of the uncertainties is elaborated in Sec.\,\ref{sec:eval-of-uncertainties}. Experimental results are presented in Sec.\,\ref{sec:discussion}, followed by conclusion and outlook in Sec.\,\ref{sec:conclusion}.

\section{Measurement principle }
\label{sec:pom}

\begin{figure*}[tb]
\centering
\includegraphics[width=1\textwidth]{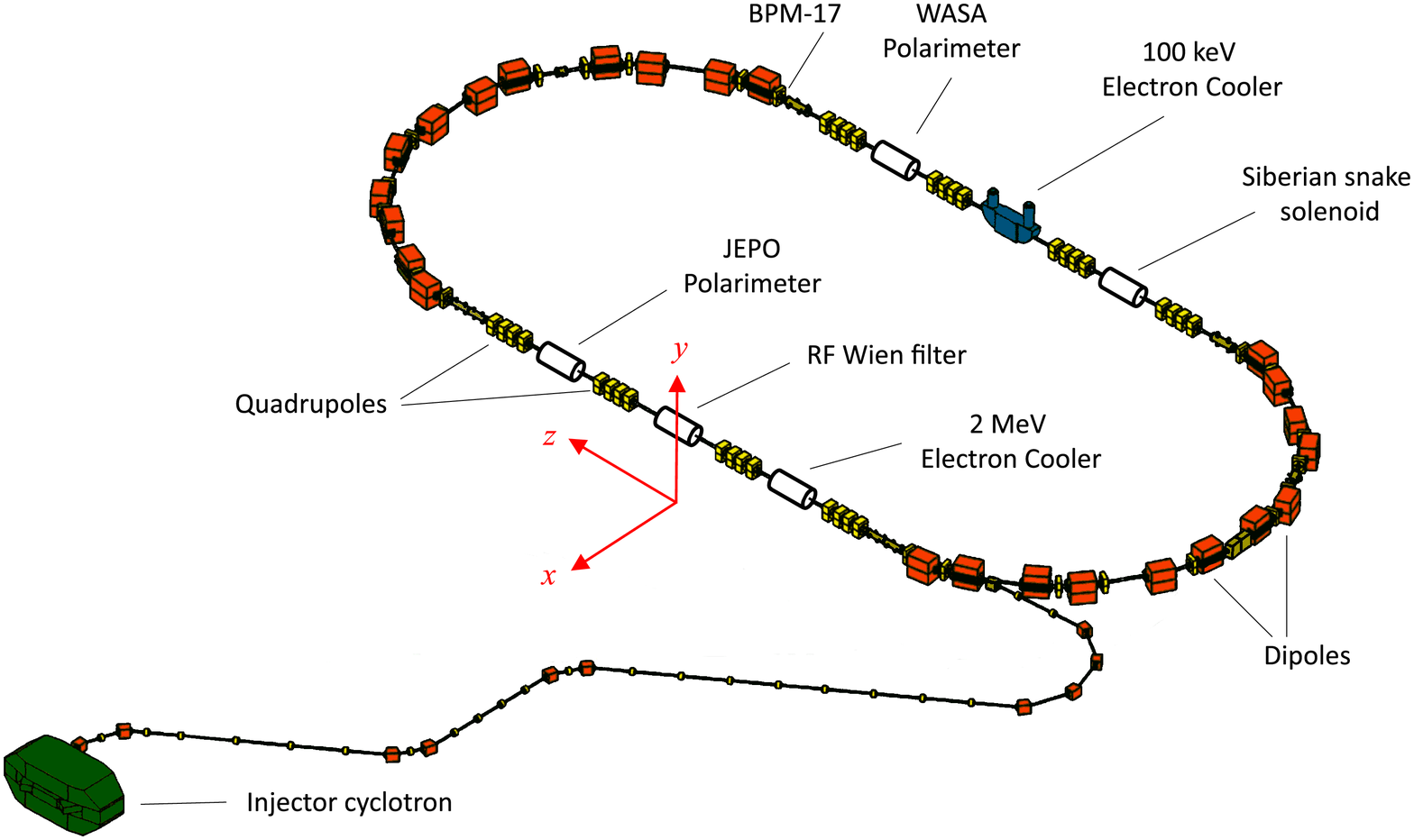}
\caption{Schematic diagram of the cooler synchrotron and storage ring (COSY) with the main components, especially the focusing/defocusing magnets (quadrupoles) and the bending magnets (dipoles). Indicated are the position of the RF Wien filter and the location of the beam position monitor\,\cite{Bohme:2018sjy} (BPM\,17), used to observe the beam oscillations. Further components such as the $\SI{2}{MeV}$ electron cooler\,\cite{Dietrich:2014uma}, the WASA\,\cite{Adam:2004ch} and the JEPO\,\cite{Muller:2019oxc,Muller_2020} polarimeters, and the Siberian snake\,\cite{Lehrach:2001db} are also shown. The coordinate system used is indicated. }
\label{fig:cosy}
\end{figure*}

The Cooler Synchrotron (COSY)\,\cite{MARTIN1985249,Maier19971} at Forschungszentrum J\"ulich is a storage ring with a circumference of approximately $\SI{184}{m}$. Its principal elements used for the experiments are indicated in Fig.\,\ref{fig:cosy}. For the investigations presented here, the two key devices are the RF Wien filter, based on a parallel-plates waveguide\,\cite{Slim2016116}, and a conventional electrostatic beam position monitor (BPM) that is used to monitor the beam oscillations\,\cite{Forck:1213277}.  The Wien filter generates orthogonal and highly-homogeneous electric and magnetic fields. In the present experiment, the Wien filter was operated in the mode with the electric field pointing vertically upward ($y$-direction), whereas the magnetic field points radially outward ($x$-direction), and the beam moves in $z$-direction (see coordinate system in Fig.\,\ref{fig:cosy}). The effective length of the Wien filter is $\ell = \SI{1.16}{m} $ (see Refs.\,\cite{slim201752, Slim2016116} for further technical details). 

As a spin rotator for the forthcoming deuteron EDM (precursor) experiment\,\cite{Rathmann:2013rqa,Morse:2013hoa,PhysRevAccelBeams.23.024601}, the Wien filter is designed to operate in resonance with the spin precession of the orbiting deuterons\,\cite{Slim2016116,slim201752,Slim:748558,Slim_2020}, and at a vanishing Lorentz force, given by
\begin{equation}
\vec F = q \left(\vec E + \vec v \times \vec B\right)\,,
\label{eq:lorentzForceRed}
\end{equation}
where $q$ denotes the elementary charge and $\vec v$ represents the velocity of the beam particles. Unlike in conventional DC Wien filters, the crossed electric field and magnetic fields ($\vec E$ and $\vec B$) of the RF Wien filter are generated \textit{simultaneously} by exciting the transverse electromagnetic (TEM) mode. The spin resonance tune mapping technique, developed for the Wien filter operation in the deuteron EDM experiment at COSY, is described in\,\cite{PhysRevAccelBeams.23.024601}.

When the electric and magnetic fields in the Wien filter are mismatched, \textit{i.e.}, when the electric and magnetic forces no longer cancel each other, the RF fields excite collective beam oscillations at the frequency at which the Wien filter is operated. In the present experimental set up, a mismatch between  electric and  magnetic fields provides a \textit{vertically} mismatched Lorentz force (see coordinate system in Fig.\,\ref{fig:cosy}). 

With a vanishing Lorentz force, the beam performs idle vertical (and horizontal) betatron oscillations 
\begin{equation} 
y(t) = y(0)\sqrt{\frac{\beta_y(t)}{\beta_y(0)}} \cos\left[ \psi_y(t) \right]\,,
\label{eq:beta-function-scaling}
\end{equation}
where $\beta_y(t)$ is the betatron amplitude function. With the beam revolution period of  $T = 2\pi/\omega_\text{rev}$, the betatron phase advance $\psi_y(t)$ satisfies $\psi_y(t+T) - \psi_y(t)=\omega_y T =2\pi \nu_y$, where $\nu_y$ is  the vertical betatron tune  given by $\nu_y = \omega_y/\omega_\text{rev}$. 

On the other hand, a mismatched Wien filter  exerts  stroboscopically, \textit{i.e.}, once per turn,  a vertical force $F_y(n)= F_y \cos(n\, \omega_\text{WF} T)$ on the stored particle, where $n$ is the turn number and  $\omega_\text{WF}$ denotes the angular velocity of the RF in the Wien filter (see discussion of Fig.\,\ref{fig:FLSim} in Sec.\,\ref{sec:discussion}). The  change of the vertical velocity of the stored particle,  accumulated during the time interval $\Delta t= \ell/v_z$ the particle spends per turn $n$ inside the Wien filter, is given by
	\begin{equation}
	\Delta v_y(nT) = \frac{ F_y (n) \Delta t } {\gamma m } = -\zeta \omega_y \cos(n\, \omega_\text{WF} T)  \,, 
	\label{eq:velKick}
	\end{equation}
where $\gamma$ and $m$ are the Lorentz-factor and the mass of the particle, respectively. The change $\Delta y$ of the vertical position $y$ in the Wien filter can be neglected. The coupling of vertical and radial beam oscillations is negligible (see Sec.\,\ref{sec:measurements}) and it is sufficient here to treat driven oscillations in a one-dimensional approximation. Due to the very strong disparity of synchrotron  and fractional WF  frequencies, synchro-betatron coupling can be neglected (see discussion in\,\cite{PhysRevLett.71.719}). Furthermore, beam attenuation either by intrabeam scattering or by interaction with residual gas during the data acquisition cycle is very small (see Appendix\,\ref{app:technique}), justifying the rarefied gas approximation.
	
According to Eq.\,(\ref{eq:beta-function-scaling}),  the stroboscopic signal of the betatron motion observed at any point in the ring, follows the harmonic law with angular velocity $\omega_y$, 	and we invoke the familiar description of the oscillatory motion in terms of the complex variable $z = y - iv_y/\omega_y$. With the initial condition $z(0)=0$, summing $\Delta v_y(kT)$ after $n$ turns, the solution for $z(n)$ behind the Wien filter reads
\begin{equation}
	\begin{split}
	z(n)  =  \frac{i \zeta}{2} &\cdot \left[\frac{ \exp(i n\omega_y T) - \exp(i n \,\omega_\text{WF} T) }{\exp(i(\omega_y -\omega_\text{WF}) T)-1  } \right.\\
	&+
	\left. \frac{\exp(i n\omega_y T) - \exp(-i n \,\omega_\text{WF} T) }{\exp(i(\omega_y +\omega_\text{WF}) T)-1  }\right] \,.
	\end{split}
	\label{eq:DrivenSolution}
	\end{equation}
This expression serves as the initial condition for the idle betatron motion during the subsequent $(n+1)$ turn and so forth. A similar analytic result holds also for generic AC dipole-driven betatron oscillations, discussed in a very different context of machine diagnostics in Ref.\,\cite{MiyamotoACdipole} (see also references therein).

Driven by the mismatched Wien filter, all beam particles participate in one and the same collective and coherent oscillation, and according to Eq.\,(\ref{eq:DrivenSolution}), the beam as a whole exhibits oscillations at the Wien filter frequency $\omega_\text{WF}$. A lock-in amplifier may  be used to selectively measure the corresponding Fourier component of the beam oscillation $y = \xi_y \cos (n \, \omega_\text{WF} T)$ from the output of a beam position monitor. Its amplitude is given by
\begin{equation}
\xi_y = \frac{\zeta}{2}\cdot \frac{\sin (2\pi \nu_y)}{\cos (2\pi \nu_\text{WF})  - \cos(2\pi \nu_y)}\, ,\label{eq:FourierAmplitude}
\end{equation}
where the vertical betatron tune $\nu_y$, and the  Wien filter tune $\nu_\text{WF}$, are given by $\nu_y = \omega_{y}/\omega_\text{rev}$ and  $\nu_\text{WF}=\omega_\text{WF}/\omega_\text{rev}$, respectively. When the Wien filter tune is close to the vertical betatron tune, a resonant enhancement of the beam oscillation amplitude $\xi_y$ occurs. Equation (\ref{eq:FourierAmplitude}) describes Hooke's law, $F_y= k_\text{H} \xi_y$, and Hooke's constant is given by
\begin{equation}
k_\text{H} =  \left|\frac{2\gamma m \omega_y}{\Delta t}\cdot \frac{\cos (2\pi \nu_\text{WF})  - \cos(2\pi \nu_y)}
{\sin (2\pi \nu_y)}\right|	\, . 
\label{eq:Hooke} 
\end{equation}

We invoke an approximate description of the betatron motion by a harmonic oscillator with constant betatron function and evaluate the  Heisenberg uncertainty limit $Q$ for the betatron oscillation amplitude $\xi_y$ in terms of the zero-point oscillator energy $\tfrac{1}{2} \hbar \omega_y$, which yields 
\begin{equation}
	Q^2 = \frac{\hbar}{m \gamma \omega_ y}\, \label{eq:Heisenberg}	\,.
\end{equation}

For the present experiment, we obtain
\begin{equation}
Q=\frac{82}{\sqrt{\gamma \nu_y}}\ {\rm nm}\, .
\label{eq:quantum-limit-1}
\end{equation}
With the actual COSY values for the  betatron tune $\nu_y$ and the Lorentz-factor $\gamma$ of the beam (see Table\,\ref{tab:num-val-sim}, Appendix\,\ref{app:beam-simulations}), the quantum limit of the vertical betatron oscillations amounts to  
\begin{equation}
Q \approx \SI{41}{nm}\,. 
\label{eq:quantum-limit-2}
\end{equation}
 
The interpretation of the measured oscillation amplitudes in terms of the Wien filter parameters requires numerical simulations of the performance of the Wien filter as an element of  the storage ring\,\cite{Slim2016116}. The details relevant to the present study are described below; the corresponding beam simulations carried out are consistent with the available experimental results on the properties of COSY\,\cite{PAX}.

\section{Wien filter operation}\label{sec:WF}

The control of the Lorentz force of the waveguide RF Wien filter is based on the wave-mismatch principle\,\cite{Slim:748558}. An impedance mismatch is introduced  at the load part of the device to deliberately create reflections that generate a standing wave pattern inside the Wien filter\,\cite{Slim_2020}. These standing waves can be represented by the complex-valued field quotient $Z_q$, defined as the ratio of the total electric to the total magnetic field strength,
\begin{equation}
\begin{split}
Z_q & = \frac{E^\text{total}}{H^\text{total}} = \frac{E^+ +E^-} {H^+ - H^-} 
= \frac{E^+ +\Gamma \cdot E^+ }{H^+ - \Gamma\cdot H^+} \\ & = Z_\text{w} \frac{1 + \Gamma}{1-\Gamma} =  Z_0 \frac{d}{W}\frac{1 +  \Gamma}{1 - \Gamma}\,,
\end{split}
\label{eq:main_gamma}
\end{equation}
where the superscripts '$+$' and '$-$' refer to the forward and backward direction of propagation, $Z_\text{w}$ is the wave impedance, $Z_0 \approx \SI{377}{\Omega}$ is the vacuum wave impedance, $d =\SI{100}{\milli \meter}$ is the distance between the electrodes, $W = \SI{182}{\milli \meter}$ is their width\,\cite{Slim2016116}, and $\Gamma$ is the reflection coefficient that controls the amplitude and phase of the reflected wave. During the measurements described here, the Wien filter was typically operated at a net input RF power of \SI{600}{W}.

The field quotient $Z_q$ is controlled via a specially designed RF circuit\,\cite{Slim_2020}. By altering $\Gamma$ via two variable vacuum  capacitors, called $C_\text{L}$ and $C_\text{T}$, a wide range of $Z_q$ values can be covered, and the matching point corresponding to the minimum induced vertical beam oscillation amplitude may be determined.

\section{Beam oscillations}
\label{sec:measurements}

\begin{figure}[htb]
\centering
\includegraphics[width=\columnwidth]{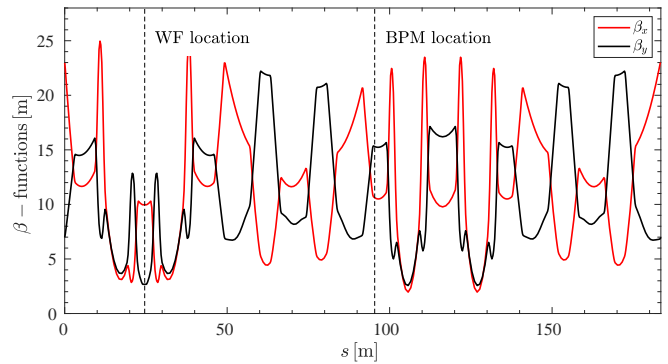}
\caption{Vertical and horizontal beta-functions along the circumference of COSY\,\cite{PAX}. The vertical dashed lines mark the location of the Wien filter and of the beam position monitor used during the measurement of the beam oscillations.}\label{fig:vertBetaFunc}
\end{figure}

In this experiment, the electric field of the Wien filter is oriented vertically and the magnetic field horizontally. This implies that the oscillations mainly take place along the $y$-axis [see Eq.\,(\ref{eq:lorentzForceRed})]. For the detection of the vertical beam oscillations, a conventional beam position monitor has been employed. In order to be most sensitive, BPM\,17 located in the  straight section opposite to the Wien filter (see Fig.\,\ref{fig:cosy}) with a large vertical $\beta$ function was used, $\beta_y^{\rm BPM} \approx \SI{15.3049}{m}$, while at the Wien filter location, $\beta_y^{\rm WF} \approx \SI{2.6784}{m}$, as shown in Fig.\,\ref{fig:vertBetaFunc}. The arguments to pick BPM\,17 are further discussed below in Sec.\,\ref{app:sec:pce}. 

In order to measure small beam oscillations, a technique based on lock-in amplifiers\footnote{HF2LI 50 MHz Lock-in Amplifier, Zurich Instruments AG,  8005 Zurich, Switzerland, \url{https://www.zhinst.com/others/products/hf2li-lock-amplifier}.} was developed\,\cite{meade1983lock}. These devices operate in the frequency domain and lock onto a signal whose frequency is set as a reference, which is particularly useful in an electromagnetically noisy environment. Each measurement consisted of two subsequent machine cycles of $\SI{3}{\minute}$ duration, as depicted in Fig.\,\ref{fig:lockin_data} in Appendix\,\ref{app:technique}. 
\begin{figure}[tb]
\centering
\includegraphics[width=\columnwidth]{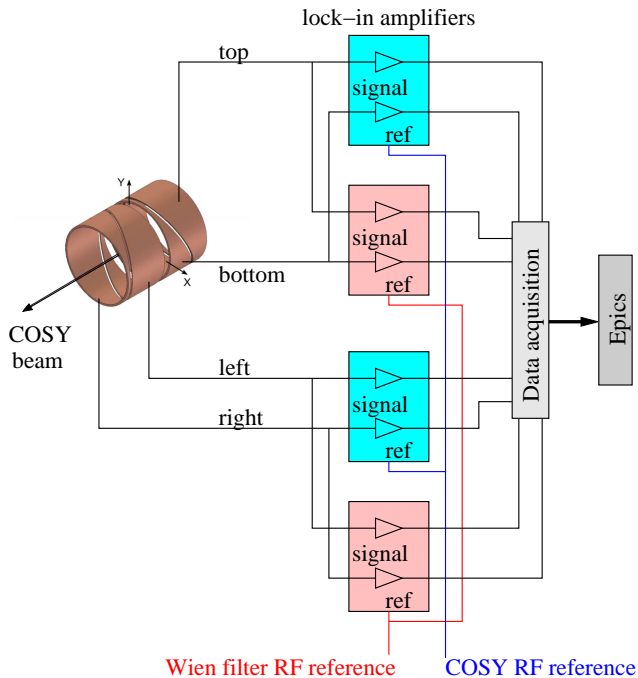}
\caption{Readout scheme of the COSY BPM\,17. The signals of the four electrodes are fed into lock-in amplifiers. The differential signal of each electrode is analyzed at the two reference frequencies given by the COSY RF and the Wien filter frequency. The resulting Fourier amplitudes of the signals are recorded in the EPICS\protect\footnote{Experimental Physics and Industrial Control System, \url{https://epics.anl.gov/index.php}.} archiving system of COSY.}\label{fig:bpm_schematic}
\end{figure}

A stored beam bunch circulating at a revolution frequency of $f_\text{rev}$ that passes through a beam position monitor induces a voltage signal on all its four electrodes, as indicated in the readout scheme of the BPM, shown in Fig.\,\ref{fig:bpm_schematic}. For the detection of vertical beam oscillations, only the voltage signals $U_{\text t,\,b}$ from the top (t) and bottom (b) electrodes are considered. These signals are trains of short pulses with the repetition frequency $f_\text{rev}$. In view of Eq.\,(\ref{eq:beta-function-scaling}) and as far as the Fourier spectrum of the beam oscillations is concerned, without loss of generality, the BPM can be considered to be located right behind the Wien filter, and the induced voltages can be represented by
\begin{equation}
\vspace{0.3cm} 
U_\text{t,\,b} = \left[ U_0 \pm \Delta U\left(\Delta y \right) \right] \cos(\omega_\text{rev}t) \,,
\label{eq:bmp_main}
\end{equation}
where the index 't' refers to the $+$ sign and the index 'b' to the $-$ sign, respectively. The harmonic factor $\cos(\omega_\text{rev}t)$ emphasizes the pulse repetition frequency, although $\cos(\omega_\text{rev}t) = 1$ for  $t=nT$. The voltage $U_\text{t,\,b}$ is non-zero only at the time the beam passes through the BPM. Here, $U_0$ denotes the voltage proportional to the beam current that is induced when the beam passes exactly through the center of the BPM, and $\Delta U\left(\Delta y\right)$ represents the voltage variation induced by a beam that is vertically displaced by $\Delta y$. 


For small beam displacements, the beam position monitor operates in its linear regime, which implies that the induced voltages take the form
\begin{equation}
\Delta U\left(\Delta y\right) = \kappa\cdot \Delta y\cdot  U_0\,,
\label{eq:bpm_lin}
\end{equation}
where $\kappa$ is a calibration factor that needs to be determined. 
At a momentum of $\SI{970}{MeV/c}$, the revolution frequency of deuterons orbiting in COSY is  $f_{\text{rev}} \approx \SI{750}{kHz}$. The Wien filter is operated at the $k^{\text{th}}$ sideband of the spin-precession frequency $f_s$, given by 
\begin{equation}
f_{\text{WF}} = \frac{\omega_\text{WF}}{2\pi} = (k + \nu_s)f_\text{rev}  =  k \cdot f_{\text{rev}} + f_s  \,.
\label{eq:wf_freq}
\end{equation}
Here $\nu_s = G\gamma$ denotes the spin tune, \textit{i.e.}, the number of spin precessions per revolution, $G \approx - \num{0.1430}$ is the magnetic anomaly of the deuteron, and the spin precession frequency $f_s = \nu_s f_\text{rev}$\,\cite{Slim2016116}. 
It should be noted that in view of $\omega_{\rm rev}T =2\pi$, trains of beam oscillation pulses do not depend on the actual choice of the sideband. In the present experiment the Wien filter was operated at $k = -1$, which corresponds to $f_{\text{WF}} \approx \SI{871}{kHz}$ \footnote{For the considerations presented in this paper, negative and positive frequencies are considered equivalent.}.

The induced oscillations of amplitude $\xi_y$ contribute to Eq.\,(\ref{eq:bmp_main}) the harmonic voltage variation $\Delta U\left( y(t) \right)$. The BPM in conjunction with the lock-in amplifiers is used to measure at times $t=nT$ the beam positions at the reference frequencies, \textit{i.e.},  at $f_\text{WF}$ and at $f_\text{rev}$. Given that $y(t)$ can be evaluated at the spin precession frequency, the BPM signals of the upper and lower electrodes can be written as follows 
\begin{widetext}
\begin{equation}
\begin{split}
U_{\text{t,\,b}}(t)  = & \left[ U_0 \pm \Delta U\left(\Delta y\right) \pm  \Delta U\left( y(t) \right) \right] \cos \left( \omega_\text{rev} t\right) =  \left[ U_0 \pm \Delta U\left(\Delta y\right) \pm  
 \kappa \xi_y U_0 \cos \left(\omega_{\text s} t\right) \right] \cos \left( \omega_\text{rev} t\right)  \\
= & \left[U_0 \pm \kappa \Delta y U_0 \right]\cos \left(\omega_\text{rev}t \right)  \pm \frac{1}{2} \kappa \xi_y U_0 \cos \left( \omega_\Delta t\right) \pm  \frac{1}{2} \kappa \xi_y U_0 \cos \left( \omega_\Sigma t \right)\,. 
\end{split}
\label{eq:bpm_top_2}
\end{equation}
\end{widetext}
Here $\omega_\Delta$ and $\omega_\Sigma$ represent sidebands of the Wien filter frequency at 
\begin{equation}
\begin{split}
\omega_\Delta & = \omega_\text{rev} - \omega_\text{s} = \omega_\text{WF}\rvert_{k = 1}\,, \text{and} \\ 
\omega_\Sigma & = \omega_\text{rev} + \omega_\text{s}  = \omega_\text{WF}\rvert_{k = -1}\,.
\end{split}
\end{equation}

\begin{figure*}[tb]
\centering
\subfigure[Magnitude of the field quotient $|Z_q|$, evaluated integrally, where $|Z_q|^\text{int} = \int |Z_q| \dd \ell$. Ideally, with $|Z_q|$ close to $\SI{176}{\ohm}$, the electric and magnetic forces are equal.]
{\includegraphics[width=.49\textwidth]{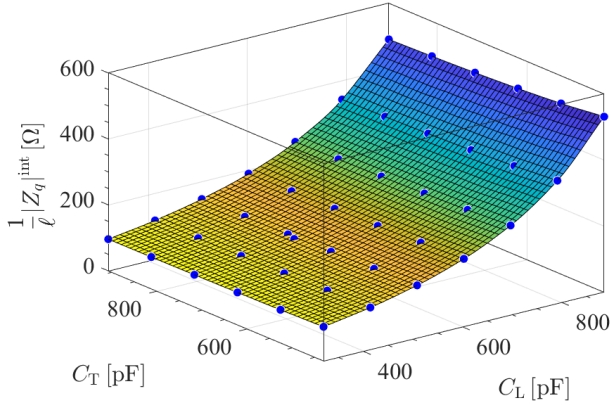}}\hspace{.1cm}
\subfigure[Phase of the field quotient $\protect\angle Z_q$ evaluated integrally, where $\protect\angle Z_q^\text{int} = \int \protect\angle Z_q \dd \ell$. A non-vanishing $\protect\angle Z_q$ implies a phase shift between the electric and magnetic fields.]
{\includegraphics[width=.49\textwidth]{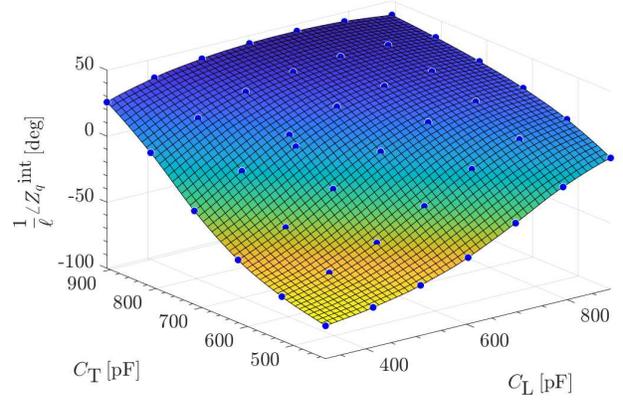}}
\caption{Simulated \textit{integral} magnitude (a) and phase of the field quotient $Z_q$ (b) at each point of the $C_\text{L}$ and $C_\text{T}$ grid, indicated by the blue points, $\ell$ denotes the effective length of the Wien filter. Besides the matching point [see Eq.\,(\ref{eq:matched-point-Cl-CT})], $\left(7 \times 6\right)$ grid points were investigated. }
\label{fig:grid}
\end{figure*}

In order to measure the beam oscillations, four lock-in amplifiers\,\cite{meade1983lock} were used, two for the horizontal and two for the vertical direction. For each direction, one lock-in amplifier detects the Fourier amplitudes at $f_\text{rev} \approx \SI{750}{kHz}$ and a second one at $f_\Sigma  = f_\text{rev} + f_s \approx \SI{871}{kHz}$. The lock-in amplifiers receive reference frequencies from the signal generator of the Wien filter and from the master oscillator of COSY. The four Fourier amplitudes of the top and bottom electrodes are determined practically in real-time, yielding
\begin{equation}
\begin{split}
	A_{\text{t,\,b}}^{\text{rev}}     & = U_0 \pm \kappa \Delta y U_0 \,, \text{and}\\
	A_{\text{t,\,b}}^\Sigma           & = \mp \frac{1}{2} \kappa \xi_y U_0 \,.
\end{split}
\label{eq:A-sum}
\end{equation}
The amplitude of the vertical oscillation $\xi_y$ can then be determined from
\begin{equation}
\frac{A_{\text{t}}^{\Sigma} - A_{\text{b}}^{\Sigma}}{A_{\text{t}}^{\text{rev}} + A_{\text{b}}^{\text{rev}}}= \hat{\xi}_y= \kappa \frac{U_0}{2U_0} \xi_y = \frac{1}{2}\kappa \xi_y\,.
\label{eq:epsilon}
\end{equation}
The uncalibrated raw asymmetry of the four Fourier amplitudes is denoted by $\hat{\xi}_y$. 

The readout scheme, shown in Fig.\,\ref{fig:bpm_schematic}, was used to concurrently record radial beam oscillations, and the above analysis has been repeated for the corresponding $\hat{\xi}_x$. The main result is that the coupling of vertical and radial betatron oscillations is negligibly weak, $|\hat{\xi}_x/\hat{\xi}_y| < \num{2e-2}$, which justifies treating the  RF-driven beam oscillations as one-dimensional.

The determination of the calibration constant $\kappa$, required to calibrate the vertical oscillation amplitude, is described in detail in Appendix\,\ref{app:technique}. It amounts to
\begin{equation}
\kappa = \left(5.82 \pm 0.43 \right) \cdot \SI{e-6}{\micro \meter}^{-1} \,.\label{eq:kappa}
\end{equation}

During the experiments, the vertical betatron tune of the machine amounted to about $\nu_y \approx 3.6040$.\footnote{The numerical values used for the simulation calculations are listed in Table\,\ref{tab:num-val-sim} of Appendix\,\ref{app:beam-simulations}.} The  frequency  $f_\Sigma  \approx \SI{871}{kHz}$, at which the Wien filter is operated, is well separated from the lowest intrinsic spin resonances\footnote{An intrinsic depolarizing resonance is encountered, when the betatron motion of the particles is in sync with the spin motion, hence, when the condition $f_s = \nu_s f_\text{rev} = f_y = (nP \pm \nu_y')f_\text{rev}$ is fulfilled\,\cite{PhysRevSTAB.7.071001}, where $n \in \mathbb{N}$, $P$ denotes the superperiodicity of the lattice, and $\nu_y'$ the fractional tune. During the experiments described here, $P = 1$ (see also Fig.\,\ref{fig:vertBetaFunc}).} at $\SI{297}{kHz}$, $\SI{453}{kHz}$, $\SI{1048}{kHz}$, and $\SI{1204}{kHz}$.

\begin{figure}[tb]
\centering
\includegraphics[width=\columnwidth]{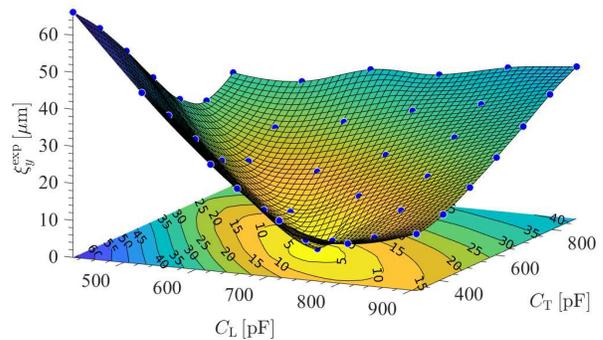}
\caption{Measured amplitudes of beam oscillations $\xi_y^\text{exp}$ at BMP\,17, plotted on a grid as a function of the variable capacitor values $C_{\text{L}}$ and $C_{\text{T}}$. To avoid crowding up the map, the error bars of the data points were omitted,  these are shown in Fig.\,\ref{fig:simMeasBeamOscillations} instead. The parameters of the matching point are given in Eq.\,(\ref{eq:matched-point-Cl-CT}).}
\label{fig:measBeamOscillMap}
\end{figure}

The two variable and highly accurate capacitors, $C_\text{L}$ and $C_\text{T}$, are driven by stepper motors. They constitute the main dynamical elements of the driving circuit. Each pair of capacitor values yields a well-defined field quotient $|Z_q|$, as shown in Fig.\,\ref{fig:grid}\,(a).  Away from the matching point, a phase shift $\protect\angle Z_q$ occurs between electric and magnetic fields, as shown in Fig.\,\,\ref{fig:grid}\,(b). The corresponding Lorentz force leads to the measured beam oscillations, \textit{i.e.}, the function $\xi_y = f\left( C_\text{L}, C_\text{T} \right)$, which can be visualized in the form of a 2D map, as shown in Fig.\,\ref{fig:measBeamOscillMap}.  The experimental data were taken on a grid of  $(7\times 6)$ points  of $C_\text{L}$ and $C_\text{T}$, with corresponding grid spacings of $\left(94.5 \pm 1.0 \right)\,\si{pF}$  for $C_\text{L}$ and $\left(95.8 \pm 1.0 \right)\,\si{pF}$  for $C_\text{T}$. Each grid spacing corresponds to 1000 steps of the corresponding stepper motors. The calibration of the capacitances  $C_\text{L}$ and  $C_\text{T}$ as a function of step number  is discussed in detail in\,\cite{Slim_2020}. The grid spans over $C_{\text L} \in [318.88,\, 885.58]\,\si{\pico \farad}$ and  $C_{\text T} \in [428.99,\, 907.79]\,\si{\pico \farad}$. The uncertainties of the grid spacings are systematic ones.\footnote{The individually measured uncertainties of the capacitors are actually much smaller than the stated uncertainty of \SI{1.0}{pF}. However, other factors, such as the capacitances and inductances of the connectors and cables and their power dependencies, also contribute to the aforementioned uncertainties.}

The map of the measured and calibrated vertical beam oscillations $\xi_y$ is shown in Fig.\,\ref{fig:measBeamOscillMap}.  The parameters of the matching point are given by 
\begin{equation}
\begin{split}
C_{\textnormal{L}} & = \left( 697.1\pm 1.0 \right)\,\si{\pico \farad}\,, \text{ and} \\
C_{\textnormal{T}} & = \left( 503.0\pm 1.0 \right)\,\si{\pico \farad }\,,
\end{split}
\label{eq:matched-point-Cl-CT}
\end{equation}
and the corresponding minimal detected beam oscillation amplitude at the location of BPM\,17 amounts to
\begin{equation}
\xi_y^\text{min}\big\rvert_\text{BPM} = (1.08 \pm 0.52)\,\si{\micro \meter} \,.
\label{eq:eps_y-minimum-BPM17}
\end{equation} 

The above accuracy of $\delta\xi_y^{\rm min}\big\rvert_\text{BPM} = \SI{0.52}{\micro \meter}$ can be compared to the accuracy of measurements of static distortions of the beam orbit, which is about $\SI{20}{\micro \meter}$ (see Table\,\ref{tab:bump_current}). As expected, the accuracy of the beam oscillation amplitudes is better by a factor of about 40  compared to the static amplitudes measured using the same BPM. 	
	
Upon rescaling the oscillation amplitudes using  Eq.\,(\ref{eq:beta-function-scaling}) with $\beta$ functions listed in  Table\,\ref{tab:num-val-sim} to the WF location, we obtain 
\begin{equation}
\xi_y^\text{min}\big\rvert_{\rm WF} = (0.45 \pm 0.22)\,\si{\micro \meter} \,,
\label{eq:eps_y-minimum-WF}
\end{equation} 
which should be compared to the value of $Q\approx \SI{41}{nm}$, given in Eq.\,(\ref{eq:quantum-limit-2}). The largest measured amplitude of driven beam oscillations at a strongly mismatched point with $\SI{600}{W}$ of input RF power amounts to
\begin{equation}
\xi_y^\text{max}\big\rvert_\text{BPM} = (66.2\pm 3.1)\,\si{\micro \meter} \,.
\end{equation}
Here one must bear in mind that the sensitivity to a periodic signal scales inversely with the square root of the observation time.\footnote{The relevant discussion is found in Ref.\,\cite{PhysRevSTAB.17.052803}. See also the observation of white noise suppression by two orders in magnitude when using \SI{5}{\hour} signal averaging in a test bench experiment with SQUID BPMs\,\cite{Haciomeroglu:2018son}.}  The frozen-spin proton EDM experiment aims at the accumulation of the EDM signal for a duration of about $\SI{e7}{\second}$\,\cite{CPEDM}.  The accuracy  $\delta\xi_y^\text{min}\big\rvert_\text{BPM} = \SI{0.52}{\micro \meter}$ [Eq.\,(\ref{eq:eps_y-minimum-BPM17})] corresponds to an averaging time of \SI{96}{s} [see Fig.\,(\ref{fig:lockin_data}) in Appendix\,(\ref{app:technique})]. With the currently used BPMs and their readout electronics, together with an extension of the averaging time to $\SI{e7}{\second}$, there would be a factor of 320 improvement in sensitivity to coherent beam oscillations, leading to an accuracy of $\SI{1.6}{\nano \meter}$.
		
\begin{figure*}[tb]
\centering
\subfigure[\label{fig:epsilonMeasMatched-a}   Measured oscillation amplitudes $\xi_y$ using data samples of \SI{0.5}{\second} duration, each sample reflects the average of \num{5000} measurements of the lock-in amplifiers.]{\includegraphics[height=.27\textheight]{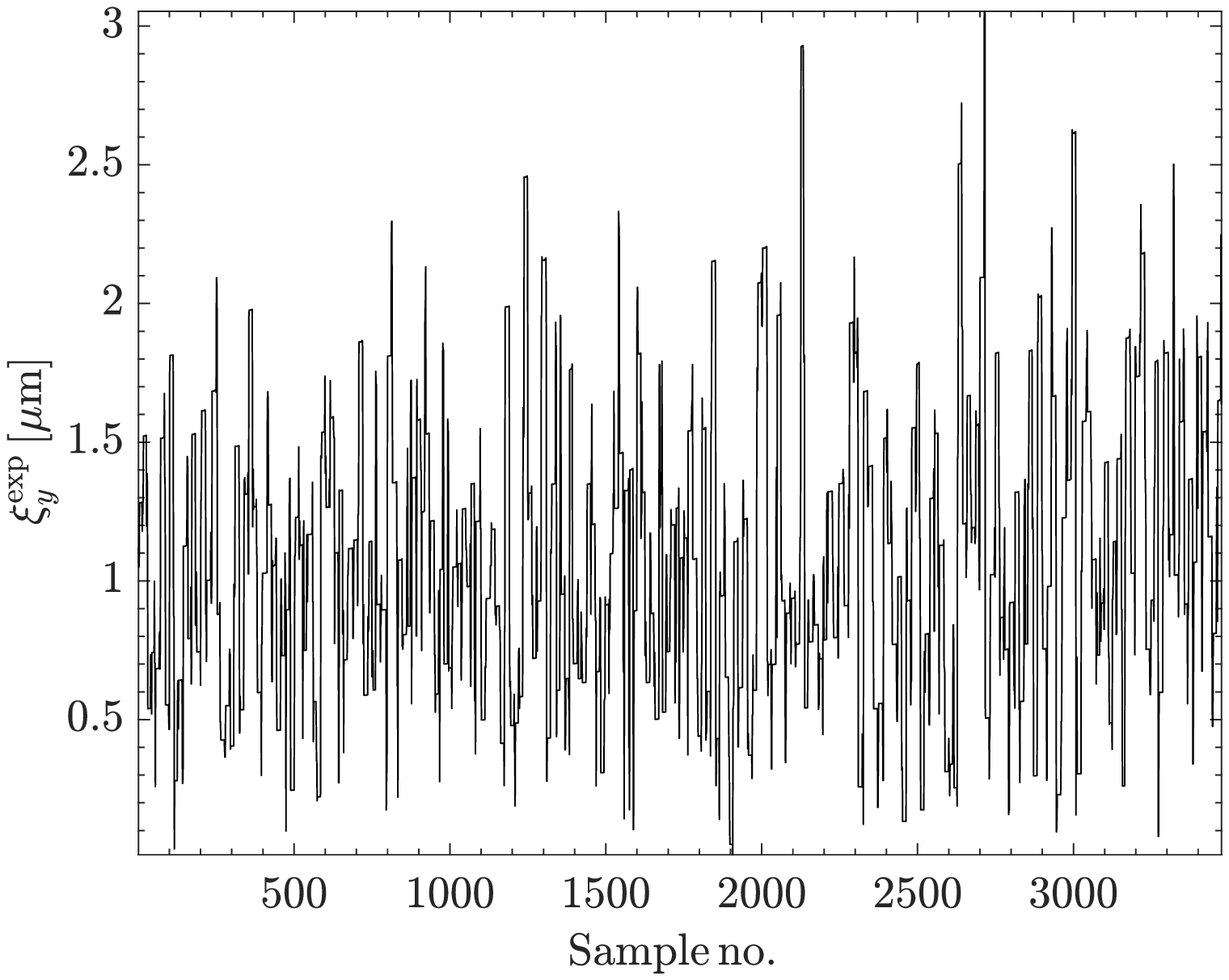}}
\hspace{.3cm}
\subfigure[\label{fig:epsilonMeasMatched-b}   Probability density distribution $f_{\xi_y}$ of the measured data, fitted with a Gaussian to determine mean and standard deviation.]{\includegraphics[height=.27\textheight]{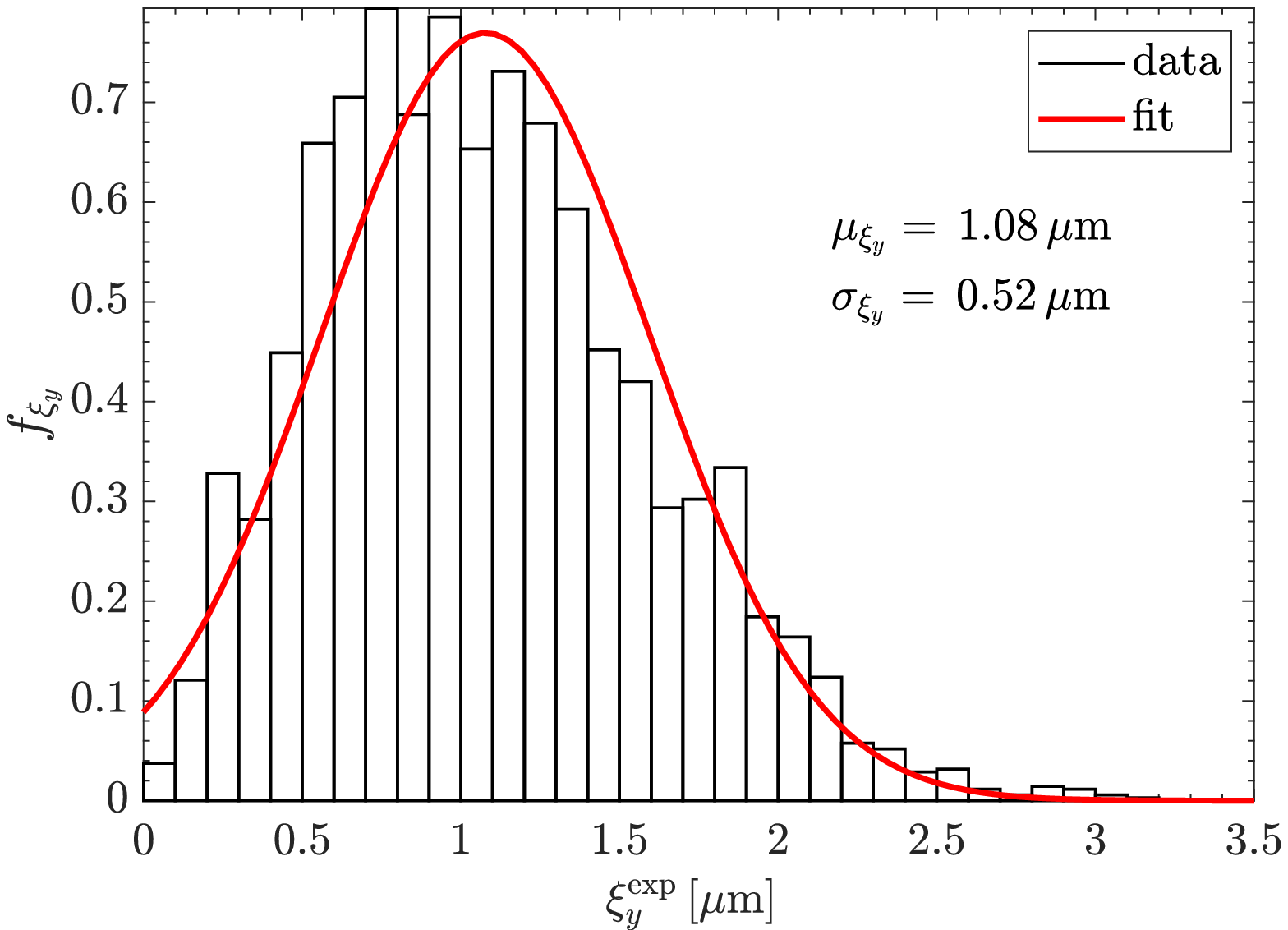}}
\caption{Measured beam oscillations at the matching point [Eq.\,(\ref{eq:matched-point-Cl-CT})] of the map shown in Fig.\,\ref{fig:measBeamOscillMap}. The samples shown in panel (a) were acquired during a data taking period of \SI{108}{\minute}, using 36 machine fills (cycles).}
\label{fig:epsilonMeasMatched}
\end{figure*}

In Fig.\,\ref{fig:epsilonMeasMatched}\,(a), the data measured at the matching point [Eq.\,(\ref{eq:matched-point-Cl-CT})] are shown. Each sample was recorded by the lock-in amplifiers with an integration time set to \SI{0.5}{\second}, corresponding to an average of \num{5000} measurements. A Monte Carlo error propagation model was applied to treat the uncertainties of the still uncalibrated raw position asymmetries $\hat{\xi}_y$ and the calibration coefficient $\kappa$\,\cite{aster2018parameter}. The results are fitted with a normal distribution, as shown in Fig.\,\ref{fig:epsilonMeasMatched}\,(b), from which the mean value $\mu_{{\xi}_y}$ and the error of the measured beam oscillations $\sigma_{{\xi}_y}$ are estimated. The latter represents the \textit{systematic} error of the measurement. It should be noted that the map shown in Fig.\,\ref{fig:measBeamOscillMap} is actually a function of \textit{all} the circuit elements. The uncertainties of $\xi_y$ are influenced by  the uncertainties of all circuit elements and also by the ones of the BPM itself, which include its readout electronics, \textit{i.e.}, the lock-in amplifiers. 

To appreciate the result given in Eq.\,(\ref{eq:eps_y-minimum-BPM17}), one can compare the oscillation amplitude to the $1\sigma$ vertical beam  size. The latter has been deduced from the $1\sigma$ beam emittance $\epsilon_y$ and the amplitude of the $\beta$ function at the position of the BPM, yielding
\begin{equation}
\sigma_y^\text{BPM} = \sqrt{\beta_y^\text{BPM}\epsilon_y} \approx \SI{1.4}{mm} \,.
\end{equation}
In the present experiment, the beam emittance was not monitored.  The above  numerical estimate of $\sigma_y^\text{BPM}$ is based on rescaling the experimental result for the $2\sigma$ beam emittance of \SI{49.3}{MeV} protons in COSY of $\epsilon_y=(0.92\pm 0.15) \, \si{\micro\meter}$\,\cite{PAX} to the conditions of the present experiment. It is noteworthy that with the present equipment it is possible to access coherent beam oscillations with amplitudes that are more than three orders of magnitude smaller than the beam size.

\section{Beam dynamics simulations}
\label{app:sec:pce}

To improve our understanding of the measured results, a computer code was developed to model the beam dynamics in the COSY storage ring. The modeled storage ring consists of a sequence of drift regions, quadrupole and dipole magnets, the Wien filter, and beam position monitors. These elements are represented by transfer matrices, which are well understood and documented in the literature\,\cite{doi:10.1142/p899}. In the model of the ring, the actual settings of the beam optics elements of COSY were those used at the time when the experiment took place. Simulations are based on the Hamiltonian formulation as presented in Ref.\,\cite{doi:10.1142/p899}. The Wien filter is modeled by a time-dependent matrix that also takes into account the arrival time of the particles. 

\subsection{Time-dependent Wien filter field maps}
\label{sec:time-dependent-field-maps}

In order to be able to perform reliable beam simulations, we have placed great emphasis on good spatial resolution and the accuracy of the 3D field maps inside the Wien filter\footnote{Each field map consists of \num{2e6}points, 200 points along the $x$ axis $\left( x\in[\SI{-5}{mm}, \SI{+5}{mm}] \right)$, 200 points along the $y$ axis $\left(y\in[\SI{-5}{mm}, \SI{+5}{mm}] \right)$, and 50 points along the $z$ (Wien filter) axis $\left(z \in [-\ell/2, +\ell/2] \right)$, where $\ell = \SI{1.16}{m}$ is the effective length of the Wien filter.}, computed using a 3D electromagnetic simulation tool\footnote{Electromagnetic and circuit simulations were performed using CST, from Dassault Syst\`emes, V\'elizy-Villacoublay, France, \url{https://www.3ds.com}.}.
The fringe fields of the Wien filter are included, because they are of particular importance for the beam oscillations, as will be discussed later. 
\begin{figure*}[tb]
\centering
\subfigure[3D electric field distribution of the component $E_y$.]{\includegraphics[width=.49\textwidth]{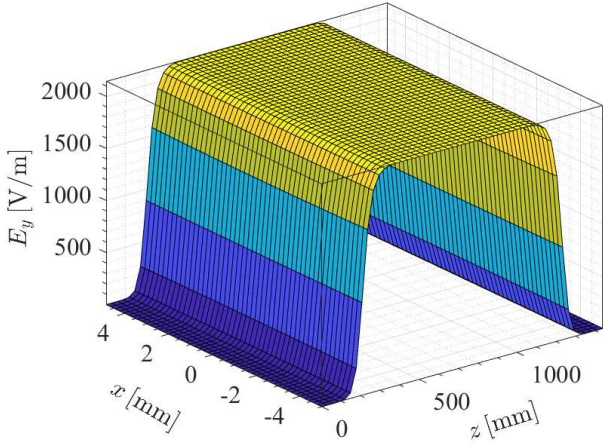}}
\hspace{.05cm}
\subfigure[3D magnetic field distribution of the flux density component $B_x$.]{\includegraphics[width=.49\textwidth]{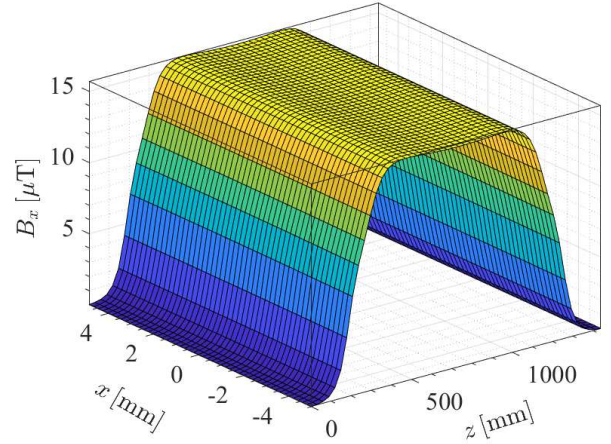}}
\caption{Examples of the main electric and magnetic field components inside the waveguide RF Wien filter at the matching point [see Eq.\,(\ref{eq:matched-point-Cl-CT})] with an input RF power of $\SI{600}{W}$. The electric field component in (a) points vertically upward ($y$-direction), while the component of the magnetic flux density in (b) points radially outward ($x$-direction).}
\label{fig:3DFields}
\end{figure*}
An example of the computed 3D fields of the Wien filter at the experimentally determined matching point is shown in Fig.\,\ref{fig:3DFields}. The beam-tracking simulations use the three vector components of the electric and magnetic fields. The Wien filter is implemented as an RF kicker, as described by Eq.\,(\ref{eq:velKick}). 


Inside COSY there are 32 BPMs available to control the horizontal and vertical beam position during operation. In order to select one of them with a good sensitivity to determine the beam oscillations induced by the Wien filter, a number of particles were tracked, as described above, and the orbit response induced by a field change at the location of the Wien filter was calculated at each BPM location\footnote{In the preparatory stage, simulations were carried out using the Software Toolkit for Charged-Particle and X-Ray Simulations BMAD\,\cite{Sagan:Bmad2006}.}. As a result, BPM\,17,  located about $\SI{70}{m}$ downstream of the Wien filter (see Fig.\,\ref{fig:cosy}), was chosen because it offered good sensitivity to both radial and vertical beam oscillations.

Figure\,\ref{fig:3DFields} shows the main field components $E_y$ and $B_x$. When mismatched, the Wien filter generates periodic transverse perturbations of the trajectory. Switching off the Wien filter eliminates such oscillations. The maximum amplitude of the observed oscillations in the simulation is then taken for  $\xi_y$. The two simulated vertical beam oscillation amplitudes of BPM\,17 and Wien filter read
\begin{equation}
\begin{split}
\xi_y^\text{BPM} & =  \left(1.086 \pm 0.082\right)\,\si{\micro \meter}\,, \text{ and} \\
\xi_y^\text{WF}  & =  \left(0.435 \pm 0.031\right)\,\si{\micro \meter}\,, 
\end{split}
\label{simulated_Amplitudes}
\end{equation}
which agree well with the experimentally measured results, given in Eqs.\,(\ref{eq:eps_y-minimum-BPM17}) and (\ref{eq:eps_y-minimum-WF}). A detailed description of the determination of the uncertainties of the beam simulations is discussed in the next section.

\begin{figure}[tb]
\centering
\includegraphics[width=\columnwidth]{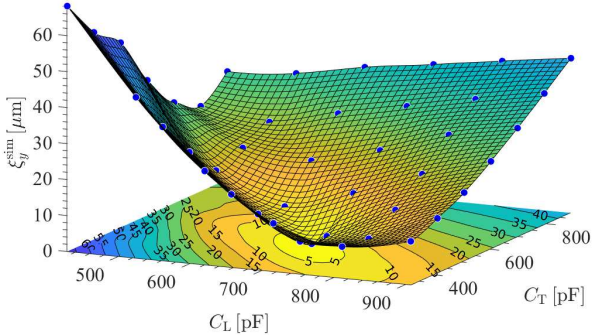}
\caption{Simulated amplitudes of beam oscillations $\xi_y^\text{sim}$ as a function of the variable capacitor values $C_\text{L}$ and $C_\text{T}$. To avoid overcrowding the map, the error bars of the data points were omitted here and are shown in Fig.\,\ref{fig:simMeasBeamOscillations}.}
\label{fig:simBeamOscillMap}
\end{figure}

For each and every \textit{measured} point on the $C_\text{L}$ versus $C_\text{T}$ grid, a beam dynamics simulation was carried out. For each of these points, a 3D field map of the Wien filter was generated and then used for the beam tracking simulations. 
The results of these simulations are shown in Fig.\,\ref{fig:simBeamOscillMap}, and are later compared with the results of the measurements. 

\subsection{Uncertainty evaluation}
\label{sec:eval-of-uncertainties}

\begin{figure*}[tb]
\centering
\subfigure[\label{fig:fieldsUncMatched-a} Electric field component $E_y(z)$ under circuit uncertainties.]{\includegraphics[width=.49\textwidth]{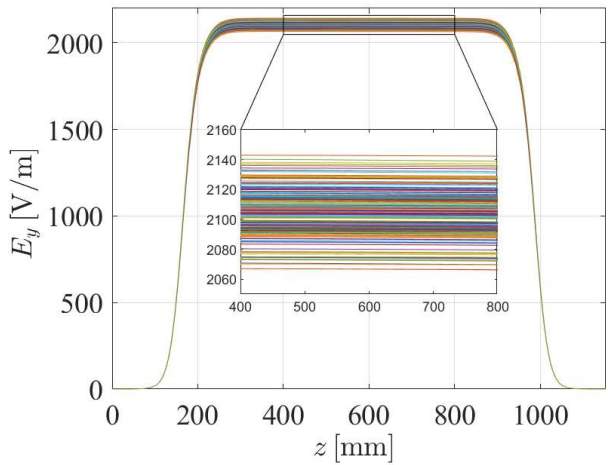}}
\hspace{.05cm}
\subfigure[\label{fig:fieldsUncMatched-b} Magnetic field component $B_x(z)$ under circuit uncertainties.]{\includegraphics[width=.49\textwidth]{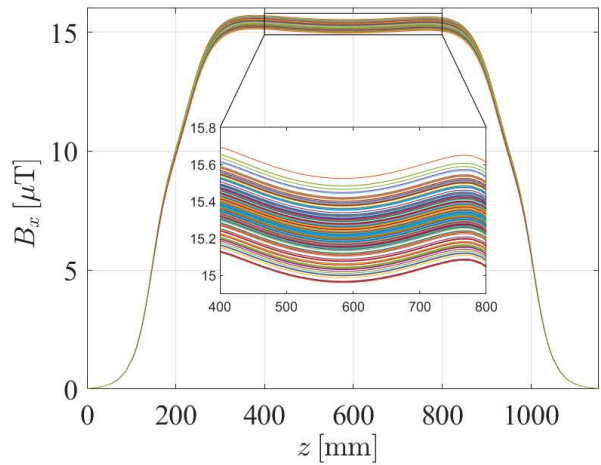}}
\caption{\num{200} examples of the electric and magnetic fields as a function of $z$ along the beam axis under the circuit uncertainties, specified in the list of uncertainties in Table\,6 of\,\cite{Slim_2020}. 
}
\label{fig:fieldsUncMatched}
\end{figure*}

The accuracy with which the Lorentz force and the resulting amplitudes of the beam oscillations can be tuned depends on the accuracy with which the field quotient $Z_q$  can be \textit{integrally} set to the desired value. $Z_q$ depends on the hardware elements in the driving circuit. In order to evaluate the effects of uncertainties of these elements, extensive coupled circuit electromagnetic simulations have been conducted, as discussed in Ref.\,\cite{Slim_2020}. The uncertainties involved are listed in Table\,6 and shown in Fig.\,16 of Ref.\,\cite{Slim_2020}. As far as the Lorentz force is concerned, most important are the uncertainties of the fixed inductance $L_\text{f}$ and the fixed resistance $R_\text{f}$. Once these uncertainties are known, one can compute the electric and magnetic fields  and the corresponding Lorentz force, \textit{including} their corresponding errors. Figure\,\ref{fig:fieldsUncMatched} shows a few examples of the main components of the electric and magnetic fields, computed with the above mentioned circuit uncertainties. As will be explained below, these 3D fields, together with their uncertainties, are subsequently used as input to the beam simulations.
%

The algorithm used to compute the uncertainties of the beam simulations is the polynomial chaos expansion (PCE), as explained in Refs.\,\cite{slim201752,Slim_2020} and in Appendix\,\ref{app:pce}. 
The PCE has been proven in many applications in science and engineering to be just as accurate as the computationally much more expensive Monte-Carlo counterpart\,\cite{10.5555/2568154,7093523,adelmann2018uncertainty, slim201752}.

To compute the uncertainties $\sigma_{\xi_y}$, the PCE algorithm requires a random set of the simulated $\xi_y$, alongside a set of randomized input parameters according to their uncertainties to generate the output. The set of $\xi_y$ is produced using  a number of beam-tracking simulations, where for each instance, a 3D field map of the Wien filter is generated, according to the randomized input parameters. An example of the electric and magnetic fields evaluated at the center of the Wien filter for the matching case [see Eq.\,(\ref{eq:matched-point-Cl-CT})] is shown in Fig.\,\ref{fig:fieldsUncMatched}. The magnitudes of the fields vary as a function of the uncertainties of the driving circuit\,\cite{Slim_2020}. The numerical tracking of the particles through these fields generates a collection of different $\xi_y$ values that the PCE algorithm can use to project the output onto orthogonal polynomial functions. These functions serve as basis functions, from which the expansion coefficients are determined that are used to generate a large sample of outputs to compute the uncertainties of the beam simulations.

\begin{figure*}[tb]
\centering
\subfigure[\label{fig:epsilonUncMatched-a}  Simulated oscillation amplitudes under uncertainties at the matching point (Eq.\,(\ref{eq:matched-point-Cl-CT})). Of the $\protect\num{e6}$ simulations that were carried out, only $\protect\num{e4}$ are shown here.]{\includegraphics[height=.26\textheight]{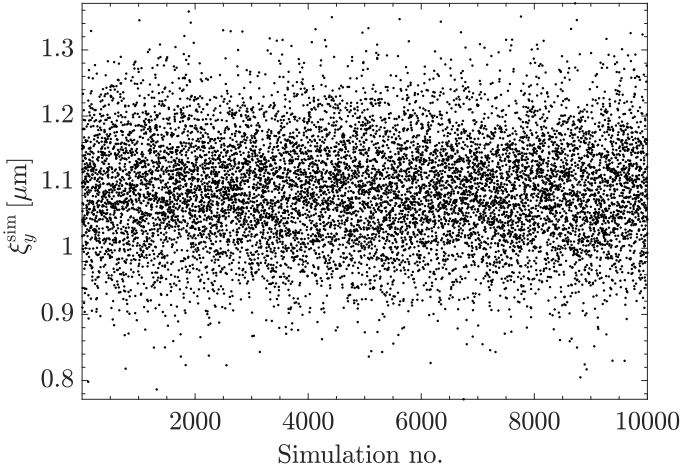}} \hspace{.3cm}
\subfigure[\label{fig:epsilonUncMatched-b}   Probability density distribution $f_{\xi_y}$ of the $\protect\num{e6}$ simulations from panel (a), fitted by a Gaussian to determine mean and standard deviation.]
{\includegraphics[height=.26\textheight]{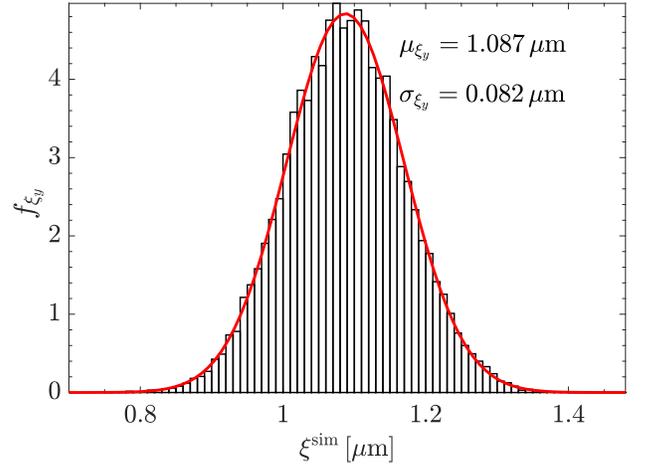}}
\caption{Results of the sparce PCE algorithm to compute the uncertainties of the simulated vertical beam oscillations at BPM\,17.}
\label{fig:epsilonUncMatched}
\end{figure*}

In Fig.\,\ref{fig:epsilonUncMatched}\,(a), the simulated values of $\xi_y$ are shown for the matching case. The detailed steps to achieve this result are discussed in Appendix\,\ref{app:pce}. As shown in Fig.\,\ref{fig:epsilonUncMatched-b}, fitting these data to a Gaussian yields a standard deviation of $\sigma_{\xi_y} = \SI{0.082}{\micro \meter}$. This number is of considerable importance, because, given the uncertainties of the driving circuit, it sets the lower limit that can be achieved by minimizing the amplitude of the vertical beam oscillations when more sharply tuning the driving circuit of the Wien filter. The same procedure is performed on each point of the map shown in Fig.\,\ref{fig:simBeamOscillMap}.   

\section{Comparison of simulation and experimental results}
\label{sec:discussion}

\begin{figure*}[tb]
\centering
\subfigure[\label{fig:FLSim-a} Local Lorentz force $F_y(z)$ exerted on a single deuteron for different passes though the Wien filter. The turn numbers used here were randomly selected between 1 to 100. The fields were evaluated at the crosses and the interconnecting lines are to guide the eye.]{\includegraphics[width=0.48\textwidth]{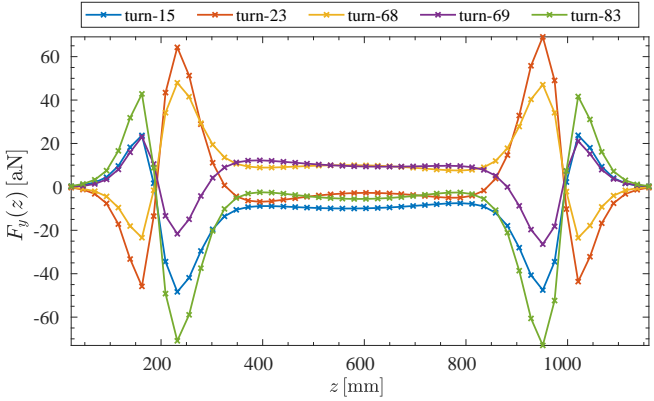}}
\hspace{0.2cm}
\subfigure[\label{fig:FLSim-b} Integral Lorentz force $F_y(n)$ evaluated along the trajectory. Each point represents an overall kick exerted  per turn $n$. The points marked in blue correspond to the integrated local Lorentz force of the individual turns shown in panel (a).]{\includegraphics[width=0.48\textwidth]{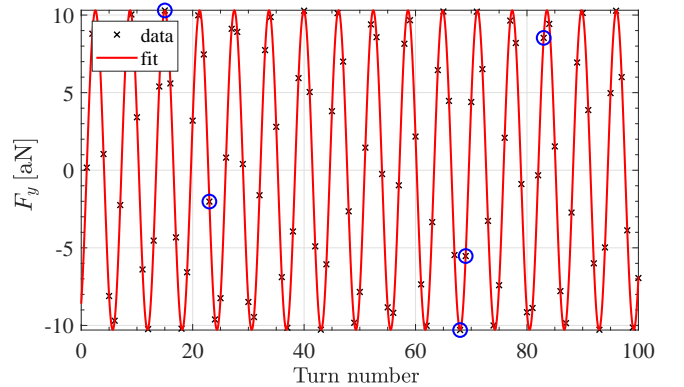}}
\caption{  Simulation of the local and integrated Lorentz force in the Wien filter at the matching point of Eq.\,(\ref{eq:matched-point-Cl-CT}). Depending on the initial coordinates in the vertical and horizontal phase space, the particle travels along different trajectories, and therefore picks up different field components $F_y$.}
\label{fig:FLSim}
\end{figure*}

The simulations yield the net Lorentz force exerted by the Wien filter on beam particles and the corresponding oscillation amplitudes for each measured  point of the $C_\text{L}$ versus $C_\text{T}$ map, shown in Fig.\,\ref{fig:measBeamOscillMap}. The only variables in this case are the field maps of the Wien filter itself. After \num{1000} turns, the beam position is computed at the same location in the ring, where the measurement using BPM\,17 took place (see Fig.\,\ref{fig:cosy}). The net Lorentz force is a result of local cancellations between the electric and  magnetic field components, as illustrated in Fig.\,\ref{fig:FLSim} for the matching point given in Eq.\,(\ref{eq:matched-point-Cl-CT}) with the minimal measured oscillation amplitude.

In Fig.\,\ref{fig:FLSim-a},  the \textit{local} Lorentz force is shown  along the trajectory for 5 randomly chosen passes though the Wien filter. The trajectory of the same particle changes from pass to pass, thereby different Wien filter fields and consequently different values of the Lorentz force $F_y$  will be picked up. As shown in Fig.\,\ref{fig:FLSim}, even at the matching point, the matching is still imperfect, and the largest local $F_y$ contributions are caused by the fringe fields at the entrance and exit of the Wien filter. Despite the different location of the particle in the vertical and horizontal phase space at the entrance of the Wien filter upon subsequent passes, the integration of these local forces along the particle trajectories exhibits nevertheless a perfectly harmonic time dependence with the frequency $f_s$, as shown in Fig.\,\ref{fig:FLSim-b}. The points encircled in blue correspond to the randomly selected  passes through the Wien filter, shown in Fig.\,\ref{fig:FLSim-a}.  

\begin{figure*}[tb]
\centering
\subfigure[\label{fig:simMeasBeamOscillations-a} Simulated Lorentz force {$F_y^\text{sim}$} at the Wien filter location as function of the oscillation amplitude $\xi_y^\text{sim}$, fitted with the function $F_y^\text{sim} = a \cdot \xi_y^\text{sim} + b$.]{\includegraphics[height=.34\textheight]{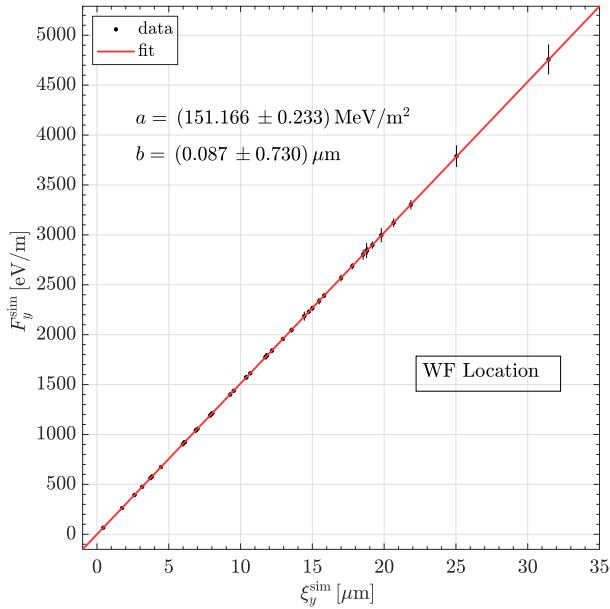}} \hspace{0.2cm}
\subfigure[\label{fig:simMeasBeamOscillations-b} Simulated beam oscillation amplitude $\xi_y^\text{sim}$ versus the measured oscillation amplitude $\xi_y^\text{exp}$ at the BPM, fitted with the function $\xi_y^\text{sim} = c \cdot \xi_y^\text{exp} + d$.]{\includegraphics[height=.34\textheight]{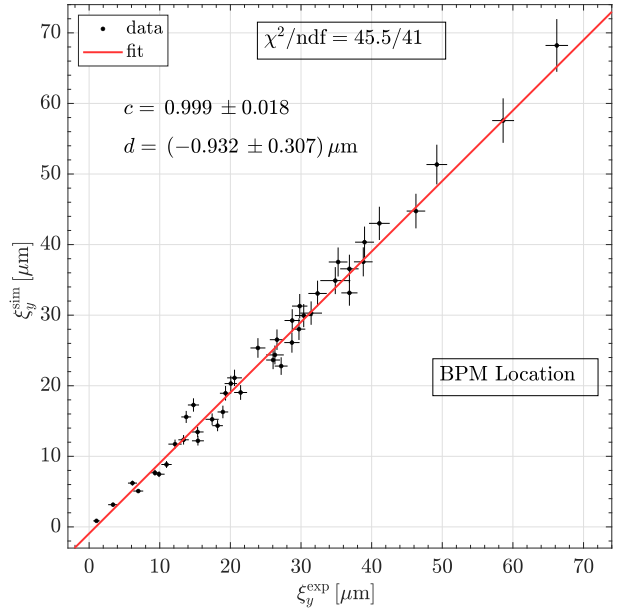}} \hspace{0.2cm}
\caption{(a): Simulated amplitude of the Lorentz force at the Wien filter location as function of the simulated beam oscillation amplitudes  $\xi_y^\text{sim}$. (b): Simulated versus measured vertical beam oscillation amplitudes at the location of BPM\,17. The horizontal error bars of the measured amplitudes $\xi_y^\text{exp}$ originate from the readout electronics of BPM\,17 and the calibration factor $\kappa$ (see Appendix\,\ref{app:technique}), whereas the vertical ones are determined by the circuit uncertainties using the PCE method, as described in Appendix\,\ref{app:pce}. }
\label{fig:simMeasBeamOscillations}
\end{figure*}

In the left panel of Fig.\,\ref{fig:simMeasBeamOscillations}, the amplitude of the simulated Lorentz force $F_y^\text{sim}$ is plotted versus the simulated oscillation amplitude $\xi_y^\text{sim}$ at the Wien filter position. As expected, it exactly follows Hooke's law with a spring constant of 
$k_\text{H} = (151.2 \pm 0.2)\,\si{MeV/m^2}$. In Fig.\,\ref{fig:simMeasBeamOscillations-b}, the measured amplitudes are compared with the ones simulated for the location of BPM 17. The two sets $\xi_y^\text{exp}$ and $\xi_y^\text{sim}$ are in very good agreement with each other. The horizontal and vertical error bars are derived from the uncertainties  of the measurements and simulations, represented by the width of the distributions, as shown in Figs.\,\ref{fig:epsilonMeasMatched-b} and \ref{fig:epsilonUncMatched-b}. It is important to note that the error bars refer to \textit{systematic uncertainties} and should not be confused with statistical ones. This implies that repetitions of either the measurements or the simulations will neither reduce the systematic error of the readout electronics of BPM\,17, nor will it affect the uncertainties of the elements of the driving circuit. 

The fit shown in Fig.\,\ref{fig:simMeasBeamOscillations} yields $\chi^2/\text{ndf} = 45.5 / 41$, very close to unity\,\cite{doi:10.1119/1.1632486}. The linear fit yields a slope of $0.999\pm 0.018$, which is perfectly consistent with unity.  The intercept parameter of the fit yields 
$(\num{-0.93} \pm 0.31) \, \si{\micro \meter}$, 
and within three standard deviations, it agrees with zero.

The very good agreement between measurements and simulations reflects our good understanding of both the electromagnetic fields generated in the Wien filter and of the underlying beam dynamics in the machine. 
This point is further substantiated by comparing the simulated amplitudes at the Wien filter and at the positions of the BPMs with the estimated amplitudes expected from rescaling based on the $\beta$ functions\footnote{The uncertainty of the $\beta$ functions amounts to about 10\%, as discussed in Ref.\,\cite{PAX}.}, taking into account the numerical values, listed in Table\,\ref{tab:num-val-sim} of Appendix\,\ref{app:beam-simulations},
\begin{equation}
\begin{split}
\left. \xi_y^{\text{WF}}\right|_{\text{sim}} & = (0.435 \pm 0.031) \,\si{\micro \meter} \,,\\
\sqrt{\frac{\beta_y^\text{WF}}{\beta_y^\text{BPM}}}{ \xi_y^{\text{BPM}}}|_\text{sim} =  \left.\xi_y^{\text{WF}}\right|_\text{est}   & = (0.435\pm 0.039)\, \si{\micro \meter} \,.
\end{split}
\label{eq:amplScaling}
\end{equation} 
The good agreement between these two numerical values in Eq.\,(\ref{eq:amplScaling}) indicates that the observation of the oscillation amplitude at one location in the ring can be reliably transferred to some other place in the ring by use of Eq.\,(\ref{eq:beta-function-scaling}). The above quoted value of $\left.\xi_y^{\text{WF}}\right|_\text{sim} = 0.435\,\si{\micro \meter}$ is about a factor of 10 larger than the quantum limit of the vertical oscillation amplitude $Q$, given in Eq.\,(\ref{eq:quantum-limit-2}). 

In searches for EDMs in dedicated all-electric storage rings, a continuous monitoring of the orbits of the two counter-rotating beams is mandatory during data acquisition within the horizontal spin-coherence time\,\cite{CPEDM}. When intrabeam scattering can be neglected\,\cite{PAX}, which is arguably justified within the horizontal spin-coherence time, the beam can be described as a rarefied gas of particles, \textit{i.e.}, the zero-point oscillations of individual particles are uncorrelated. 
We repeat the point from the introduction that in a static regime, the quantum limit of the center of mass of a bunch with $N$ particles can be estimated via $Q_\text{bunch} = Q/\sqrt{N}$. For a bunch of $N = \num{e10}$ stored particles, one obtains $Q_\text{bunch} \simeq \SI{0.4}{\pico \meter}$. It follows that Heisenberg's uncertainty relation does not present an obstacle to achieving a sensitivity of \SI{5}{pm} for the vertical separation of clockwise and counter-clockwise beams -- the real challenge is to develop compact BPMs with a sensitivity improved by a factor of about 300 compared to those used here\,\cite{Bohme:2018sjy}.

Finally, a satisfactory agreement has been achieved between Hooke's constant, simulated using the electromagnetic fields in the Wien filter and the $\beta$ functions of the COSY lattice,  and the theoretical approximation of the no-lattice model assuming constant $\beta$ functions of Eq.\,(\ref{eq:Hooke}), yielding 
\begin{equation}
\begin{split}
 k_\text{H}^\text{sim} & = (151.2 \pm 0.2) \,\si{\mega \electronvolt \per \meter \squared}\,, \text{  and}\\
 k_\text{H}^\text{th} & = \SI{207}{\mega \electronvolt \per \meter \squared}\,.
\end{split}
\end{equation}
The theoretical estimate of $k_\text{H}^\text{th}$, calculated using the numerical values listed in Table\,\ref{tab:num-val-sim} of Appendix\,\ref{app:beam-simulations}, is about a factor of 1.4 larger than  the simulated one. The given uncertainty of $k_\text{H}^\text{sim}$ does not include the systematic scale uncertainty of the BPM calibration factor $\kappa$ [see Eq.\,(\ref{eq:calibration-factor-kappa}) in Appendix\,\ref{app:beam-simulations}]. At the matching point [see Eq.\,(\ref{eq:matched-point-Cl-CT})], the Lorentz force amounts to 
\begin{equation}
F_y^\text{WF} = k_\text{H}^\text{sim} \cdot \left. \xi_y^{\text{WF}}\right|_\text{sim} 
\approx 66\,\si{\electronvolt \per \meter} = {10.6}\,\si{\atto \newton} \,,
\end{equation} 
where the intercept parameter has been ignored because of its smallness.

\section{Conclusion and outlook }
\label{sec:conclusion}

As part of several studies to investigate the performance of the waveguide RF Wien filter, exploratory data were taken to provide a benchmark on the sensitivity to very weak collective vertical beam oscillations of deuterons stored in the COSY ring. To a good approximation, the beam can be viewed as a rarefied gas of uncorrelated particles, and the sensitivity limit is applicable to the classical motion of individual particles, propagating along the ring circumference in the confining oscillatory potential. Simulations of the beam dynamics in the COSY ring equipped with an RF Wien filter suggest that with the present apparatus, the sensitivity to collective beam oscillations on the sub-micron level is only a factor of about 10 larger than the amplitude of single-particle zero-point quantum oscillations of the stored deuterons. From the perspective of future EDM experiments, our finding confirms that, as far as the Heisenberg uncertainty relation is concerned, a separation of the centroids of two counter-propagating beams may be determined to sub-picometer accuracy.

The reported excellent agreement between simulated and experimentally observed vertical beam oscillations at COSY suggests that a further increase in sensitivity to collective beam oscillations is possible. Specifically, the simulation on finer capacitor grids indicates that by further optimization of the Wien filter settings to $C_{\textnormal{L}} =  \left( 692.76 \pm 1.00\right) \si{pF}$ and $C_{\textnormal{T}} = \left( 495.77 \pm 1.00\right) \si{pF}$, an oscillation amplitude at the Wien filter location of $\xi_y = \left(0.077 \pm 0.032\right) \si{\micro m}$ may be achieved. Thus in that case, the vertical oscillation amplitude would only be about a factor of 2 away from the quantum limit, with a corresponding Lorentz force of $ F_y \sim \SI{3}{\atto \newton}$.

\section*{Acknowledgments}
We would like to thank I. Bekman, B. Breitkreutz, J. Hetzel and R. Stassen for their support in setting up the COSY accelerator for the experiment. The work presented here has been performed in the framework of the JEDI collaboration and is supported by an ERC Advanced Grant of the European Union (proposal number 694340). In addition, it was supported by the Russian Fund for Basic Research (Grant No. 18-02-40092 MEGA) and by the 
Shota Rustaveli National Science Foundation of the Republic of Georgia (SRNSFG Grant No.\ DI-18-298: \textit{High precision polarimetry for charged-particle EDM searches in storage rings}).

\appendix

\section{Quantities used in beam simulations}
\label{app:beam-simulations}

In order to provide a consistent calculation of all effects in the storage ring, the beam simulations were carried out using the set of quantities given in Table\,\ref{tab:num-val-sim} as an input. The vertical machine tune $\nu_y$ is a result of simulations with the known COSY lattice, reflecting the actual currents of the magnetic elements in the machine at the time when the experiment was conducted. The simulations provide the uncalibrated parameters of the vertical beam oscillations to about per mill accuracy, and giving the kinematic, ring, and Wien filter parameters to four digits appears therefore sufficient. It should be noted that within the simulation calculations carried out in the context of the present work, all quantities have been computed to double precision (machine epsilon of \num{1.11e-16}). Of the physical quantities, the highest sensitivity to the vertical betatron tune is exhibited by the theoretical estimate for Hooke's constant, $\dd k_\text{H}^\text{th}/\dd\nu_y \approx \SI{2e3}{MeV/m^2}$. The largest uncertainty contributing to the error of the detected oscillation amplitudes arises from the  calibration factor $\kappa$ of the beam position monitor, given in Eq.\,(\ref{eq:calibration-factor-kappa}). It amounts to about $7.3\%$ and is considered a systematic scale-factor uncertainty (see Appendix\,\ref{app:technique}).
\begin{table}[tb]
\caption{\label{tab:num-val-sim} Numerical values used for the beam simulations. The genuinely independent input parameters are listed in bold face. The derived quantities are displayed in normal font and are truncated to four decimal places.} 
\centering
\begin{ruledtabular}
\renewcommand{\arraystretch}{1.2}
\begin{tabular}{rlr}
   Quantity					& Symbol	   					& Value \\\hline
   deuteron beam momentum			& $\bm{p}$						& \SI{970.0000}{MeV/c} \\
   deuteron mass				& $\bm{m}$ 						& \SI{1875.6128}{MeV/c^2} \\
   deuteron $G$ factor				& $\bm{G}$						& \num{-0.1430}\\
   Lorentz factor				& $\beta$						& \num{0.4594}\\
   Lorentz factor				& $\gamma$						& \num{1.1258} \\ \hline
   COSY circumference 				& $\bm{L_\text{COSY}}$  					& \SI{183.4728}{m} \\
   revolution frequency		 		& $f_\text{rev}$   					& \SI{750603.7600}{Hz}\\
   vertical machine tune			& $\nu_y$						& \num{3.6040}\\
   vertical $\beta$ function at BPM\,17		& $\beta_y^\text{BPM}$					& \SI{15.3049}{m}\\  
   vertical $\beta$ function at WF		& $\beta_y^\text{WF}$					& \SI{2.6784}{m} \\   \hline
   effective length  WF 			& $\bm{\ell}$ 	   					& \SI{1.1600}{m}\\
   frequency  WF 				& $\bm{f_\text{WF}}$    					& \SI{871000.0000}{Hz} \\
   tune WF					& $\nu_\text{WF}$
   & \num{1.1604} \\
%
\end{tabular}
\end{ruledtabular}
\end{table}

\section{Calibration of beam position monitor}\label{app:technique}

\begin{figure*}[tb]
\centering
\subfigure[Real ($X^\Sigma$) and imaginary part ($Y^\Sigma$) of the complex Fourier amplitudes $A^\Sigma$ at the Wien filter frequency.]{\includegraphics[width=0.49\textwidth]{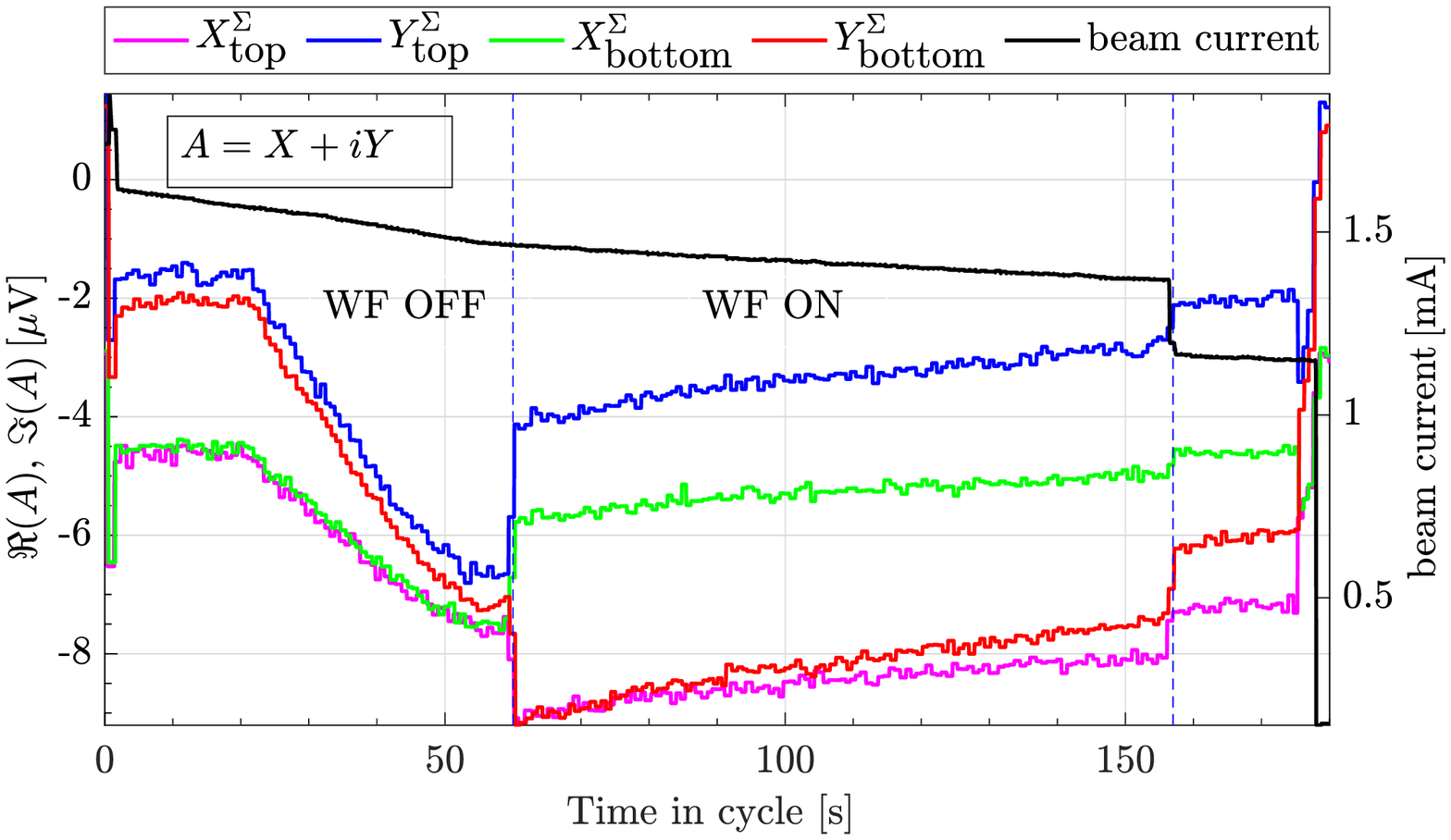}}
\hspace{0.1cm}
\subfigure[Real ($X^\text{rev}$) and imaginary part ($Y^\text{rev}$) of the complex Fourier amplitudes $A^\text{rev}$ at the revolution frequency.]{\includegraphics[width=0.49\textwidth]{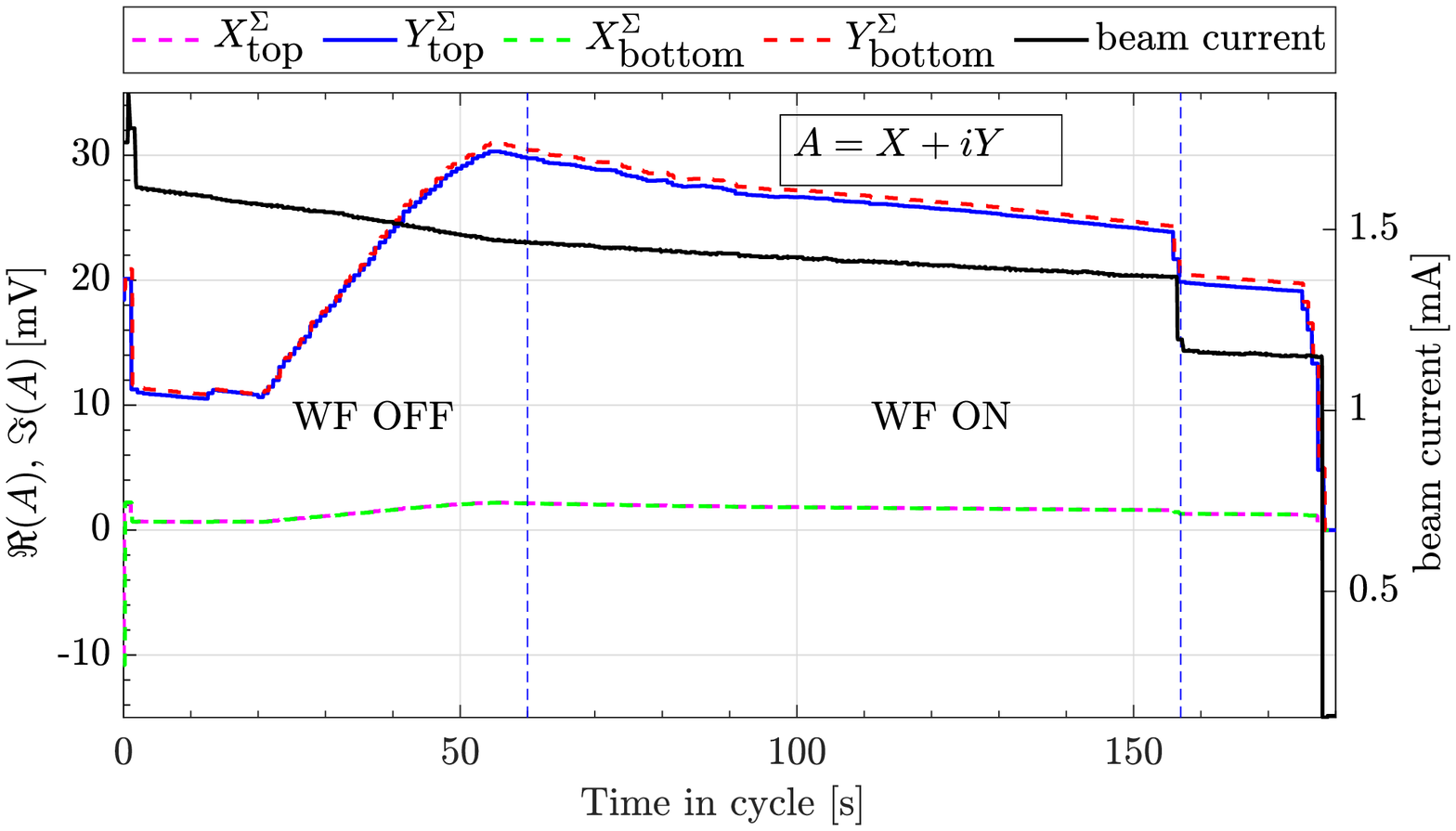}}
\caption{Fourier amplitudes $A = X + iY$ for the top and bottom electrodes of BPM\,17 recorded by the lock-in amplifier as a function of time in the cycle at a strongly mismatched point ($C_\text{L} = \SI{907.79}{pF}$ and $C_\text{T} = \SI{885.58}{pF}$), at the Wien filter frequency (a), and at the revolution frequency (b). In both panels, the stored beam current is shown in black. The cycle starts right after injection is completed at $t = \SI{0}{s}$, beam preparation continues until $t=\SI{55}{s}$, and the Wien filter is switched on and data acquisition starts at $t = \SI{60}{s}$. At $t = \SI{156}{s}$ the Wien filter is switched off and data acquisition stops.}
\label{fig:lockin_data}
\end{figure*}

The complex amplitudes measured by the lock-in amplifiers describe the magnitude and phase of each signal, and are here expressed by the corresponding real and imaginary components, denoted by $X$ and $Y$, respectively, \textit{i.e.}, $A = X +iY$. Examples of the data recorded at the sum frequency $f^\Sigma$ and at the revolution frequency $f^{\text{rev}}$ are shown in Fig.\,\ref{fig:lockin_data}. The observed weak attenuation of the beam current during a measurement cycle by less than 7\% clearly indicates a weak beam loss by intrabeam or  residual gas interactions, thus justifying our treatment of the beam as a rarefied gas. The effect of switching on the power amplifiers of the Wien filter at $t = \SI{60}{s}$ is clearly visible. In both panels, one observes a separation of the quantities recorded by the top and bottom electrodes in the \si{\micro \volt} range for both frequencies after the Wien filter is switched on. This separation is much more pronounced at the Wien filter frequency than at the revolution frequency.
\begin{table}[tb]
	\centering
	\caption{Current $I$ (in $\%$ of the maximum admissible current) in the vertical steerers to generate bumps and the corresponding position change of the vertical orbit $y$ by $\Delta y$ at the location of BPM 17.}
\begin{ruledtabular}	
\renewcommand{\arraystretch}{1.1}
\begin{tabular}{lll}
	$I$ (steerer) [\%] & $y$  [$\si{mm}$] & $\Delta y$ [$\si{mm}$] \\ \hline
	$-5$                      & $-\num{7.756} \pm \num{0.030}$ & $-\num{7.466} \pm \num{0.030}$\\
	$-4$                      & $-6.684 \pm 0.038$           & $-\num{6.395} \pm \num{0.038}$\\
	$-3$                      & $-5.629 \pm 0.016$           & $-\num{5.339} \pm \num{0.016}$\\
	$-2$                      & $-4.518 \pm 0.020$ & $-\num{4.229} \pm \num{0.020}$\\
	$-1$                      & $-3.489 \pm 0.018$ & $-\num{3.119} \pm \num{0.018}$\\
	$\phantom{+}0$            & $-2.439 \pm 0.029$ & $-\num{2.150} \pm \num{0.029}$\\
	$+1$                      & $-1.429 \pm 0.020$ & $-\num{1.140} \pm \num{0.020}$\\
	$+2$                      & $-0.288 \pm 0.028$ & $\phantom{+}\num{0.000} \pm \num{0.000}$\\
	$+3$                      & $+0.798 \pm 0.044$ & $+\num{1.085} \pm \num{0.044}$\\
	$+4$                      & $+1.872 \pm 0.014$ & $+\num{2.160} \pm \num{0.014}$\\
	$+5$                      & $+2.928 \pm 0.069$ & $+\num{3.211} \pm \num{0.069}$
\label{tab:bump_current}  
\end{tabular}
\end{ruledtabular}
\end{table}
\begin{figure}[h]
\centering
\includegraphics[width=\columnwidth]{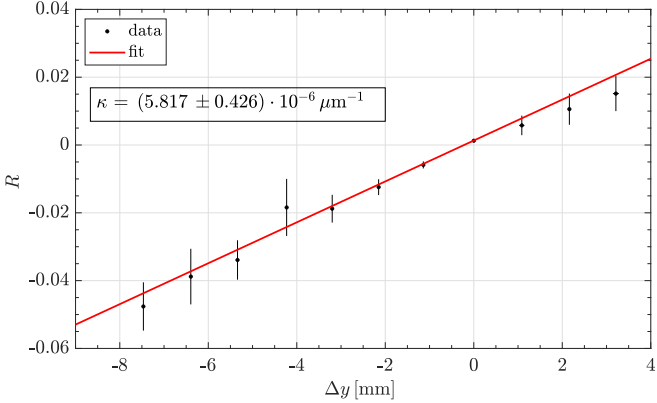}
\caption{Calibration curve of BPM\,17. The ratio $R$, defined in Eq.\,(\ref{eq:cal_ratio}), depends on the introduced vertical beam displacement $\Delta y$ at the beam position monitor.}
\label{fig:cal_ratio}
\end{figure}

The quantities $A_\text{t}^\text{rev}$ and $A_\text{b}^\text{rev}$, given in Eqs.\,(\ref{eq:A-sum}), are related to a vertical beam displacement $\Delta y$ in the following way,
\begin{eqnarray}
R = \frac{A_{\text{t}}^{\text{rev}} - A_{\text{b}}^{\text{rev}}}{A_{\text{t}}^{\text{rev}} + A_{\text{b}}^{\text{rev}}}=\kappa \frac{2 U_0 \Delta y}{2U_0} = \kappa \Delta y\,.
\label{eq:cal_ratio}
\end{eqnarray}
The calibration constant $\kappa$ is experimentally determined by introducing local \textit{vertical} beam bumps in the ring at the location of BPM\,17. The orbit positions $y$ and the orbit displacements $\Delta y$, listed in Table\,\ref{tab:bump_current}, differ by the position of the unperturbed orbit, and are generated by altering the current of a set of vertical steerers.

The steerer magnets have well-known conversion factors from current to magnetic field. The calibration factor $\kappa$ is obtained by fitting the ratio $R$ from Eq.\,(\ref{eq:cal_ratio}) as a function of the vertical orbit variation $\Delta y$, exhibiting the nearly linear relationship shown in Fig.\,\ref{fig:cal_ratio}. The slope corresponds to 
\begin{equation}
\kappa = \left(5.82 \pm 0.43 \right) \cdot \SI{e-6}{\micro m^{-1}}\,.
\label{eq:calibration-factor-kappa}
\end{equation}

\section{Simulation uncertainties}
\label{app:pce}

The uncertainties of the simulated amplitudes of the beam oscillations are computed using the Polynomial Chaos Expansion (PCE) algorithm. The functionality of the algorithm is explained below for one of the simulated data points of the map shown in Fig.\,\ref{fig:simBeamOscillMap}.

The PCE algorithm offers an alternative to the well-known Monte-Carlo (MC) method without compromising the intended accuracy. It uses orthogonal polynomials to represent randomly changing variables to describe observables by means of a finite (truncated) series (for more details, see, e.g., Ref.\,\cite{slim201752}). When the defined criteria of convergence are met, the expansion coefficients can be used to generate an arbitrarily large sample of observables, from which the uncertainties can be computed to the desired statistical accuracy.

The PCE algorithm has been compared with the MC method in many applications and has been shown to provide very reliable results\,\cite{10.5555/2568154}. The PCE requires much fewer simulations to converge compared to the MC method. For instance, for the present case, 200 beam tracking simulations per point in the 2D the map of beam oscillations, shown in Fig.\,\ref{fig:simBeamOscillMap} were sufficient to reach convergence. 
%
\begin{algorithm}[tb]
\KwData{Generates Gaussian-distributed ensemble of uncertain circuit parameters using the Latin Hypercube Sampling (LHS) scheme $X_i$}
\KwResult{Compute uncertainty of $\xi_y$ }
		Given $X_i$, run full wave simulations\;		
		Generate 3D electric and magnetic fields\;		
		Run beam tracking simulations to compute ${\xi_y}_i$\;
		Standardize input data $X_i \longrightarrow \tilde{X}_i$\;
		Guess hyperbolic truncation norm, $q$-norm\;
		Start with lowest possible expansion order $p$\;
		Generate basis functions ${H}_p({\tilde{X}_i})$ ($p^{\text{th}}$-order Hermite polynomials)\;
		Generate hyperbolically truncated set of basis functions ${H}_p^q({\tilde{X}_i})$\;
		Apply Least-Angle Regression (LAR) algorithm\;
		Estimate optimum sparse set of basis functions ${H}_p^{q*}({\tilde{X}_i})$\;
		Compute expansion coefficients $C_j$, given ${\xi_y}_i = \sum_{j} C_j {H}_p^{q*}({\tilde{X}_i})$\;
		Compute leave-one-out error $\text{LOO}_{\text{err}}$\;
		Check convergence condition( $\text{LOO}_{\text{err}} < 10^{-2}$)\;
\While{ not convergent }{Enhance model (vary $p$ and $q$)\;
\eIf{convergent}{
		Generate large sample of $\xi_y$\;
		Estimate statistical parameters\;
		Terminate algorithm\;
	}{
		Enrich input samples $X_i$\;
		Repeat algorithm\;
			}
		}
\caption{Sparse Polynomial Chaos Expansion\,\cite{slim201752}.}
\label{algo:pce}
\end{algorithm}
In cases where the number of random input variables $m$ is larger than $10$, the PCE method offers clear advantages over the MC method. The reason is that the number of basis functions in the PCE method increases enormously as a consequence of the tensor product of the  involved polynomials. Therefore, the algorithm has been improved further to allow for a reduction of the number of simulations required. Such an approach is also adopted here, as described in Algorithm\,\ref{algo:pce}. The hyperbolic truncation scheme together with the Least-Angle Regression (LAR) method form a sparse version of the original algorithm.

An $m$-dimensional set is first created, representing $N$ combinations of  simultaneous random variables. Many methods can be used to generate such sets, and here the Latin-hypercube sample scheme is adopted\,\cite{Slim_2020}. Subsequently, the set is standardized for convergence reasons. Depending on the distribution of the data, the basis functions, here Hermite polynomials, are determined. The number of basis functions restricts the lower limit of the number of simulations (full-wave and tracking) which are usually computationally expensive. As a rule of thumb, with $N$  basis functions, the PCE algorithm requires at least $1.5 \times  N$ (in this case, full-wave) simulations to converge. The number of basis functions itself can, however,  be reduced by the hyperbolic truncation scheme that eliminates higher-order terms that do not have a significant impact on the observation objects \,\cite{sudret2000stochastic,sudret2008global}. Furthermore, by applying the LAR algorithm, the number of remaining basis functions can be further reduced substantially, whereby the problem becomes computationally solvable in a very efficient fashion.  

The matching point, specified in Eq.\,(\ref{eq:matched-point-Cl-CT}), yields the minimum measurable beam oscillations, as given by Eq.\,(\ref{eq:eps_y-minimum-BPM17}). This experimental result can be estimated using the beam-tracking calculations. Subsequently, the concrete steps of the application of the PCE algorithm are discussed.   

\begin{figure*}[tb]
\centering
\subfigure[\label{fig:pceSim-1} Sample of $C_\text{L} $, $C_\text{T}$, and $R_\text{f}$ used in the PCE calculations showing a subset of the 15-dimensional input of random circuit uncertainties.]{\includegraphics[height=0.21\textheight]{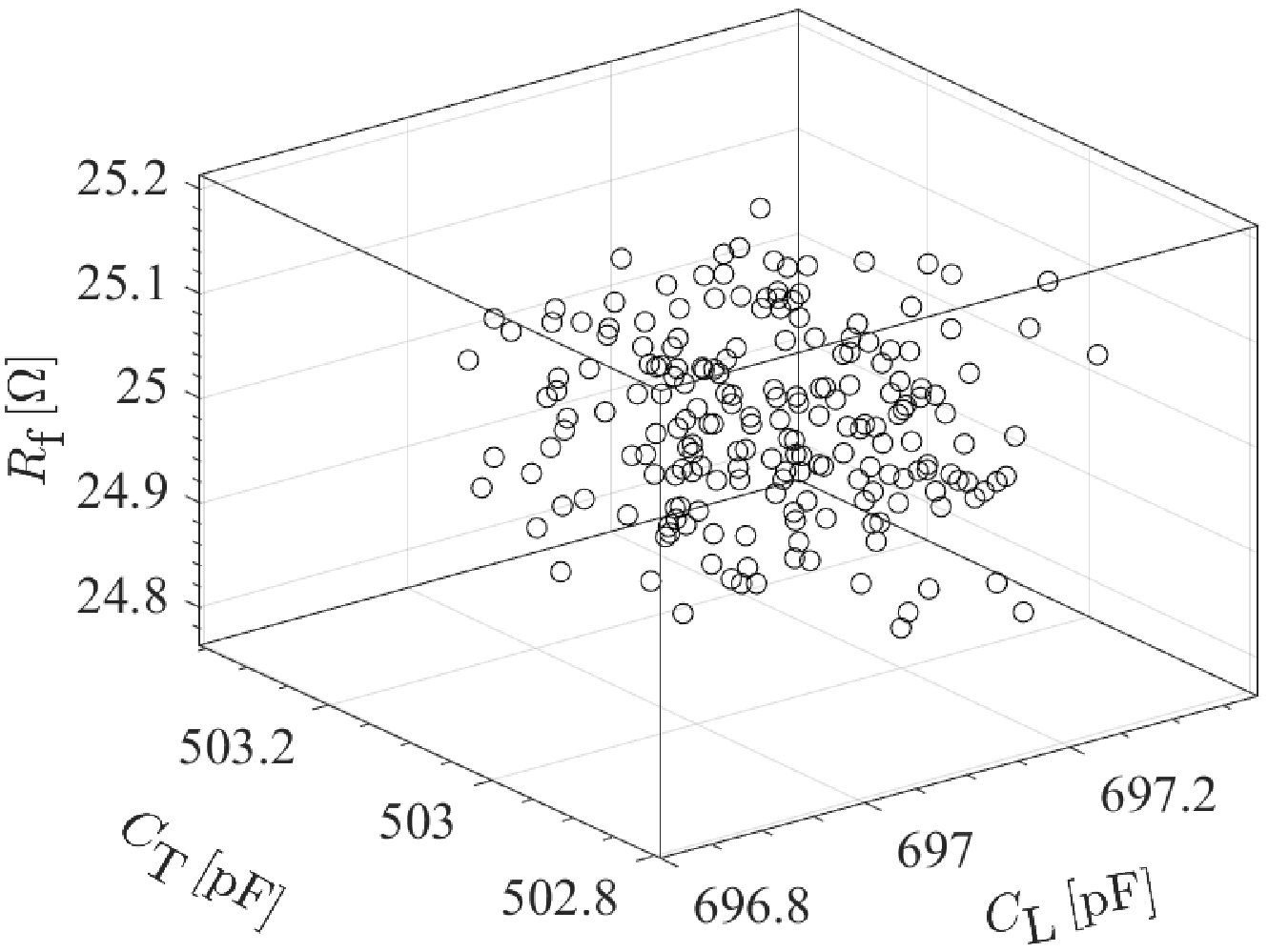}}
\hspace{0.1cm}
\subfigure[\label{fig:pceSim-2} Truncation schemes of the PCE algorithm.]{\includegraphics[height=0.21\textheight]{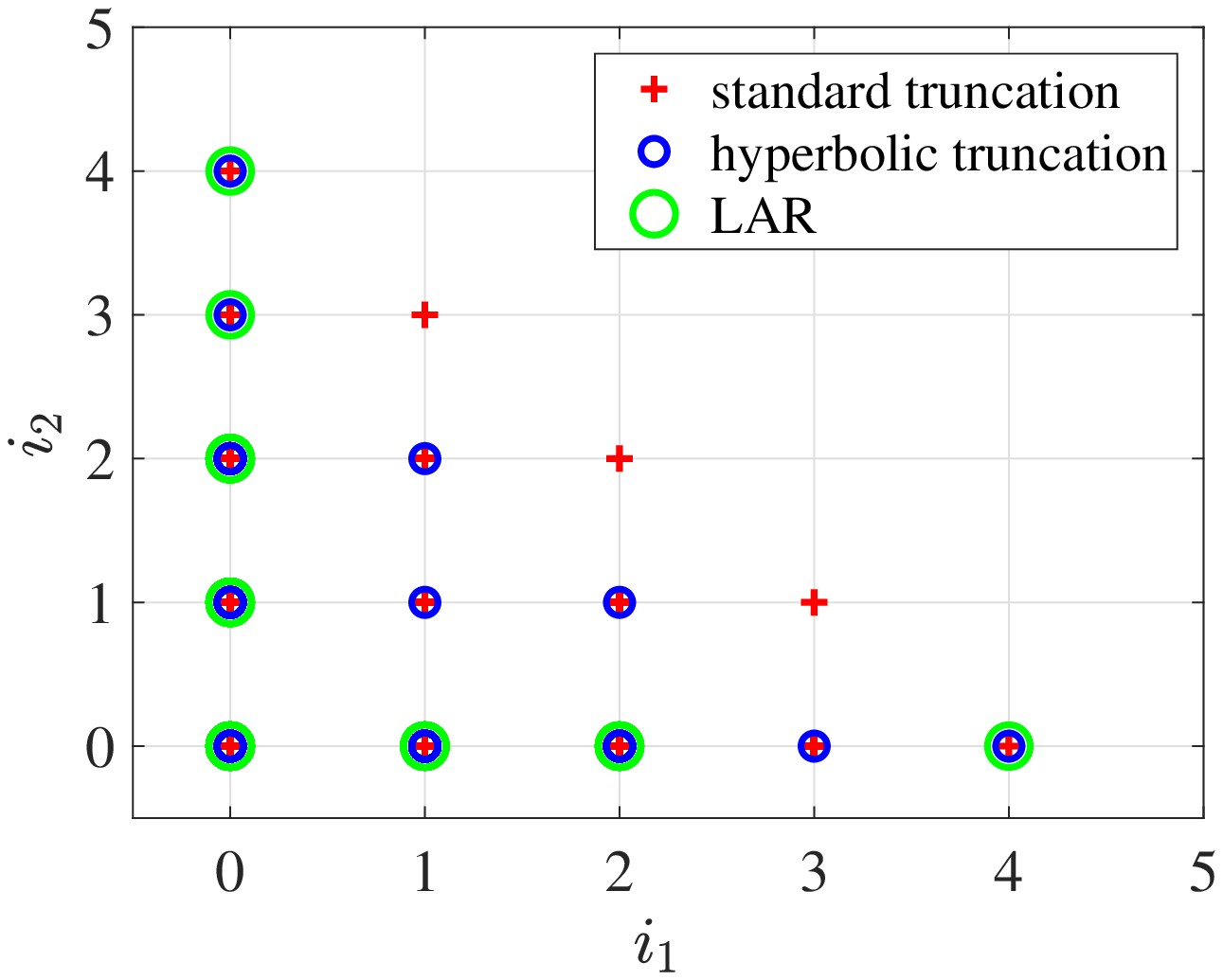}}
\subfigure[\label{fig:pceSim-3} Expansion coefficients on a semi-log scale. 91 coefficients have been selected after applying the LAR algorithm to the matching point.] {\includegraphics[height=0.21\textheight]{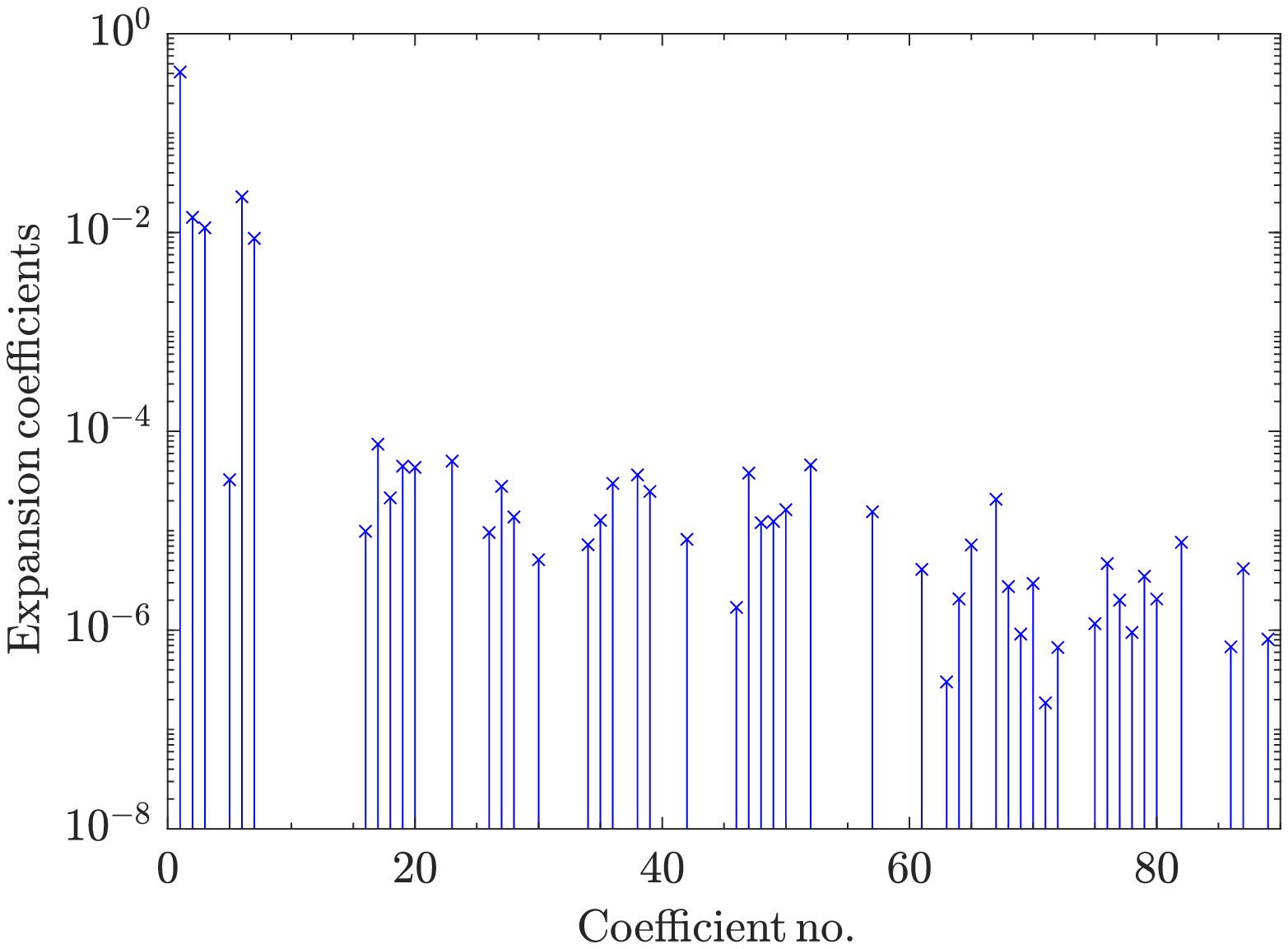}}
\hspace{0.1cm}
\subfigure[\label{fig:pceSim-4} 
Comparison between the tracking results and the PCE with respect to the oscillation amplitude, determined using the expansion coefficients of (c).
]{\includegraphics[height=0.21\textheight]{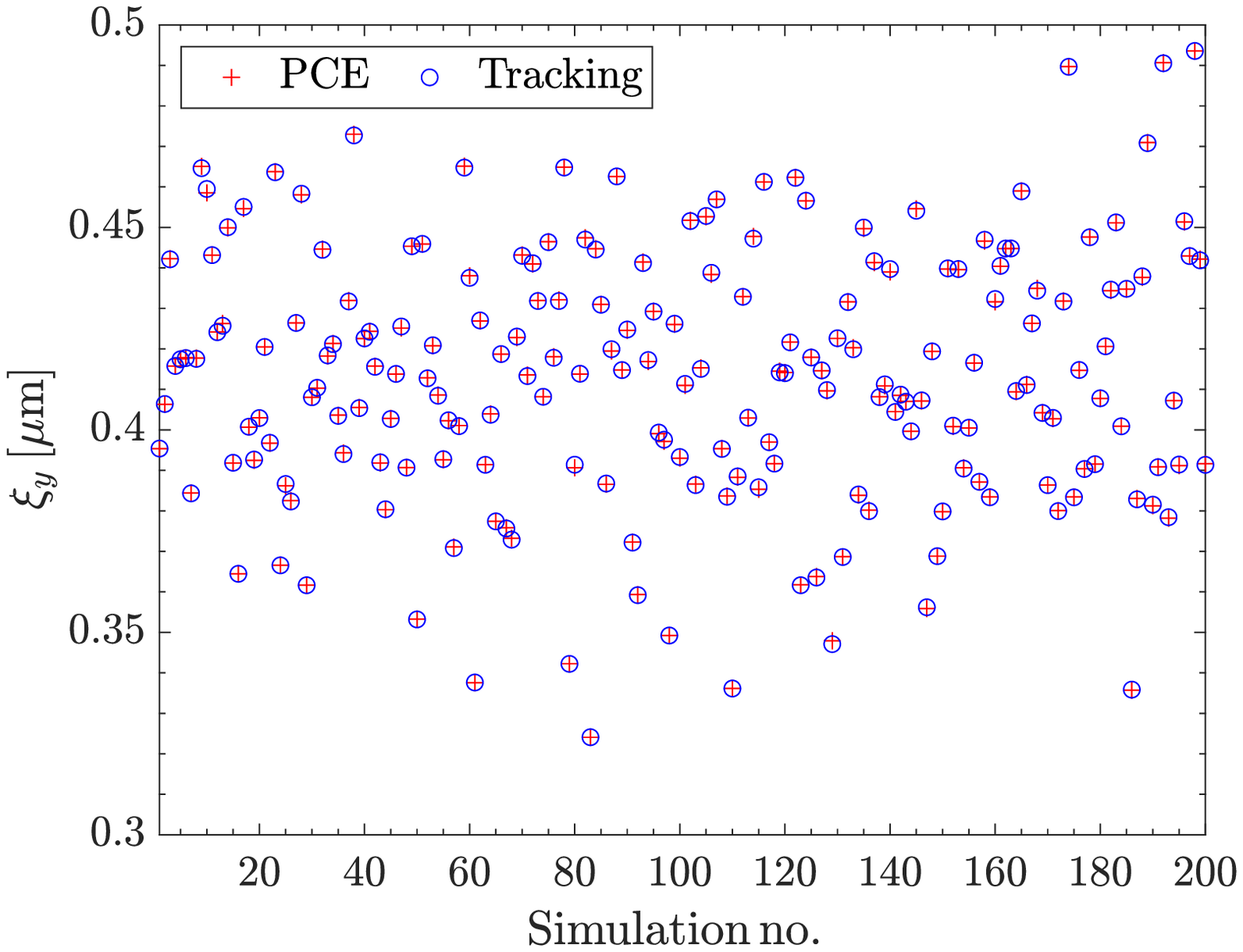}}
\caption{Intermediate results of the PCE algorithm applied at the matching point [see Eq.\,(\ref{eq:matched-point-Cl-CT})]. Quantitative results of the PCE algorithm are summarized in Table\,\ref{tab:pce_param}.}
\label{fig:pceSim}
\end{figure*}

All the reasonable sources of uncertainties of the circuit are represented by 15 random parameters that are allowed to vary simultaneously. At first, a sample of ($200 \times 15$) entries is generated using the Latin-hypercube sampling scheme. As an example of this sample, the variation of the three circuit elements $C_{\textnormal{L}}$,  $C_{\textnormal{T}}$, and the load resistor $R_{\text{f}}$ is shown in Fig.\,\ref{fig:pceSim-1}. 

All the uncertain parameters in the electromagnetic circuit simulations are used to generate the electric and magnetic fields shown in Fig.\,\ref{fig:fieldsUncMatched}. These are subsequently used in the beam-tracking calculations. For the matching point of the map [Eq.\,(\ref{eq:matched-point-Cl-CT})], $N =200$ full-wave simulations were conducted. The import of these field maps into the beam-tracking calculations resulted in a set of $N =200$ values of $\xi_y$. This set is not directly used  to conduct the statistical analysis. Instead, in conjunction with the input samples, these data are used as input to the sparse PCE algorithm. 

The optimum set of basis function is determined using the LAR algorithm, as shown in Fig.\,\ref{fig:pceSim-2}. With an expansion order of $p = 6$ and a truncation norm $q = 0.35$, executing the PCE algorithm required \num{91} basis functions to converge, reflected by the low value of the leave-one-out error $\text{LOO}_{\text{err}} = \num{1.7e-4}$. Subsequently, the expansion coefficients are computed, qualitatively depicted in Fig.\,\ref{fig:pceSim-3}. It is shown in Fig.\,\ref{fig:pceSim-4} that the PCE algorithm perfectly reproduces the tracking results using these expansion coefficients. Finally, these coefficients are used to reconstruct a larger sample of $\xi_y$ to estimate the error $\sigma_{\xi_y}$. Figure\,\ref{fig:epsilonUncMatched} shows $\num{e4}$ of the   $\num{e6}$ reconstructed samples. The PCE parameters used are summarized in Table\,\ref{tab:pce_param}. 

The fitting of these results with a Gaussian, as depicted in panel (b) of Fig.\,\ref{fig:epsilonUncMatched}, yields a standard deviation of $\sigma_{\xi_y} = \SI{0.082}{\micro m}$. The same technique is repeated for each point in the map. 
\begin{table}[b]
\caption{PCE simulation parameters of the matching point in Eq.\,(\ref{eq:matched-point-Cl-CT}).}
\begin{ruledtabular}
\renewcommand{\arraystretch}{1.1}
\begin{tabular}{rl}
	Parameter                                       & Value \\\hline
	order of expansion  $p$                         &           $6$ \\
	dimension    $m$                                &           $15$ \\
	hyperbolic truncation $q$                       &           $0.35$ \\
	leave-one-out error $\text{LOO}_{\text{err}}$   & \num{1.71e-4}\\
	number of used basis functions $P^{\text{LAR}}$ & 91\\
	number of used full-wave simulations  $N$       & 200\\
\end{tabular} 
\label{tab:pce_param} 
\end{ruledtabular}
\end{table}

\bibliography{dBase_24.11.2020}

\begin{thebibliography}{45}%
\makeatletter
\providecommand \@ifxundefined [1]{%
 \@ifx{#1\undefined}
}%
\providecommand \@ifnum [1]{%
 \ifnum #1\expandafter \@firstoftwo
 \else \expandafter \@secondoftwo
 \fi
}%
\providecommand \@ifx [1]{%
 \ifx #1\expandafter \@firstoftwo
 \else \expandafter \@secondoftwo
 \fi
}%
\providecommand \natexlab [1]{#1}%
\providecommand \enquote  [1]{``#1''}%
\providecommand \bibnamefont  [1]{#1}%
\providecommand \bibfnamefont [1]{#1}%
\providecommand \citenamefont [1]{#1}%
\providecommand \href@noop [0]{\@secondoftwo}%
\providecommand \href [0]{\begingroup \@sanitize@url \@href}%
\providecommand \@href[1]{\@@startlink{#1}\@@href}%
\providecommand \@@href[1]{\endgroup#1\@@endlink}%
\providecommand \@sanitize@url [0]{\catcode `\\12\catcode `\$12\catcode
  `\&12\catcode `\#12\catcode `\^12\catcode `\_12\catcode `\%12\relax}%
\providecommand \@@startlink[1]{}%
\providecommand \@@endlink[0]{}%
\providecommand \url  [0]{\begingroup\@sanitize@url \@url }%
\providecommand \@url [1]{\endgroup\@href {#1}{\urlprefix }}%
\providecommand \urlprefix  [0]{URL }%
\providecommand \Eprint [0]{\href }%
\providecommand \doibase [0]{http://dx.doi.org/}%
\providecommand \selectlanguage [0]{\@gobble}%
\providecommand \bibinfo  [0]{\@secondoftwo}%
\providecommand \bibfield  [0]{\@secondoftwo}%
\providecommand \translation [1]{[#1]}%
\providecommand \BibitemOpen [0]{}%
\providecommand \bibitemStop [0]{}%
\providecommand \bibitemNoStop [0]{.\EOS\space}%
\providecommand \EOS [0]{\spacefactor3000\relax}%
\providecommand \BibitemShut  [1]{\csname bibitem#1\endcsname}%
\let\auto@bib@innerbib\@empty
\bibitem [{\citenamefont {Schreppler}\ \emph {et~al.}(2014)\citenamefont
  {Schreppler}, \citenamefont {Spethmann}, \citenamefont {Brahms},
  \citenamefont {Botter}, \citenamefont {Barrios},\ and\ \citenamefont
  {Stamper-Kurn}}]{Schreppler1486}%
  \BibitemOpen
  \bibfield  {author} {\bibinfo {author} {\bibfnamefont {Sydney}\ \bibnamefont
  {Schreppler}}, \bibinfo {author} {\bibfnamefont {Nicolas}\ \bibnamefont
  {Spethmann}}, \bibinfo {author} {\bibfnamefont {Nathan}\ \bibnamefont
  {Brahms}}, \bibinfo {author} {\bibfnamefont {Thierry}\ \bibnamefont
  {Botter}}, \bibinfo {author} {\bibfnamefont {Maryrose}\ \bibnamefont
  {Barrios}}, \ and\ \bibinfo {author} {\bibfnamefont {Dan~M.}\ \bibnamefont
  {Stamper-Kurn}},\ }\bibfield  {title} {\enquote {\bibinfo {title} {Optically
  measuring force near the standard quantum limit},}\ }\href {\doibase
  10.1126/science.1249850} {\bibfield  {journal} {\bibinfo  {journal}
  {Science}\ }\textbf {\bibinfo {volume} {344}},\ \bibinfo {pages} {1486--1489}
  (\bibinfo {year} {2014})}\BibitemShut {NoStop}%
\bibitem [{\citenamefont {Abbott}\ \emph {et~al.}(2009)\citenamefont {Abbott}
  \emph {et~al.}}]{Abbott:2009zz}%
  \BibitemOpen
  \bibfield  {author} {\bibinfo {author} {\bibfnamefont {B.}~\bibnamefont
  {Abbott}} \emph {et~al.} (\bibinfo {collaboration} {LIGO Scientific}),\
  }\bibfield  {title} {\enquote {\bibinfo {title} {{Observation of a
  kilogram-scale oscillator near its quantum ground state}},}\ }\href {\doibase
  10.1088/1367-2630/11/7/073032} {\bibfield  {journal} {\bibinfo  {journal}
  {New J. Phys.}\ }\textbf {\bibinfo {volume} {11}},\ \bibinfo {pages} {073032}
  (\bibinfo {year} {2009})}\BibitemShut {NoStop}%
\bibitem [{\citenamefont {Murch}\ \emph {et~al.}(2008)\citenamefont {Murch},
  \citenamefont {Moore}, \citenamefont {Gupta},\ and\ \citenamefont
  {Stamper-Kurn}}]{Murch2008}%
  \BibitemOpen
  \bibfield  {author} {\bibinfo {author} {\bibfnamefont {Kater~W.}\
  \bibnamefont {Murch}}, \bibinfo {author} {\bibfnamefont {Kevin~L.}\
  \bibnamefont {Moore}}, \bibinfo {author} {\bibfnamefont {Subhadeep}\
  \bibnamefont {Gupta}}, \ and\ \bibinfo {author} {\bibfnamefont {Dan~M.}\
  \bibnamefont {Stamper-Kurn}},\ }\bibfield  {title} {\enquote {\bibinfo
  {title} {Observation of quantum-measurement backaction with an ultracold
  atomic gas},}\ }\href {\doibase 10.1038/nphys965} {\bibfield  {journal}
  {\bibinfo  {journal} {Nature Physics}\ }\textbf {\bibinfo {volume} {4}},\
  \bibinfo {pages} {561--564} (\bibinfo {year} {2008})}\BibitemShut {NoStop}%
\bibitem [{\citenamefont {Biercuk}\ \emph {et~al.}(2010)\citenamefont
  {Biercuk}, \citenamefont {Uys}, \citenamefont {Britton}, \citenamefont
  {VanDevender},\ and\ \citenamefont {Bollinger}}]{Biercuk2010}%
  \BibitemOpen
  \bibfield  {author} {\bibinfo {author} {\bibfnamefont {Michael~J.}\
  \bibnamefont {Biercuk}}, \bibinfo {author} {\bibfnamefont {Hermann}\
  \bibnamefont {Uys}}, \bibinfo {author} {\bibfnamefont {Joe~W.}\ \bibnamefont
  {Britton}}, \bibinfo {author} {\bibfnamefont {Aaron~P.}\ \bibnamefont
  {VanDevender}}, \ and\ \bibinfo {author} {\bibfnamefont {John~J.}\
  \bibnamefont {Bollinger}},\ }\bibfield  {title} {\enquote {\bibinfo {title}
  {Ultrasensitive detection of force and displacement using trapped ions},}\
  }\href {\doibase 10.1038/nnano.2010.165} {\bibfield  {journal} {\bibinfo
  {journal} {Nature Nanotechnology}\ }\textbf {\bibinfo {volume} {5}},\
  \bibinfo {pages} {646--650} (\bibinfo {year} {2010})}\BibitemShut {NoStop}%
\bibitem [{\citenamefont {Rugar}\ \emph {et~al.}(2004)\citenamefont {Rugar},
  \citenamefont {Budakian}, \citenamefont {Mamin},\ and\ \citenamefont
  {Chui}}]{Rugar2004}%
  \BibitemOpen
  \bibfield  {author} {\bibinfo {author} {\bibfnamefont {D.}~\bibnamefont
  {Rugar}}, \bibinfo {author} {\bibfnamefont {R.}~\bibnamefont {Budakian}},
  \bibinfo {author} {\bibfnamefont {H.~J.}\ \bibnamefont {Mamin}}, \ and\
  \bibinfo {author} {\bibfnamefont {B.~W.}\ \bibnamefont {Chui}},\ }\bibfield
  {title} {\enquote {\bibinfo {title} {Single spin detection by magnetic
  resonance force microscopy},}\ }\href {\doibase 10.1038/nature02658}
  {\bibfield  {journal} {\bibinfo  {journal} {Nature}\ }\textbf {\bibinfo
  {volume} {430}},\ \bibinfo {pages} {329--332} (\bibinfo {year}
  {2004})}\BibitemShut {NoStop}%
\bibitem [{\citenamefont {Abbott}\ \emph {et~al.}(2016)\citenamefont {Abbott}
  \emph {et~al.}}]{Abbott:2016izl}%
  \BibitemOpen
  \bibfield  {author} {\bibinfo {author} {\bibfnamefont {Thomas~D.}\
  \bibnamefont {Abbott}} \emph {et~al.} (\bibinfo {collaboration} {LIGO
  Scientific, Virgo}),\ }\bibfield  {title} {\enquote {\bibinfo {title}
  {{Improved analysis of GW150914 using a fully spin-precessing waveform
  Model}},}\ }\href {\doibase 10.1103/PhysRevX.6.041014} {\bibfield  {journal}
  {\bibinfo  {journal} {Phys. Rev. X}\ }\textbf {\bibinfo {volume} {6}},\
  \bibinfo {pages} {041014} (\bibinfo {year} {2016})},\ \Eprint
  {http://arxiv.org/abs/1606.01210} {arXiv:1606.01210 [gr-qc]} \BibitemShut
  {NoStop}%
\bibitem [{\citenamefont {Abusaif}\ \emph {et~al.}(2021)\citenamefont
  {Abusaif}, \citenamefont {Aksentev}, \citenamefont {Aggarwal}, \citenamefont
  {Alberdi-Esuain}, \citenamefont {Andres}, \citenamefont {Atanasov},
  \citenamefont {Barion}, \citenamefont {Basile}, \citenamefont {Berz},
  \citenamefont {Böhme}, \citenamefont {Böker}, \citenamefont {Borburgh},
  \citenamefont {Canale}, \citenamefont {Carli}, \citenamefont {Ciepał},
  \citenamefont {Ciullo}, \citenamefont {Contalbrigo}, \citenamefont
  {De~Conto}, \citenamefont {Dymov}, \citenamefont {Felden}, \citenamefont
  {Gaisser}, \citenamefont {Gebel}, \citenamefont {Giese}, \citenamefont
  {Gooding}, \citenamefont {Grigoryev}, \citenamefont {Grzonka}, \citenamefont
  {{Haj Tahar}}, \citenamefont {Hahnraths}, \citenamefont {Heberling},
  \citenamefont {Hejny}, \citenamefont {Hetzel}, \citenamefont {Hölscher},
  \citenamefont {Javakhishvili}, \citenamefont {Jorat}, \citenamefont
  {Kacharava}, \citenamefont {Kamerdzhiev}, \citenamefont {Karanth},
  \citenamefont {Keshelashvili}, \citenamefont {Koop}, \citenamefont {Kulikov},
  \citenamefont {Laihem}, \citenamefont {Lamont}, \citenamefont {Lehrach},
  \citenamefont {Lenisa}, \citenamefont {Lomidze}, \citenamefont {Lomidze},
  \citenamefont {Lorentz}, \citenamefont {Macharashvili}, \citenamefont
  {Magiera}, \citenamefont {Makino}, \citenamefont {Martin}, \citenamefont
  {Mchedlishvili}, \citenamefont {Meißner}, \citenamefont {Metreveli},
  \citenamefont {Michaud}, \citenamefont {Müller}, \citenamefont {Nass},
  \citenamefont {Natour}, \citenamefont {Nikolaev}, \citenamefont {Nogga},
  \citenamefont {Okropiridze}, \citenamefont {Pesce}, \citenamefont {Poncza},
  \citenamefont {Prasuhn}, \citenamefont {Pretz}, \citenamefont {Rathmann},
  \citenamefont {Ritman}, \citenamefont {Rosenthal}, \citenamefont {Saleev},
  \citenamefont {Schott}, \citenamefont {Sefzick}, \citenamefont {Senichev},
  \citenamefont {Shankar}, \citenamefont {Shergelashvili}, \citenamefont
  {Shmakova}, \citenamefont {Siddique}, \citenamefont {Silenko}, \citenamefont
  {Simon}, \citenamefont {Slim}, \citenamefont {Soltner}, \citenamefont
  {Stahl}, \citenamefont {Stassen}, \citenamefont {Stephenson}, \citenamefont
  {Straatmann}, \citenamefont {Ströher}, \citenamefont {Tabidze},
  \citenamefont {Tagliente}, \citenamefont {Talman}, \citenamefont {Uzikov},
  \citenamefont {Valdau}, \citenamefont {Valetov}, \citenamefont {Vilella},
  \citenamefont {Vitz}, \citenamefont {Vossebeld}, \citenamefont {Wagner},
  \citenamefont {Weidemann}, \citenamefont {Wirzba}, \citenamefont
  {Wro\ifmmode~\acute{n}\else \'{n}\fi{}ska}, \citenamefont {Wüstner},
  \citenamefont {Zupra\ifmmode~\acute{n}\else \'{n}\fi{}ski},\ and\
  \citenamefont {\.{Z}urek}}]{CPEDM}%
  \BibitemOpen
  \bibfield  {author} {\bibinfo {author} {\bibfnamefont {F.}~\bibnamefont
  {Abusaif}}, \bibinfo {author} {\bibfnamefont {A.}~\bibnamefont {Aksentev}},
  \bibinfo {author} {\bibfnamefont {A.}~\bibnamefont {Aggarwal}}, \bibinfo
  {author} {\bibfnamefont {B.}~\bibnamefont {Alberdi-Esuain}}, \bibinfo
  {author} {\bibfnamefont {A.}~\bibnamefont {Andres}}, \bibinfo {author}
  {\bibfnamefont {A.}~\bibnamefont {Atanasov}}, \bibinfo {author}
  {\bibfnamefont {L.}~\bibnamefont {Barion}}, \bibinfo {author} {\bibfnamefont
  {S.}~\bibnamefont {Basile}}, \bibinfo {author} {\bibfnamefont
  {M.}~\bibnamefont {Berz}}, \bibinfo {author} {\bibfnamefont {C.}~\bibnamefont
  {Böhme}}, \bibinfo {author} {\bibfnamefont {J.}~\bibnamefont {Böker}},
  \bibinfo {author} {\bibfnamefont {J.}~\bibnamefont {Borburgh}}, \bibinfo
  {author} {\bibfnamefont {N.}~\bibnamefont {Canale}}, \bibinfo {author}
  {\bibfnamefont {C.}~\bibnamefont {Carli}}, \bibinfo {author} {\bibfnamefont
  {I.}~\bibnamefont {Ciepał}}, \bibinfo {author} {\bibfnamefont
  {G.}~\bibnamefont {Ciullo}}, \bibinfo {author} {\bibfnamefont
  {M.}~\bibnamefont {Contalbrigo}}, \bibinfo {author} {\bibfnamefont {{J.-M.}}\
  \bibnamefont {De~Conto}}, \bibinfo {author} {\bibfnamefont {S.}~\bibnamefont
  {Dymov}}, \bibinfo {author} {\bibfnamefont {O.}~\bibnamefont {Felden}},
  \bibinfo {author} {\bibfnamefont {M.}~\bibnamefont {Gaisser}}, \bibinfo
  {author} {\bibfnamefont {R.}~\bibnamefont {Gebel}}, \bibinfo {author}
  {\bibfnamefont {N.}~\bibnamefont {Giese}}, \bibinfo {author} {\bibfnamefont
  {J.}~\bibnamefont {Gooding}}, \bibinfo {author} {\bibfnamefont
  {K.}~\bibnamefont {Grigoryev}}, \bibinfo {author} {\bibfnamefont
  {D.}~\bibnamefont {Grzonka}}, \bibinfo {author} {\bibfnamefont
  {M.}~\bibnamefont {{Haj Tahar}}}, \bibinfo {author} {\bibfnamefont
  {T.}~\bibnamefont {Hahnraths}}, \bibinfo {author} {\bibfnamefont
  {D.}~\bibnamefont {Heberling}}, \bibinfo {author} {\bibfnamefont
  {V.}~\bibnamefont {Hejny}}, \bibinfo {author} {\bibfnamefont
  {J.}~\bibnamefont {Hetzel}}, \bibinfo {author} {\bibfnamefont
  {D.}~\bibnamefont {Hölscher}}, \bibinfo {author} {\bibfnamefont
  {O.}~\bibnamefont {Javakhishvili}}, \bibinfo {author} {\bibfnamefont
  {L.}~\bibnamefont {Jorat}}, \bibinfo {author} {\bibfnamefont
  {A.}~\bibnamefont {Kacharava}}, \bibinfo {author} {\bibfnamefont
  {V.}~\bibnamefont {Kamerdzhiev}}, \bibinfo {author} {\bibfnamefont
  {S.}~\bibnamefont {Karanth}}, \bibinfo {author} {\bibfnamefont
  {I.}~\bibnamefont {Keshelashvili}}, \bibinfo {author} {\bibfnamefont
  {I.}~\bibnamefont {Koop}}, \bibinfo {author} {\bibfnamefont {A.}~\bibnamefont
  {Kulikov}}, \bibinfo {author} {\bibfnamefont {K.}~\bibnamefont {Laihem}},
  \bibinfo {author} {\bibfnamefont {M.}~\bibnamefont {Lamont}}, \bibinfo
  {author} {\bibfnamefont {A.}~\bibnamefont {Lehrach}}, \bibinfo {author}
  {\bibfnamefont {P.}~\bibnamefont {Lenisa}}, \bibinfo {author} {\bibfnamefont
  {I.}~\bibnamefont {Lomidze}}, \bibinfo {author} {\bibfnamefont
  {N.}~\bibnamefont {Lomidze}}, \bibinfo {author} {\bibfnamefont
  {B.}~\bibnamefont {Lorentz}}, \bibinfo {author} {\bibfnamefont
  {G.}~\bibnamefont {Macharashvili}}, \bibinfo {author} {\bibfnamefont
  {A.}~\bibnamefont {Magiera}}, \bibinfo {author} {\bibfnamefont
  {K.}~\bibnamefont {Makino}}, \bibinfo {author} {\bibfnamefont
  {S.}~\bibnamefont {Martin}}, \bibinfo {author} {\bibfnamefont
  {D.}~\bibnamefont {Mchedlishvili}}, \bibinfo {author} {\bibfnamefont
  {{U.-G.}}\ \bibnamefont {Meißner}}, \bibinfo {author} {\bibfnamefont
  {Z.}~\bibnamefont {Metreveli}}, \bibinfo {author} {\bibfnamefont
  {J.}~\bibnamefont {Michaud}}, \bibinfo {author} {\bibfnamefont
  {F.}~\bibnamefont {Müller}}, \bibinfo {author} {\bibfnamefont
  {A.}~\bibnamefont {Nass}}, \bibinfo {author} {\bibfnamefont {G.}~\bibnamefont
  {Natour}}, \bibinfo {author} {\bibfnamefont {N.}~\bibnamefont {Nikolaev}},
  \bibinfo {author} {\bibfnamefont {A.}~\bibnamefont {Nogga}}, \bibinfo
  {author} {\bibfnamefont {D.}~\bibnamefont {Okropiridze}}, \bibinfo {author}
  {\bibfnamefont {A.}~\bibnamefont {Pesce}}, \bibinfo {author} {\bibfnamefont
  {V.}~\bibnamefont {Poncza}}, \bibinfo {author} {\bibfnamefont
  {D.}~\bibnamefont {Prasuhn}}, \bibinfo {author} {\bibfnamefont
  {J.}~\bibnamefont {Pretz}}, \bibinfo {author} {\bibfnamefont
  {F.}~\bibnamefont {Rathmann}}, \bibinfo {author} {\bibfnamefont
  {J.}~\bibnamefont {Ritman}}, \bibinfo {author} {\bibfnamefont
  {M.}~\bibnamefont {Rosenthal}}, \bibinfo {author} {\bibfnamefont
  {A.}~\bibnamefont {Saleev}}, \bibinfo {author} {\bibfnamefont
  {M.}~\bibnamefont {Schott}}, \bibinfo {author} {\bibfnamefont
  {T.}~\bibnamefont {Sefzick}}, \bibinfo {author} {\bibfnamefont
  {Y.}~\bibnamefont {Senichev}}, \bibinfo {author} {\bibfnamefont
  {R.}~\bibnamefont {Shankar}}, \bibinfo {author} {\bibfnamefont
  {D.}~\bibnamefont {Shergelashvili}}, \bibinfo {author} {\bibfnamefont
  {V.}~\bibnamefont {Shmakova}}, \bibinfo {author} {\bibfnamefont
  {S.}~\bibnamefont {Siddique}}, \bibinfo {author} {\bibfnamefont
  {A.}~\bibnamefont {Silenko}}, \bibinfo {author} {\bibfnamefont
  {M.}~\bibnamefont {Simon}}, \bibinfo {author} {\bibfnamefont
  {J.}~\bibnamefont {Slim}}, \bibinfo {author} {\bibfnamefont {H.}~\bibnamefont
  {Soltner}}, \bibinfo {author} {\bibfnamefont {A.}~\bibnamefont {Stahl}},
  \bibinfo {author} {\bibfnamefont {R.}~\bibnamefont {Stassen}}, \bibinfo
  {author} {\bibfnamefont {E.}~\bibnamefont {Stephenson}}, \bibinfo {author}
  {\bibfnamefont {H.}~\bibnamefont {Straatmann}}, \bibinfo {author}
  {\bibfnamefont {H.}~\bibnamefont {Ströher}}, \bibinfo {author}
  {\bibfnamefont {M.}~\bibnamefont {Tabidze}}, \bibinfo {author} {\bibfnamefont
  {G.}~\bibnamefont {Tagliente}}, \bibinfo {author} {\bibfnamefont
  {R.}~\bibnamefont {Talman}}, \bibinfo {author} {\bibfnamefont
  {Y.}~\bibnamefont {Uzikov}}, \bibinfo {author} {\bibfnamefont
  {Y.}~\bibnamefont {Valdau}}, \bibinfo {author} {\bibfnamefont
  {E.}~\bibnamefont {Valetov}}, \bibinfo {author} {\bibfnamefont
  {E.}~\bibnamefont {Vilella}}, \bibinfo {author} {\bibfnamefont
  {M.}~\bibnamefont {Vitz}}, \bibinfo {author} {\bibfnamefont {J.}~\bibnamefont
  {Vossebeld}}, \bibinfo {author} {\bibfnamefont {T.}~\bibnamefont {Wagner}},
  \bibinfo {author} {\bibfnamefont {C.}~\bibnamefont {Weidemann}}, \bibinfo
  {author} {\bibfnamefont {A.}~\bibnamefont {Wirzba}}, \bibinfo {author}
  {\bibfnamefont {A.}~\bibnamefont {Wro\ifmmode~\acute{n}\else \'{n}\fi{}ska}},
  \bibinfo {author} {\bibfnamefont {P.}~\bibnamefont {Wüstner}}, \bibinfo
  {author} {\bibfnamefont {P.}~\bibnamefont {Zupra\ifmmode~\acute{n}\else
  \'{n}\fi{}ski}}, \ and\ \bibinfo {author} {\bibfnamefont {M.}~\bibnamefont
  {\.{Z}urek}},\ }\bibfield  {title} {\enquote {\bibinfo {title} {Storage ring
  to search for electric dipole moments of charged particles: {{Feasibility}}
  study},}\ }\href {\doibase 10.23731/CYRM-2021-003} {\bibfield  {journal}
  {\bibinfo  {journal} {CERN Yellow Report}\ ,\ \bibinfo {pages} {257}}
  (\bibinfo {year} {2021})}\BibitemShut {NoStop}%
\bibitem [{ari()}]{ariesWS:2021}%
  \BibitemOpen
  \href@noop {} {}\bibinfo {note} {{see, e.g., the presentations at the ARIES
  WP6 Workshop: Storage Rings and Gravitational Waves "SRGW2021", 2 February -
  11 March 2021, available from
  \url{https://indico.cern.ch/event/982987}}}\BibitemShut {NoStop}%
\bibitem [{\citenamefont {Saleev}\ \emph {et~al.}(2017)\citenamefont {Saleev},
  \citenamefont {Nikolaev}, \citenamefont {Rathmann}, \citenamefont
  {Augustyniak}, \citenamefont {Bagdasarian}, \citenamefont {Bai},
  \citenamefont {Barion}, \citenamefont {Berz}, \citenamefont {Chekmenev},
  \citenamefont {Ciullo}, \citenamefont {Dymov}, \citenamefont {Eversmann},
  \citenamefont {Gaisser}, \citenamefont {Gebel}, \citenamefont {Grigoryev},
  \citenamefont {Grzonka}, \citenamefont {Guidoboni}, \citenamefont
  {Heberling}, \citenamefont {Hejny}, \citenamefont {Hempelmann}, \citenamefont
  {Hetzel}, \citenamefont {Hinder}, \citenamefont {Kacharava}, \citenamefont
  {Kamerdzhiev}, \citenamefont {Keshelashvili}, \citenamefont {Koop},
  \citenamefont {Kulikov}, \citenamefont {Lehrach}, \citenamefont {Lenisa},
  \citenamefont {Lomidze}, \citenamefont {Lorentz}, \citenamefont {Maanen},
  \citenamefont {Macharashvili}, \citenamefont {Magiera}, \citenamefont
  {Mchedlishvili}, \citenamefont {Mey}, \citenamefont {M\"uller}, \citenamefont
  {Nass}, \citenamefont {Pesce}, \citenamefont {Prasuhn}, \citenamefont
  {Pretz}, \citenamefont {Rosenthal}, \citenamefont {Schmidt}, \citenamefont
  {Semertzidis}, \citenamefont {Senichev}, \citenamefont {Shmakova},
  \citenamefont {Silenko}, \citenamefont {Slim}, \citenamefont {Soltner},
  \citenamefont {Stahl}, \citenamefont {Stassen}, \citenamefont {Stephenson},
  \citenamefont {Stockhorst}, \citenamefont {Str\"oher}, \citenamefont
  {Tabidze}, \citenamefont {Tagliente}, \citenamefont {Talman}, \citenamefont
  {Eng\-blom}, \citenamefont {Trinkel}, \citenamefont {Uzikov}, \citenamefont
  {Valdau}, \citenamefont {Valetov}, \citenamefont {Vassiliev}, \citenamefont
  {Weidemann}, \citenamefont {Wro\ifmmode~\acute{n}\else \'{n}\fi{}ska},
  \citenamefont {W\"ustner}, \citenamefont {Zupra\ifmmode~\acute{n}\else
  \'{n}\fi{}ski},\ and\ \citenamefont
  {\.{Z}urek}}]{PhysRevAccelBeams.20.072801}%
  \BibitemOpen
  \bibfield  {author} {\bibinfo {author} {\bibfnamefont {A.}~\bibnamefont
  {Saleev}}, \bibinfo {author} {\bibfnamefont {N.~N.}\ \bibnamefont
  {Nikolaev}}, \bibinfo {author} {\bibfnamefont {F.}~\bibnamefont {Rathmann}},
  \bibinfo {author} {\bibfnamefont {W.}~\bibnamefont {Augustyniak}}, \bibinfo
  {author} {\bibfnamefont {Z.}~\bibnamefont {Bagdasarian}}, \bibinfo {author}
  {\bibfnamefont {M.}~\bibnamefont {Bai}}, \bibinfo {author} {\bibfnamefont
  {L.}~\bibnamefont {Barion}}, \bibinfo {author} {\bibfnamefont
  {M.}~\bibnamefont {Berz}}, \bibinfo {author} {\bibfnamefont {S.}~\bibnamefont
  {Chekmenev}}, \bibinfo {author} {\bibfnamefont {G.}~\bibnamefont {Ciullo}},
  \bibinfo {author} {\bibfnamefont {S.}~\bibnamefont {Dymov}}, \bibinfo
  {author} {\bibfnamefont {D.}~\bibnamefont {Eversmann}}, \bibinfo {author}
  {\bibfnamefont {M.}~\bibnamefont {Gaisser}}, \bibinfo {author} {\bibfnamefont
  {R.}~\bibnamefont {Gebel}}, \bibinfo {author} {\bibfnamefont
  {K.}~\bibnamefont {Grigoryev}}, \bibinfo {author} {\bibfnamefont
  {D.}~\bibnamefont {Grzonka}}, \bibinfo {author} {\bibfnamefont
  {G.}~\bibnamefont {Guidoboni}}, \bibinfo {author} {\bibfnamefont
  {D.}~\bibnamefont {Heberling}}, \bibinfo {author} {\bibfnamefont
  {V.}~\bibnamefont {Hejny}}, \bibinfo {author} {\bibfnamefont
  {N.}~\bibnamefont {Hempelmann}}, \bibinfo {author} {\bibfnamefont
  {J.}~\bibnamefont {Hetzel}}, \bibinfo {author} {\bibfnamefont
  {F.}~\bibnamefont {Hinder}}, \bibinfo {author} {\bibfnamefont
  {A.}~\bibnamefont {Kacharava}}, \bibinfo {author} {\bibfnamefont
  {V.}~\bibnamefont {Kamerdzhiev}}, \bibinfo {author} {\bibfnamefont
  {I.}~\bibnamefont {Keshelashvili}}, \bibinfo {author} {\bibfnamefont
  {I.}~\bibnamefont {Koop}}, \bibinfo {author} {\bibfnamefont {A.}~\bibnamefont
  {Kulikov}}, \bibinfo {author} {\bibfnamefont {A.}~\bibnamefont {Lehrach}},
  \bibinfo {author} {\bibfnamefont {P.}~\bibnamefont {Lenisa}}, \bibinfo
  {author} {\bibfnamefont {N.}~\bibnamefont {Lomidze}}, \bibinfo {author}
  {\bibfnamefont {B.}~\bibnamefont {Lorentz}}, \bibinfo {author} {\bibfnamefont
  {P.}~\bibnamefont {Maanen}}, \bibinfo {author} {\bibfnamefont
  {G.}~\bibnamefont {Macharashvili}}, \bibinfo {author} {\bibfnamefont
  {A.}~\bibnamefont {Magiera}}, \bibinfo {author} {\bibfnamefont
  {D.}~\bibnamefont {Mchedlishvili}}, \bibinfo {author} {\bibfnamefont
  {S.}~\bibnamefont {Mey}}, \bibinfo {author} {\bibfnamefont {F.}~\bibnamefont
  {M\"uller}}, \bibinfo {author} {\bibfnamefont {A.}~\bibnamefont {Nass}},
  \bibinfo {author} {\bibfnamefont {A.}~\bibnamefont {Pesce}}, \bibinfo
  {author} {\bibfnamefont {D.}~\bibnamefont {Prasuhn}}, \bibinfo {author}
  {\bibfnamefont {J.}~\bibnamefont {Pretz}}, \bibinfo {author} {\bibfnamefont
  {M.}~\bibnamefont {Rosenthal}}, \bibinfo {author} {\bibfnamefont
  {V.}~\bibnamefont {Schmidt}}, \bibinfo {author} {\bibfnamefont
  {Y.}~\bibnamefont {Semertzidis}}, \bibinfo {author} {\bibfnamefont
  {Y.}~\bibnamefont {Senichev}}, \bibinfo {author} {\bibfnamefont
  {V.}~\bibnamefont {Shmakova}}, \bibinfo {author} {\bibfnamefont
  {A.}~\bibnamefont {Silenko}}, \bibinfo {author} {\bibfnamefont
  {J.}~\bibnamefont {Slim}}, \bibinfo {author} {\bibfnamefont {H.}~\bibnamefont
  {Soltner}}, \bibinfo {author} {\bibfnamefont {A.}~\bibnamefont {Stahl}},
  \bibinfo {author} {\bibfnamefont {R.}~\bibnamefont {Stassen}}, \bibinfo
  {author} {\bibfnamefont {E.}~\bibnamefont {Stephenson}}, \bibinfo {author}
  {\bibfnamefont {H.}~\bibnamefont {Stockhorst}}, \bibinfo {author}
  {\bibfnamefont {H.}~\bibnamefont {Str\"oher}}, \bibinfo {author}
  {\bibfnamefont {M.}~\bibnamefont {Tabidze}}, \bibinfo {author} {\bibfnamefont
  {G.}~\bibnamefont {Tagliente}}, \bibinfo {author} {\bibfnamefont
  {R.}~\bibnamefont {Talman}}, \bibinfo {author} {\bibfnamefont
  {P.~Th\"orngren}\ \bibnamefont {Eng\-blom}}, \bibinfo {author} {\bibfnamefont
  {F.}~\bibnamefont {Trinkel}}, \bibinfo {author} {\bibfnamefont {Yu.}\
  \bibnamefont {Uzikov}}, \bibinfo {author} {\bibfnamefont {Yu.}\ \bibnamefont
  {Valdau}}, \bibinfo {author} {\bibfnamefont {E.}~\bibnamefont {Valetov}},
  \bibinfo {author} {\bibfnamefont {A.}~\bibnamefont {Vassiliev}}, \bibinfo
  {author} {\bibfnamefont {C.}~\bibnamefont {Weidemann}}, \bibinfo {author}
  {\bibfnamefont {A.}~\bibnamefont {Wro\ifmmode~\acute{n}\else \'{n}\fi{}ska}},
  \bibinfo {author} {\bibfnamefont {P.}~\bibnamefont {W\"ustner}}, \bibinfo
  {author} {\bibfnamefont {P.}~\bibnamefont {Zupra\ifmmode~\acute{n}\else
  \'{n}\fi{}ski}}, \ and\ \bibinfo {author} {\bibfnamefont {M.}~\bibnamefont
  {\.{Z}urek}} (\bibinfo {collaboration} {JEDI}),\ }\bibfield  {title}
  {\enquote {\bibinfo {title} {Spin tune mapping as a novel tool to probe the
  spin dynamics in storage rings},}\ }\href {\doibase
  10.1103/PhysRevAccelBeams.20.072801} {\bibfield  {journal} {\bibinfo
  {journal} {Phys. Rev. Accel. Beams}\ }\textbf {\bibinfo {volume} {20}},\
  \bibinfo {pages} {072801} (\bibinfo {year} {2017})}\BibitemShut {NoStop}%
\bibitem [{\citenamefont {Wagner}\ \emph {et~al.}(2021)\citenamefont {Wagner}
  \emph {et~al.}}]{Wagner:2020akw}%
  \BibitemOpen
  \bibfield  {author} {\bibinfo {author} {\bibfnamefont {T.}~\bibnamefont
  {Wagner}} \emph {et~al.} (\bibinfo {collaboration} {JEDI}),\ }\bibfield
  {title} {\enquote {\bibinfo {title} {{Beam-based alignment at the Cooler
  Synchrotron COSY as a prerequisite for an electric dipole moment
  measurement}},}\ }\href {\doibase 10.1088/1748-0221/16/02/t02001} {\bibfield
  {journal} {\bibinfo  {journal} {JINST}\ }\textbf {\bibinfo {volume} {16}},\
  \bibinfo {pages} {T02001} (\bibinfo {year} {2021})},\ \Eprint
  {http://arxiv.org/abs/2009.02058} {arXiv:2009.02058 [physics.acc-ph]}
  \BibitemShut {NoStop}%
\bibitem [{\citenamefont {Pospelov}\ and\ \citenamefont
  {Ritz}(2005)}]{POSPELOV2005119}%
  \BibitemOpen
  \bibfield  {author} {\bibinfo {author} {\bibfnamefont {Maxim}\ \bibnamefont
  {Pospelov}}\ and\ \bibinfo {author} {\bibfnamefont {Adam}\ \bibnamefont
  {Ritz}},\ }\bibfield  {title} {\enquote {\bibinfo {title} {Electric dipole
  moments as probes of new physics},}\ }\href {\doibase
  https://doi.org/10.1016/j.aop.2005.04.002} {\bibfield  {journal} {\bibinfo
  {journal} {Annals of Physics}\ }\textbf {\bibinfo {volume} {318}},\ \bibinfo
  {pages} {119 -- 169} (\bibinfo {year} {2005})},\ \bibinfo {note} {special
  Issue}\BibitemShut {NoStop}%
\bibitem [{\citenamefont {Bernreuther}(2002)}]{Bernreuther:2002uj}%
  \BibitemOpen
  \bibfield  {author} {\bibinfo {author} {\bibfnamefont {Werner}\ \bibnamefont
  {Bernreuther}},\ }\bibfield  {title} {\enquote {\bibinfo {title} {{CP
  violation and baryogenesis}},}\ }\href@noop {} {\bibfield  {journal}
  {\bibinfo  {journal} {Lect. Notes Phys.}\ }\textbf {\bibinfo {volume}
  {591}},\ \bibinfo {pages} {237--293} (\bibinfo {year} {2002})},\ \Eprint
  {http://arxiv.org/abs/hep-ph/0205279} {arXiv:hep-ph/0205279} \BibitemShut
  {NoStop}%
\bibitem [{\citenamefont {Anastassopoulos}\ \emph {et~al.}(2016)\citenamefont
  {Anastassopoulos}, \citenamefont {Andrianov}, \citenamefont {Baartman},
  \citenamefont {Baessler}, \citenamefont {Bai}, \citenamefont {Benante},
  \citenamefont {Berz}, \citenamefont {Blaskiewicz}, \citenamefont {Bowcock},
  \citenamefont {Brown}, \citenamefont {Casey}, \citenamefont {Conte},
  \citenamefont {Crnkovic}, \citenamefont {D’Imperio}, \citenamefont
  {Fanourakis}, \citenamefont {Fedotov}, \citenamefont {Fierlinger},
  \citenamefont {Fischer}, \citenamefont {Gaisser}, \citenamefont {Giomataris},
  \citenamefont {Grosse-Perdekamp}, \citenamefont {Guidoboni}, \citenamefont
  {Hacıömeroğlu}, \citenamefont {Hoffstaetter}, \citenamefont {Huang},
  \citenamefont {Incagli}, \citenamefont {Ivanov}, \citenamefont {Kawall},
  \citenamefont {Kim}, \citenamefont {King}, \citenamefont {Koop},
  \citenamefont {Lazarus}, \citenamefont {Lebedev}, \citenamefont {Lee},
  \citenamefont {Lee}, \citenamefont {Lee}, \citenamefont {Lehrach},
  \citenamefont {Lenisa}, \citenamefont {Levi~Sandri}, \citenamefont {Luccio},
  \citenamefont {Lyapin}, \citenamefont {MacKay}, \citenamefont {Maier},
  \citenamefont {Makino}, \citenamefont {Malitsky}, \citenamefont {Marciano},
  \citenamefont {Meng}, \citenamefont {Meot}, \citenamefont {Metodiev},
  \citenamefont {Miceli}, \citenamefont {Moricciani}, \citenamefont {Morse},
  \citenamefont {Nagaitsev}, \citenamefont {Nayak}, \citenamefont {Orlov},
  \citenamefont {Ozben}, \citenamefont {Park}, \citenamefont {Pesce},
  \citenamefont {Petrakou}, \citenamefont {Pile}, \citenamefont {Podobedov},
  \citenamefont {Polychronakos}, \citenamefont {Pretz}, \citenamefont
  {Ptitsyn}, \citenamefont {Ramberg}, \citenamefont {Raparia}, \citenamefont
  {Rathmann}, \citenamefont {Rescia}, \citenamefont {Roser}, \citenamefont
  {Kamal~Sayed}, \citenamefont {Semertzidis}, \citenamefont {Senichev},
  \citenamefont {Sidorin}, \citenamefont {Silenko}, \citenamefont {Simos},
  \citenamefont {Stahl}, \citenamefont {Stephenson}, \citenamefont {Ströher},
  \citenamefont {Syphers}, \citenamefont {Talman}, \citenamefont {Talman},
  \citenamefont {Tishchenko}, \citenamefont {Touramanis}, \citenamefont
  {Tsoupas}, \citenamefont {Venanzoni}, \citenamefont {Vetter}, \citenamefont
  {Vlassis}, \citenamefont {Won}, \citenamefont {Zavattini}, \citenamefont
  {Zelenski},\ and\ \citenamefont {Zioutas}}]{doi:10.1063/1.4967465}%
  \BibitemOpen
  \bibfield  {author} {\bibinfo {author} {\bibfnamefont {V.}~\bibnamefont
  {Anastassopoulos}}, \bibinfo {author} {\bibfnamefont {S.}~\bibnamefont
  {Andrianov}}, \bibinfo {author} {\bibfnamefont {R.}~\bibnamefont {Baartman}},
  \bibinfo {author} {\bibfnamefont {S.}~\bibnamefont {Baessler}}, \bibinfo
  {author} {\bibfnamefont {M.}~\bibnamefont {Bai}}, \bibinfo {author}
  {\bibfnamefont {J.}~\bibnamefont {Benante}}, \bibinfo {author} {\bibfnamefont
  {M.}~\bibnamefont {Berz}}, \bibinfo {author} {\bibfnamefont {M.}~\bibnamefont
  {Blaskiewicz}}, \bibinfo {author} {\bibfnamefont {T.}~\bibnamefont
  {Bowcock}}, \bibinfo {author} {\bibfnamefont {K.}~\bibnamefont {Brown}},
  \bibinfo {author} {\bibfnamefont {B.}~\bibnamefont {Casey}}, \bibinfo
  {author} {\bibfnamefont {M.}~\bibnamefont {Conte}}, \bibinfo {author}
  {\bibfnamefont {J.~D.}\ \bibnamefont {Crnkovic}}, \bibinfo {author}
  {\bibfnamefont {N.}~\bibnamefont {D’Imperio}}, \bibinfo {author}
  {\bibfnamefont {G.}~\bibnamefont {Fanourakis}}, \bibinfo {author}
  {\bibfnamefont {A.}~\bibnamefont {Fedotov}}, \bibinfo {author} {\bibfnamefont
  {P.}~\bibnamefont {Fierlinger}}, \bibinfo {author} {\bibfnamefont
  {W.}~\bibnamefont {Fischer}}, \bibinfo {author} {\bibfnamefont {M.~O.}\
  \bibnamefont {Gaisser}}, \bibinfo {author} {\bibfnamefont {Y.}~\bibnamefont
  {Giomataris}}, \bibinfo {author} {\bibfnamefont {M.}~\bibnamefont
  {Grosse-Perdekamp}}, \bibinfo {author} {\bibfnamefont {G.}~\bibnamefont
  {Guidoboni}}, \bibinfo {author} {\bibfnamefont {S.}~\bibnamefont
  {Hacıömeroğlu}}, \bibinfo {author} {\bibfnamefont {G.}~\bibnamefont
  {Hoffstaetter}}, \bibinfo {author} {\bibfnamefont {H.}~\bibnamefont {Huang}},
  \bibinfo {author} {\bibfnamefont {M.}~\bibnamefont {Incagli}}, \bibinfo
  {author} {\bibfnamefont {A.}~\bibnamefont {Ivanov}}, \bibinfo {author}
  {\bibfnamefont {D.}~\bibnamefont {Kawall}}, \bibinfo {author} {\bibfnamefont
  {Y.~I.}\ \bibnamefont {Kim}}, \bibinfo {author} {\bibfnamefont
  {B.}~\bibnamefont {King}}, \bibinfo {author} {\bibfnamefont {I.~A.}\
  \bibnamefont {Koop}}, \bibinfo {author} {\bibfnamefont {D.~M.}\ \bibnamefont
  {Lazarus}}, \bibinfo {author} {\bibfnamefont {V.}~\bibnamefont {Lebedev}},
  \bibinfo {author} {\bibfnamefont {M.~J.}\ \bibnamefont {Lee}}, \bibinfo
  {author} {\bibfnamefont {S.}~\bibnamefont {Lee}}, \bibinfo {author}
  {\bibfnamefont {Y.~H.}\ \bibnamefont {Lee}}, \bibinfo {author} {\bibfnamefont
  {A.}~\bibnamefont {Lehrach}}, \bibinfo {author} {\bibfnamefont
  {P.}~\bibnamefont {Lenisa}}, \bibinfo {author} {\bibfnamefont
  {P.}~\bibnamefont {Levi~Sandri}}, \bibinfo {author} {\bibfnamefont {A.~U.}\
  \bibnamefont {Luccio}}, \bibinfo {author} {\bibfnamefont {A.}~\bibnamefont
  {Lyapin}}, \bibinfo {author} {\bibfnamefont {W.}~\bibnamefont {MacKay}},
  \bibinfo {author} {\bibfnamefont {R.}~\bibnamefont {Maier}}, \bibinfo
  {author} {\bibfnamefont {K.}~\bibnamefont {Makino}}, \bibinfo {author}
  {\bibfnamefont {N.}~\bibnamefont {Malitsky}}, \bibinfo {author}
  {\bibfnamefont {W.~J.}\ \bibnamefont {Marciano}}, \bibinfo {author}
  {\bibfnamefont {W.}~\bibnamefont {Meng}}, \bibinfo {author} {\bibfnamefont
  {F.}~\bibnamefont {Meot}}, \bibinfo {author} {\bibfnamefont {E.~M.}\
  \bibnamefont {Metodiev}}, \bibinfo {author} {\bibfnamefont {L.}~\bibnamefont
  {Miceli}}, \bibinfo {author} {\bibfnamefont {D.}~\bibnamefont {Moricciani}},
  \bibinfo {author} {\bibfnamefont {W.~M.}\ \bibnamefont {Morse}}, \bibinfo
  {author} {\bibfnamefont {S.}~\bibnamefont {Nagaitsev}}, \bibinfo {author}
  {\bibfnamefont {S.~K.}\ \bibnamefont {Nayak}}, \bibinfo {author}
  {\bibfnamefont {Y.~F.}\ \bibnamefont {Orlov}}, \bibinfo {author}
  {\bibfnamefont {C.~S.}\ \bibnamefont {Ozben}}, \bibinfo {author}
  {\bibfnamefont {S.~T.}\ \bibnamefont {Park}}, \bibinfo {author}
  {\bibfnamefont {A.}~\bibnamefont {Pesce}}, \bibinfo {author} {\bibfnamefont
  {E.}~\bibnamefont {Petrakou}}, \bibinfo {author} {\bibfnamefont
  {P.}~\bibnamefont {Pile}}, \bibinfo {author} {\bibfnamefont {B.}~\bibnamefont
  {Podobedov}}, \bibinfo {author} {\bibfnamefont {V.}~\bibnamefont
  {Polychronakos}}, \bibinfo {author} {\bibfnamefont {J.}~\bibnamefont
  {Pretz}}, \bibinfo {author} {\bibfnamefont {V.}~\bibnamefont {Ptitsyn}},
  \bibinfo {author} {\bibfnamefont {E.}~\bibnamefont {Ramberg}}, \bibinfo
  {author} {\bibfnamefont {D.}~\bibnamefont {Raparia}}, \bibinfo {author}
  {\bibfnamefont {F.}~\bibnamefont {Rathmann}}, \bibinfo {author}
  {\bibfnamefont {S.}~\bibnamefont {Rescia}}, \bibinfo {author} {\bibfnamefont
  {T.}~\bibnamefont {Roser}}, \bibinfo {author} {\bibfnamefont
  {H.}~\bibnamefont {Kamal~Sayed}}, \bibinfo {author} {\bibfnamefont {Y.~K.}\
  \bibnamefont {Semertzidis}}, \bibinfo {author} {\bibfnamefont
  {Y.}~\bibnamefont {Senichev}}, \bibinfo {author} {\bibfnamefont
  {A.}~\bibnamefont {Sidorin}}, \bibinfo {author} {\bibfnamefont
  {A.}~\bibnamefont {Silenko}}, \bibinfo {author} {\bibfnamefont
  {N.}~\bibnamefont {Simos}}, \bibinfo {author} {\bibfnamefont
  {A.}~\bibnamefont {Stahl}}, \bibinfo {author} {\bibfnamefont {E.~J.}\
  \bibnamefont {Stephenson}}, \bibinfo {author} {\bibfnamefont
  {H.}~\bibnamefont {Ströher}}, \bibinfo {author} {\bibfnamefont {M.~J.}\
  \bibnamefont {Syphers}}, \bibinfo {author} {\bibfnamefont {J.}~\bibnamefont
  {Talman}}, \bibinfo {author} {\bibfnamefont {R.~M.}\ \bibnamefont {Talman}},
  \bibinfo {author} {\bibfnamefont {V.}~\bibnamefont {Tishchenko}}, \bibinfo
  {author} {\bibfnamefont {C.}~\bibnamefont {Touramanis}}, \bibinfo {author}
  {\bibfnamefont {N.}~\bibnamefont {Tsoupas}}, \bibinfo {author} {\bibfnamefont
  {G.}~\bibnamefont {Venanzoni}}, \bibinfo {author} {\bibfnamefont
  {K.}~\bibnamefont {Vetter}}, \bibinfo {author} {\bibfnamefont
  {S.}~\bibnamefont {Vlassis}}, \bibinfo {author} {\bibfnamefont
  {E.}~\bibnamefont {Won}}, \bibinfo {author} {\bibfnamefont {G.}~\bibnamefont
  {Zavattini}}, \bibinfo {author} {\bibfnamefont {A.}~\bibnamefont {Zelenski}},
  \ and\ \bibinfo {author} {\bibfnamefont {K.}~\bibnamefont {Zioutas}},\
  }\bibfield  {title} {\enquote {\bibinfo {title} {A storage ring experiment to
  detect a proton electric dipole moment},}\ }\href {\doibase
  10.1063/1.4967465} {\bibfield  {journal} {\bibinfo  {journal} {Review of
  Scientific Instruments}\ }\textbf {\bibinfo {volume} {87}},\ \bibinfo {pages}
  {115116} (\bibinfo {year} {2016})},\ \Eprint
  {http://arxiv.org/abs/https://aip.scitation.org/doi/pdf/10.1063/1.4967465}
  {https://aip.scitation.org/doi/pdf/10.1063/1.4967465} \BibitemShut {NoStop}%
\bibitem [{\citenamefont {Maier}(1997)}]{Maier19971}%
  \BibitemOpen
  \bibfield  {author} {\bibinfo {author} {\bibfnamefont {R.}~\bibnamefont
  {Maier}},\ }\bibfield  {title} {\enquote {\bibinfo {title} {Cooler
  synchrotron {COSY} — performance and perspectives},}\ }\href {\doibase
  http://dx.doi.org/10.1016/S0168-9002(97)00324-0} {\bibfield  {journal}
  {\bibinfo  {journal} {Nuclear Instruments and Methods in Physics Research
  Section A: Accelerators, Spectrometers, Detectors and Associated Equipment}\
  }\textbf {\bibinfo {volume} {390}},\ \bibinfo {pages} {1 -- 8} (\bibinfo
  {year} {1997})}\BibitemShut {NoStop}%
\bibitem [{\citenamefont {Rathmann}\ \emph {et~al.}(2013)\citenamefont
  {Rathmann}, \citenamefont {Saleev},\ and\ \citenamefont
  {Nikolaev}}]{Rathmann:2013rqa}%
  \BibitemOpen
  \bibfield  {author} {\bibinfo {author} {\bibfnamefont {Frank}\ \bibnamefont
  {Rathmann}}, \bibinfo {author} {\bibfnamefont {Artem}\ \bibnamefont
  {Saleev}}, \ and\ \bibinfo {author} {\bibfnamefont {N.~N.}\ \bibnamefont
  {Nikolaev}} (\bibinfo {collaboration} {JEDI, srEDM}),\ }\bibfield  {title}
  {\enquote {\bibinfo {title} {{The search for electric dipole moments of light
  ions in storage rings}},}\ }\href {\doibase 10.1088/1742-6596/447/1/012011}
  {\bibfield  {journal} {\bibinfo  {journal} {J.\ Phys.\ Conf.\ Ser.}\ }\textbf
  {\bibinfo {volume} {447}},\ \bibinfo {pages} {012011} (\bibinfo {year}
  {2013})}\BibitemShut {NoStop}%
\bibitem [{\citenamefont {Morse}\ \emph {et~al.}(2013)\citenamefont {Morse},
  \citenamefont {Orlov},\ and\ \citenamefont {Semertzidis}}]{Morse:2013hoa}%
  \BibitemOpen
  \bibfield  {author} {\bibinfo {author} {\bibfnamefont {William~M.}\
  \bibnamefont {Morse}}, \bibinfo {author} {\bibfnamefont {Yuri~F.}\
  \bibnamefont {Orlov}}, \ and\ \bibinfo {author} {\bibfnamefont {Yannis~K.}\
  \bibnamefont {Semertzidis}},\ }\bibfield  {title} {\enquote {\bibinfo {title}
  {{rf Wien filter in an electric dipole moment storage ring: The ``partially
  frozen spin'' effect}},}\ }\href {\doibase 10.1103/PhysRevSTAB.16.114001}
  {\bibfield  {journal} {\bibinfo  {journal} {Phys.\ Rev.\ ST Accel.\ Beams}\
  }\textbf {\bibinfo {volume} {16}},\ \bibinfo {pages} {114001} (\bibinfo
  {year} {2013})}\BibitemShut {NoStop}%
\bibitem [{\citenamefont {Rathmann}\ \emph {et~al.}(2020)\citenamefont
  {Rathmann}, \citenamefont {Nikolaev},\ and\ \citenamefont
  {Slim}}]{PhysRevAccelBeams.23.024601}%
  \BibitemOpen
  \bibfield  {author} {\bibinfo {author} {\bibfnamefont {F.}~\bibnamefont
  {Rathmann}}, \bibinfo {author} {\bibfnamefont {N.~N.}\ \bibnamefont
  {Nikolaev}}, \ and\ \bibinfo {author} {\bibfnamefont {J.}~\bibnamefont
  {Slim}},\ }\bibfield  {title} {\enquote {\bibinfo {title} {Spin dynamics
  investigations for the electric dipole moment experiment},}\ }\href {\doibase
  10.1103/PhysRevAccelBeams.23.024601} {\bibfield  {journal} {\bibinfo
  {journal} {Phys. Rev. Accel. Beams}\ }\textbf {\bibinfo {volume} {23}},\
  \bibinfo {pages} {024601} (\bibinfo {year} {2020})}\BibitemShut {NoStop}%
\bibitem [{\citenamefont {Slim}\ \emph {et~al.}(2016)\citenamefont {Slim},
  \citenamefont {Gebel}, \citenamefont {Heberling}, \citenamefont {Hinder},
  \citenamefont {H{\"o}lscher}, \citenamefont {Lehrach}, \citenamefont
  {Lorentz}, \citenamefont {Mey}, \citenamefont {Nass}, \citenamefont
  {Rathmann}, \citenamefont {Reifferscheidt}, \citenamefont {Soltner},
  \citenamefont {Straatmann}, \citenamefont {Trinkel},\ and\ \citenamefont
  {Wolters}}]{Slim2016116}%
  \BibitemOpen
  \bibfield  {author} {\bibinfo {author} {\bibfnamefont {J.}~\bibnamefont
  {Slim}}, \bibinfo {author} {\bibfnamefont {R.}~\bibnamefont {Gebel}},
  \bibinfo {author} {\bibfnamefont {D.}~\bibnamefont {Heberling}}, \bibinfo
  {author} {\bibfnamefont {F.}~\bibnamefont {Hinder}}, \bibinfo {author}
  {\bibfnamefont {D.}~\bibnamefont {H{\"o}lscher}}, \bibinfo {author}
  {\bibfnamefont {A.}~\bibnamefont {Lehrach}}, \bibinfo {author} {\bibfnamefont
  {B.}~\bibnamefont {Lorentz}}, \bibinfo {author} {\bibfnamefont
  {S.}~\bibnamefont {Mey}}, \bibinfo {author} {\bibfnamefont {A.}~\bibnamefont
  {Nass}}, \bibinfo {author} {\bibfnamefont {F.}~\bibnamefont {Rathmann}},
  \bibinfo {author} {\bibfnamefont {L.}~\bibnamefont {Reifferscheidt}},
  \bibinfo {author} {\bibfnamefont {H.}~\bibnamefont {Soltner}}, \bibinfo
  {author} {\bibfnamefont {H.}~\bibnamefont {Straatmann}}, \bibinfo {author}
  {\bibfnamefont {F.}~\bibnamefont {Trinkel}}, \ and\ \bibinfo {author}
  {\bibfnamefont {J.}~\bibnamefont {Wolters}},\ }\bibfield  {title} {\enquote
  {\bibinfo {title} {{Electromagnetic Simulation and Design of a Novel
  Waveguide {RF} Wien Filter for Electric Dipole Moment Measurements of Protons
  and Deuterons}},}\ }\href {\doibase
  http://dx.doi.org/10.1016/j.nima.2016.05.012} {\bibfield  {journal} {\bibinfo
   {journal} {Nuclear Instruments and Methods in Physics Research Section A:
  Accelerators, Spectrometers, Detectors and Associated Equipment}\ }\textbf
  {\bibinfo {volume} {828}},\ \bibinfo {pages} {116 -- 124} (\bibinfo {year}
  {2016})}\BibitemShut {NoStop}%
\bibitem [{\citenamefont {Slim}\ \emph {et~al.}(2017)\citenamefont {Slim},
  \citenamefont {Rathmann}, \citenamefont {Nass}, \citenamefont {Soltner},
  \citenamefont {Gebel}, \citenamefont {Pretz},\ and\ \citenamefont
  {Heberling}}]{slim201752}%
  \BibitemOpen
  \bibfield  {author} {\bibinfo {author} {\bibfnamefont {J.}~\bibnamefont
  {Slim}}, \bibinfo {author} {\bibfnamefont {F.}~\bibnamefont {Rathmann}},
  \bibinfo {author} {\bibfnamefont {A.}~\bibnamefont {Nass}}, \bibinfo {author}
  {\bibfnamefont {H.}~\bibnamefont {Soltner}}, \bibinfo {author} {\bibfnamefont
  {R.}~\bibnamefont {Gebel}}, \bibinfo {author} {\bibfnamefont
  {J.}~\bibnamefont {Pretz}}, \ and\ \bibinfo {author} {\bibfnamefont
  {D.}~\bibnamefont {Heberling}},\ }\bibfield  {title} {\enquote {\bibinfo
  {title} {Polynomial chaos expansion method as a tool to evaluate and quantify
  field homogeneities of a novel waveguide rf wien filter},}\ }\href {\doibase
  https://doi.org/10.1016/j.nima.2017.03.040} {\bibfield  {journal} {\bibinfo
  {journal} {Nuclear Instruments and Methods in Physics Research Section A:
  Accelerators, Spectrometers, Detectors and Associated Equipment}\ }\textbf
  {\bibinfo {volume} {859}},\ \bibinfo {pages} {52 -- 62} (\bibinfo {year}
  {2017})}\BibitemShut {NoStop}%
\bibitem [{\citenamefont {Slim}(2018)}]{Slim:748558}%
  \BibitemOpen
  \bibfield  {author} {\bibinfo {author} {\bibfnamefont {Jamal}\ \bibnamefont
  {Slim}},\ }\emph {\bibinfo {title} {{A} novel waveguide {RF} {W}ien filter
  for electric dipole moment measurements of deuterons and protons at the
  {CO}oler {SY}nchrotron ({COSY})/{J}ülich}},\ \href {\doibase
  10.18154/RWTH-2018-229484} {\bibinfo {type} {Dissertation}},\ \bibinfo
  {school} {Rheinisch-Westfälische Technische Hochschule Aachen}, \bibinfo
  {address} {Aachen} (\bibinfo {year} {2018}),\ \bibinfo {note} {published on
  the publication server of RWTH Aachen University. Awarded the Borchers
  Plakette and the Friedrich Wilhelm Prize 2019. Dissertation,
  Rheinisch-Westfälische Technische Hochschule Aachen, 2018,
  \url{https://publications.rwth-aachen.de/record/748558}}\BibitemShut
  {NoStop}%
\bibitem [{\citenamefont {Weidemann}\ \emph {et~al.}(2015)\citenamefont
  {Weidemann} \emph {et~al.}}]{PAX}%
  \BibitemOpen
  \bibfield  {author} {\bibinfo {author} {\bibfnamefont {C.}~\bibnamefont
  {Weidemann}} \emph {et~al.},\ }\bibfield  {title} {\enquote {\bibinfo {title}
  {{Toward polarized antiprotons: Machine development for spin-filtering
  experiments}},}\ }\href {\doibase 10.1103/PhysRevSTAB.18.020101} {\bibfield
  {journal} {\bibinfo  {journal} {Phys. Rev. ST Accel. Beams}\ }\textbf
  {\bibinfo {volume} {18}},\ \bibinfo {pages} {020101} (\bibinfo {year}
  {2015})},\ \Eprint {http://arxiv.org/abs/1407.6724} {arXiv:1407.6724
  [physics.acc-ph]} \BibitemShut {NoStop}%
\bibitem [{\citenamefont {Hacıömeroğlu}\ \emph {et~al.}(2019)\citenamefont
  {Hacıömeroğlu}, \citenamefont {Kawall}, \citenamefont {Lee}, \citenamefont
  {Matlashov}, \citenamefont {Omarov},\ and\ \citenamefont
  {Semertzidis}}]{Haciomeroglu:2018son}%
  \BibitemOpen
  \bibfield  {author} {\bibinfo {author} {\bibfnamefont {Selcuk}\ \bibnamefont
  {Hacıömeroğlu}}, \bibinfo {author} {\bibfnamefont {David}\ \bibnamefont
  {Kawall}}, \bibinfo {author} {\bibfnamefont {Yong-Ho}\ \bibnamefont {Lee}},
  \bibinfo {author} {\bibfnamefont {Andrei}\ \bibnamefont {Matlashov}},
  \bibinfo {author} {\bibfnamefont {Zhanibek}\ \bibnamefont {Omarov}}, \ and\
  \bibinfo {author} {\bibfnamefont {Yannis~K.}\ \bibnamefont {Semertzidis}},\
  }\bibfield  {title} {\enquote {\bibinfo {title} {{SQUID-based beam position
  monitor}},}\ }\href {\doibase 10.22323/1.340.0279} {\bibfield  {journal}
  {\bibinfo  {journal} {PoS}\ }\textbf {\bibinfo {volume} {ICHEP2018}},\
  \bibinfo {pages} {279} (\bibinfo {year} {2019})}\BibitemShut {NoStop}%
\bibitem [{\citenamefont {Martin}\ \emph {et~al.}(1985)\citenamefont {Martin},
  \citenamefont {Prasuhn}, \citenamefont {Schott},\ and\ \citenamefont
  {Wiedner}}]{MARTIN1985249}%
  \BibitemOpen
  \bibfield  {author} {\bibinfo {author} {\bibfnamefont {S.A.}\ \bibnamefont
  {Martin}}, \bibinfo {author} {\bibfnamefont {D.}~\bibnamefont {Prasuhn}},
  \bibinfo {author} {\bibfnamefont {W.}~\bibnamefont {Schott}}, \ and\ \bibinfo
  {author} {\bibfnamefont {C.A.}\ \bibnamefont {Wiedner}},\ }\bibfield  {title}
  {\enquote {\bibinfo {title} {A storage ring for the julic cyclotron},}\
  }\href {\doibase https://doi.org/10.1016/0168-9002(85)90157-3} {\bibfield
  {journal} {\bibinfo  {journal} {Nuclear Instruments and Methods in Physics
  Research Section A: Accelerators, Spectrometers, Detectors and Associated
  Equipment}\ }\textbf {\bibinfo {volume} {236}},\ \bibinfo {pages} {249 --
  255} (\bibinfo {year} {1985})}\BibitemShut {NoStop}%
\bibitem [{\citenamefont {Forck}\ \emph {et~al.}(2008)\citenamefont {Forck},
  \citenamefont {Kowina},\ and\ \citenamefont {Liakin}}]{Forck:1213277}%
  \BibitemOpen
  \bibfield  {author} {\bibinfo {author} {\bibfnamefont {P.}~\bibnamefont
  {Forck}}, \bibinfo {author} {\bibfnamefont {P.}~\bibnamefont {Kowina}}, \
  and\ \bibinfo {author} {\bibfnamefont {D.}~\bibnamefont {Liakin}},\
  }\bibfield  {title} {\enquote {\bibinfo {title} {Beam position monitors},}\
  }in\ \href {\doibase 10.5170/CERN-2009-005} {\emph {\bibinfo {booktitle} {CAS
  - CERN Accelerator School: Course on Beam Diagnostics}}},\ \bibinfo {series
  and number} {CERN Accelerator School},\ \bibinfo {editor} {edited by\
  \bibinfo {editor} {\bibfnamefont {Daniel}\ \bibnamefont {Brandt}}}\ (\bibinfo
  {organization} {CERN},\ \bibinfo {year} {2008})\ p.\ \bibinfo {pages}
  {188}\BibitemShut {NoStop}%
\bibitem [{\citenamefont {B\"ohme}\ \emph {et~al.}(2018)\citenamefont
  {B\"ohme}, \citenamefont {Bekman}, \citenamefont {Kamerdzhiev}, \citenamefont
  {Lorentz}, \citenamefont {Simon},\ and\ \citenamefont
  {Weidemann}}]{Bohme:2018sjy}%
  \BibitemOpen
  \bibfield  {author} {\bibinfo {author} {\bibfnamefont {Christian}\
  \bibnamefont {B\"ohme}}, \bibinfo {author} {\bibfnamefont {Ilja}\
  \bibnamefont {Bekman}}, \bibinfo {author} {\bibfnamefont {Vsevolod}\
  \bibnamefont {Kamerdzhiev}}, \bibinfo {author} {\bibfnamefont {Bernd}\
  \bibnamefont {Lorentz}}, \bibinfo {author} {\bibfnamefont {Michael}\
  \bibnamefont {Simon}}, \ and\ \bibinfo {author} {\bibfnamefont {Christian}\
  \bibnamefont {Weidemann}},\ }\bibfield  {title} {\enquote {\bibinfo {title}
  {{COSY Orbit Control Upgrade}},}\ }in\ \href {\doibase
  10.18429/JACoW-IBIC2017-TUPCF20} {\emph {\bibinfo {booktitle} {{6th
  International Beam Instrumentation Conference}}}}\ (\bibinfo {year}
  {2018})\BibitemShut {NoStop}%
\bibitem [{\citenamefont {Dietrich}\ \emph {et~al.}(2014)\citenamefont
  {Dietrich}, \citenamefont {Kamerdzhiev}, \citenamefont {Parkhomchuk},
  \citenamefont {Reva},\ and\ \citenamefont {Bryzgunov}}]{Dietrich:2014uma}%
  \BibitemOpen
  \bibfield  {author} {\bibinfo {author} {\bibfnamefont {J\"urgen}\
  \bibnamefont {Dietrich}}, \bibinfo {author} {\bibfnamefont {V.}~\bibnamefont
  {Kamerdzhiev}}, \bibinfo {author} {\bibfnamefont {V.V.}\ \bibnamefont
  {Parkhomchuk}}, \bibinfo {author} {\bibfnamefont {V.B.}\ \bibnamefont
  {Reva}}, \ and\ \bibinfo {author} {\bibfnamefont {M.I.}\ \bibnamefont
  {Bryzgunov}},\ }\bibfield  {title} {\enquote {\bibinfo {title} {{2 MeV
  Electron Cooler at COSY Juelich}},}\ }\href@noop {} {\bibfield  {journal}
  {\bibinfo  {journal} {ICFA Beam Dyn. Newslett.}\ }\textbf {\bibinfo {volume}
  {64}},\ \bibinfo {pages} {75--86} (\bibinfo {year} {2014})}\BibitemShut
  {NoStop}%
\bibitem [{\citenamefont {Adam}\ \emph {et~al.}(2004)\citenamefont {Adam} \emph
  {et~al.}}]{Adam:2004ch}%
  \BibitemOpen
  \bibfield  {author} {\bibinfo {author} {\bibfnamefont {H.-H.}\ \bibnamefont
  {Adam}} \emph {et~al.} (\bibinfo {collaboration} {WASA-at-COSY}),\ }\bibfield
   {title} {\enquote {\bibinfo {title} {{Proposal for the wide angle shower
  apparatus (WASA) at COSY-Julich: WASA at COSY}},}\ }\href@noop {} {\
  (\bibinfo {year} {2004})},\ \Eprint {http://arxiv.org/abs/nucl-ex/0411038}
  {arXiv:nucl-ex/0411038} \BibitemShut {NoStop}%
\bibitem [{\citenamefont {M\"uller}(2019)}]{Muller:2019oxc}%
  \BibitemOpen
  \bibfield  {author} {\bibinfo {author} {\bibfnamefont {Fabian}\ \bibnamefont
  {M\"uller}},\ }\emph {\bibinfo {title} {{Polarimeter Development for Electric
  Dipole Moment Measurements in Storage Rings}}},\ \href {\doibase
  10.18154/RWTH-2019-11439} {Ph.D. thesis},\ \bibinfo  {school} {RWTH Aachen
  University} (\bibinfo {year} {2019})\BibitemShut {NoStop}%
\bibitem [{\citenamefont {Müller}\ \emph {et~al.}(2020)\citenamefont
  {Müller}, \citenamefont {Javakhishvili}, \citenamefont {Shergelashvili},
  \citenamefont {Keshelashvili}, \citenamefont {Mchedlishvili}, \citenamefont
  {Abusaif}, \citenamefont {Aggarwal}, \citenamefont {Barion}, \citenamefont
  {Basile}, \citenamefont {Böker}, \citenamefont {Canale}, \citenamefont
  {Ciullo}, \citenamefont {Dymov}, \citenamefont {Felden}, \citenamefont
  {Gagoshidze}, \citenamefont {Gebel}, \citenamefont {Demary}, \citenamefont
  {Grigoryev}, \citenamefont {Grzonka}, \citenamefont {Hahnraths},
  \citenamefont {Hejny}, \citenamefont {Kacharava}, \citenamefont
  {Kamerdzhiev}, \citenamefont {Karanth}, \citenamefont {Kulikov},
  \citenamefont {Lehrach}, \citenamefont {Lenisa}, \citenamefont {Lomidze},
  \citenamefont {Lorentz}, \citenamefont {Macharashvili}, \citenamefont
  {Magiera}, \citenamefont {Metreveli}, \citenamefont {Nass}, \citenamefont
  {Nikolaev}, \citenamefont {Nioradze}, \citenamefont {Pesce}, \citenamefont
  {Poncza}, \citenamefont {Prasuhn}, \citenamefont {Pretz}, \citenamefont
  {Rathmann}, \citenamefont {Saleev}, \citenamefont {Sef\-zick}, \citenamefont
  {Senichev}, \citenamefont {Shmakova}, \citenamefont {Slim}, \citenamefont
  {Soltner}, \citenamefont {Stephenson}, \citenamefont {Ströher},
  \citenamefont {Tabidze}, \citenamefont {Tagliente}, \citenamefont {Uzikov},
  \citenamefont {Valdau}, \citenamefont {Wagner}, \citenamefont
  {Wro{\'{n}}ska}, \citenamefont {Wüstner},\ and\ \citenamefont
  {\.{Z}urek}}]{Muller_2020}%
  \BibitemOpen
  \bibfield  {author} {\bibinfo {author} {\bibfnamefont {F.}~\bibnamefont
  {Müller}}, \bibinfo {author} {\bibfnamefont {O.}~\bibnamefont
  {Javakhishvili}}, \bibinfo {author} {\bibfnamefont {D.}~\bibnamefont
  {Shergelashvili}}, \bibinfo {author} {\bibfnamefont {I.}~\bibnamefont
  {Keshelashvili}}, \bibinfo {author} {\bibfnamefont {D.}~\bibnamefont
  {Mchedlishvili}}, \bibinfo {author} {\bibfnamefont {F.}~\bibnamefont
  {Abusaif}}, \bibinfo {author} {\bibfnamefont {A.}~\bibnamefont {Aggarwal}},
  \bibinfo {author} {\bibfnamefont {L.}~\bibnamefont {Barion}}, \bibinfo
  {author} {\bibfnamefont {S.}~\bibnamefont {Basile}}, \bibinfo {author}
  {\bibfnamefont {J.}~\bibnamefont {Böker}}, \bibinfo {author} {\bibfnamefont
  {N.}~\bibnamefont {Canale}}, \bibinfo {author} {\bibfnamefont
  {G.}~\bibnamefont {Ciullo}}, \bibinfo {author} {\bibfnamefont
  {S.}~\bibnamefont {Dymov}}, \bibinfo {author} {\bibfnamefont
  {O.}~\bibnamefont {Felden}}, \bibinfo {author} {\bibfnamefont
  {M.}~\bibnamefont {Gagoshidze}}, \bibinfo {author} {\bibfnamefont
  {R.}~\bibnamefont {Gebel}}, \bibinfo {author} {\bibfnamefont
  {N.}~\bibnamefont {Demary}}, \bibinfo {author} {\bibfnamefont
  {K.}~\bibnamefont {Grigoryev}}, \bibinfo {author} {\bibfnamefont
  {D.}~\bibnamefont {Grzonka}}, \bibinfo {author} {\bibfnamefont
  {T.}~\bibnamefont {Hahnraths}}, \bibinfo {author} {\bibfnamefont
  {V.}~\bibnamefont {Hejny}}, \bibinfo {author} {\bibfnamefont
  {A.}~\bibnamefont {Kacharava}}, \bibinfo {author} {\bibfnamefont
  {V.}~\bibnamefont {Kamerdzhiev}}, \bibinfo {author} {\bibfnamefont
  {S.}~\bibnamefont {Karanth}}, \bibinfo {author} {\bibfnamefont
  {A.}~\bibnamefont {Kulikov}}, \bibinfo {author} {\bibfnamefont
  {A.}~\bibnamefont {Lehrach}}, \bibinfo {author} {\bibfnamefont
  {P.}~\bibnamefont {Lenisa}}, \bibinfo {author} {\bibfnamefont
  {N.}~\bibnamefont {Lomidze}}, \bibinfo {author} {\bibfnamefont
  {B.}~\bibnamefont {Lorentz}}, \bibinfo {author} {\bibfnamefont
  {G.}~\bibnamefont {Macharashvili}}, \bibinfo {author} {\bibfnamefont
  {A.}~\bibnamefont {Magiera}}, \bibinfo {author} {\bibfnamefont
  {Z.}~\bibnamefont {Metreveli}}, \bibinfo {author} {\bibfnamefont
  {A.}~\bibnamefont {Nass}}, \bibinfo {author} {\bibfnamefont {N.N.}\
  \bibnamefont {Nikolaev}}, \bibinfo {author} {\bibfnamefont {M.}~\bibnamefont
  {Nioradze}}, \bibinfo {author} {\bibfnamefont {A.}~\bibnamefont {Pesce}},
  \bibinfo {author} {\bibfnamefont {V.}~\bibnamefont {Poncza}}, \bibinfo
  {author} {\bibfnamefont {D.}~\bibnamefont {Prasuhn}}, \bibinfo {author}
  {\bibfnamefont {J.}~\bibnamefont {Pretz}}, \bibinfo {author} {\bibfnamefont
  {F.}~\bibnamefont {Rathmann}}, \bibinfo {author} {\bibfnamefont
  {A.}~\bibnamefont {Saleev}}, \bibinfo {author} {\bibfnamefont
  {T.}~\bibnamefont {Sef\-zick}}, \bibinfo {author} {\bibfnamefont {Yu.}\
  \bibnamefont {Senichev}}, \bibinfo {author} {\bibfnamefont {V.}~\bibnamefont
  {Shmakova}}, \bibinfo {author} {\bibfnamefont {J.}~\bibnamefont {Slim}},
  \bibinfo {author} {\bibfnamefont {H.}~\bibnamefont {Soltner}}, \bibinfo
  {author} {\bibfnamefont {E.}~\bibnamefont {Stephenson}}, \bibinfo {author}
  {\bibfnamefont {H.}~\bibnamefont {Ströher}}, \bibinfo {author}
  {\bibfnamefont {M.}~\bibnamefont {Tabidze}}, \bibinfo {author} {\bibfnamefont
  {G.}~\bibnamefont {Tagliente}}, \bibinfo {author} {\bibfnamefont {Yu.}\
  \bibnamefont {Uzikov}}, \bibinfo {author} {\bibfnamefont {Yu.}\ \bibnamefont
  {Valdau}}, \bibinfo {author} {\bibfnamefont {T.}~\bibnamefont {Wagner}},
  \bibinfo {author} {\bibfnamefont {A.}~\bibnamefont {Wro{\'{n}}ska}}, \bibinfo
  {author} {\bibfnamefont {P.}~\bibnamefont {Wüstner}}, \ and\ \bibinfo
  {author} {\bibfnamefont {M.}~\bibnamefont {\.{Z}urek}},\ }\bibfield  {title}
  {\enquote {\bibinfo {title} {A new beam polarimeter at {COSY} to search for
  electric dipole moments of charged particles},}\ }\href {\doibase
  10.1088/1748-0221/15/12/p12005} {\bibfield  {journal} {\bibinfo  {journal}
  {Journal of Instrumentation}\ }\textbf {\bibinfo {volume} {15}},\ \bibinfo
  {pages} {P12005--P12005} (\bibinfo {year} {2020})}\BibitemShut {NoStop}%
\bibitem [{\citenamefont {Lehrach}\ and\ \citenamefont
  {Maier}(2001)}]{Lehrach:2001db}%
  \BibitemOpen
  \bibfield  {author} {\bibinfo {author} {\bibfnamefont {A.}~\bibnamefont
  {Lehrach}}\ and\ \bibinfo {author} {\bibfnamefont {R.}~\bibnamefont
  {Maier}},\ }\bibfield  {title} {\enquote {\bibinfo {title} {{Siberian Snake
  for the Cooler Synchrotron COSY}},}\ }\href@noop {} {\bibfield  {journal}
  {\bibinfo  {journal} {Conf. Proc. C}\ }\textbf {\bibinfo {volume}
  {0106181}},\ \bibinfo {pages} {2566--2568} (\bibinfo {year}
  {2001})}\BibitemShut {NoStop}%
\bibitem [{\citenamefont {Slim}\ \emph {et~al.}(2020)\citenamefont {Slim},
  \citenamefont {Nass}, \citenamefont {Rathmann}, \citenamefont {Soltner},
  \citenamefont {Tagliente},\ and\ \citenamefont {Heberling}}]{Slim_2020}%
  \BibitemOpen
  \bibfield  {author} {\bibinfo {author} {\bibfnamefont {J.}~\bibnamefont
  {Slim}}, \bibinfo {author} {\bibfnamefont {A.}~\bibnamefont {Nass}}, \bibinfo
  {author} {\bibfnamefont {F.}~\bibnamefont {Rathmann}}, \bibinfo {author}
  {\bibfnamefont {H.}~\bibnamefont {Soltner}}, \bibinfo {author} {\bibfnamefont
  {G.}~\bibnamefont {Tagliente}}, \ and\ \bibinfo {author} {\bibfnamefont
  {D.}~\bibnamefont {Heberling}},\ }\bibfield  {title} {\enquote {\bibinfo
  {title} {The driving circuit of the waveguide {RF} {Wien} filter for the
  deuteron {EDM} precursor experiment at {COSY}},}\ }\href {\doibase
  10.1088/1748-0221/15/03/p03021} {\bibfield  {journal} {\bibinfo  {journal}
  {Journal of Instrumentation}\ }\textbf {\bibinfo {volume} {15}},\ \bibinfo
  {pages} {P03021--P03021} (\bibinfo {year} {2020})}\BibitemShut {NoStop}%
\bibitem [{\citenamefont {Syphers}\ \emph {et~al.}(1993)\citenamefont
  {Syphers}, \citenamefont {Ball}, \citenamefont {Brabson}, \citenamefont
  {Budnick}, \citenamefont {Caussyn}, \citenamefont {Chao}, \citenamefont
  {Collins}, \citenamefont {Derenchuk}, \citenamefont {Dutt}, \citenamefont
  {East}, \citenamefont {Ellison}, \citenamefont {Ellison}, \citenamefont
  {Friesel}, \citenamefont {Gabella}, \citenamefont {Hamilton}, \citenamefont
  {Huang}, \citenamefont {Jones}, \citenamefont {Lee}, \citenamefont {Li},
  \citenamefont {Minty}, \citenamefont {Nagaitsev}, \citenamefont {Ng},
  \citenamefont {Pei}, \citenamefont {Rondeau}, \citenamefont {Sloan},
  \citenamefont {Teng}, \citenamefont {Tepikian}, \citenamefont {Wang},
  \citenamefont {Yan},\ and\ \citenamefont {Zhang}}]{PhysRevLett.71.719}%
  \BibitemOpen
  \bibfield  {author} {\bibinfo {author} {\bibfnamefont {M.}~\bibnamefont
  {Syphers}}, \bibinfo {author} {\bibfnamefont {M.}~\bibnamefont {Ball}},
  \bibinfo {author} {\bibfnamefont {B.}~\bibnamefont {Brabson}}, \bibinfo
  {author} {\bibfnamefont {J.}~\bibnamefont {Budnick}}, \bibinfo {author}
  {\bibfnamefont {D.~D.}\ \bibnamefont {Caussyn}}, \bibinfo {author}
  {\bibfnamefont {A.~W.}\ \bibnamefont {Chao}}, \bibinfo {author}
  {\bibfnamefont {J.}~\bibnamefont {Collins}}, \bibinfo {author} {\bibfnamefont
  {V.}~\bibnamefont {Derenchuk}}, \bibinfo {author} {\bibfnamefont
  {S.}~\bibnamefont {Dutt}}, \bibinfo {author} {\bibfnamefont {G.}~\bibnamefont
  {East}}, \bibinfo {author} {\bibfnamefont {M.}~\bibnamefont {Ellison}},
  \bibinfo {author} {\bibfnamefont {T.}~\bibnamefont {Ellison}}, \bibinfo
  {author} {\bibfnamefont {D.}~\bibnamefont {Friesel}}, \bibinfo {author}
  {\bibfnamefont {W.}~\bibnamefont {Gabella}}, \bibinfo {author} {\bibfnamefont
  {B.}~\bibnamefont {Hamilton}}, \bibinfo {author} {\bibfnamefont
  {H.}~\bibnamefont {Huang}}, \bibinfo {author} {\bibfnamefont {W.~P.}\
  \bibnamefont {Jones}}, \bibinfo {author} {\bibfnamefont {S.~Y.}\ \bibnamefont
  {Lee}}, \bibinfo {author} {\bibfnamefont {D.}~\bibnamefont {Li}}, \bibinfo
  {author} {\bibfnamefont {M.~G.}\ \bibnamefont {Minty}}, \bibinfo {author}
  {\bibfnamefont {S.}~\bibnamefont {Nagaitsev}}, \bibinfo {author}
  {\bibfnamefont {K.~Y.}\ \bibnamefont {Ng}}, \bibinfo {author} {\bibfnamefont
  {X.}~\bibnamefont {Pei}}, \bibinfo {author} {\bibfnamefont {G.}~\bibnamefont
  {Rondeau}}, \bibinfo {author} {\bibfnamefont {T.}~\bibnamefont {Sloan}},
  \bibinfo {author} {\bibfnamefont {L.}~\bibnamefont {Teng}}, \bibinfo {author}
  {\bibfnamefont {S.}~\bibnamefont {Tepikian}}, \bibinfo {author}
  {\bibfnamefont {Y.}~\bibnamefont {Wang}}, \bibinfo {author} {\bibfnamefont
  {Y.~T.}\ \bibnamefont {Yan}}, \ and\ \bibinfo {author} {\bibfnamefont
  {P.~L.}\ \bibnamefont {Zhang}},\ }\bibfield  {title} {\enquote {\bibinfo
  {title} {Experimental study of synchro-betatron coupling induced by dipole
  modulation},}\ }\href {\doibase 10.1103/PhysRevLett.71.719} {\bibfield
  {journal} {\bibinfo  {journal} {Phys. Rev. Lett.}\ }\textbf {\bibinfo
  {volume} {71}},\ \bibinfo {pages} {719--722} (\bibinfo {year}
  {1993})}\BibitemShut {NoStop}%
\bibitem [{\citenamefont {Miyamoto}\ \emph {et~al.}(2008)\citenamefont
  {Miyamoto}, \citenamefont {Kopp}, \citenamefont {Jansson},\ and\
  \citenamefont {Syphers}}]{MiyamotoACdipole}%
  \BibitemOpen
  \bibfield  {author} {\bibinfo {author} {\bibfnamefont {R.}~\bibnamefont
  {Miyamoto}}, \bibinfo {author} {\bibfnamefont {S.~E.}\ \bibnamefont {Kopp}},
  \bibinfo {author} {\bibfnamefont {A.}~\bibnamefont {Jansson}}, \ and\
  \bibinfo {author} {\bibfnamefont {M.~J.}\ \bibnamefont {Syphers}},\
  }\bibfield  {title} {\enquote {\bibinfo {title} {{Parametrization of the
  driven betatron oscillation}},}\ }\href {\doibase
  10.1103/PhysRevSTAB.11.084002} {\bibfield  {journal} {\bibinfo  {journal}
  {Phys. Rev. ST Accel. Beams}\ }\textbf {\bibinfo {volume} {11}},\ \bibinfo
  {pages} {084002} (\bibinfo {year} {2008})},\ \Eprint
  {http://arxiv.org/abs/0709.4192} {arXiv:0709.4192 [physics.acc-ph]}
  \BibitemShut {NoStop}%
\bibitem [{\citenamefont {Meade}(1983)}]{meade1983lock}%
  \BibitemOpen
  \bibfield  {author} {\bibinfo {author} {\bibfnamefont {M.L.}\ \bibnamefont
  {Meade}},\ }\href@noop {} {\emph {\bibinfo {title} {Lock-in Amplifiers:
  Principles and Applications}}},\ IEE electrical measurement series\ (\bibinfo
   {publisher} {P. Peregrinus},\ \bibinfo {year} {1983})\BibitemShut {NoStop}%
\bibitem [{\citenamefont {Huang}\ \emph {et~al.}(2004)\citenamefont {Huang},
  \citenamefont {Ahrens}, \citenamefont {Bai}, \citenamefont {Brown},
  \citenamefont {Glenn}, \citenamefont {Luccio}, \citenamefont {MacKay},
  \citenamefont {Montag}, \citenamefont {Ptitsyn}, \citenamefont {Roser},
  \citenamefont {Tsoupas}, \citenamefont {Zeno}, \citenamefont {Ranjbar},
  \citenamefont {Spinka},\ and\ \citenamefont
  {Underwood}}]{PhysRevSTAB.7.071001}%
  \BibitemOpen
  \bibfield  {author} {\bibinfo {author} {\bibfnamefont {H.}~\bibnamefont
  {Huang}}, \bibinfo {author} {\bibfnamefont {L.}~\bibnamefont {Ahrens}},
  \bibinfo {author} {\bibfnamefont {M.}~\bibnamefont {Bai}}, \bibinfo {author}
  {\bibfnamefont {K.~A.}\ \bibnamefont {Brown}}, \bibinfo {author}
  {\bibfnamefont {J.~W.}\ \bibnamefont {Glenn}}, \bibinfo {author}
  {\bibfnamefont {A.~U.}\ \bibnamefont {Luccio}}, \bibinfo {author}
  {\bibfnamefont {W.~W.}\ \bibnamefont {MacKay}}, \bibinfo {author}
  {\bibfnamefont {C.}~\bibnamefont {Montag}}, \bibinfo {author} {\bibfnamefont
  {V.}~\bibnamefont {Ptitsyn}}, \bibinfo {author} {\bibfnamefont
  {T.}~\bibnamefont {Roser}}, \bibinfo {author} {\bibfnamefont
  {N.}~\bibnamefont {Tsoupas}}, \bibinfo {author} {\bibfnamefont
  {K.}~\bibnamefont {Zeno}}, \bibinfo {author} {\bibfnamefont {V.}~\bibnamefont
  {Ranjbar}}, \bibinfo {author} {\bibfnamefont {H.}~\bibnamefont {Spinka}}, \
  and\ \bibinfo {author} {\bibfnamefont {D.}~\bibnamefont {Underwood}},\
  }\bibfield  {title} {\enquote {\bibinfo {title} {Overcoming an intrinsic
  depolarizing resonance with a partial siberian snake},}\ }\href {\doibase
  10.1103/PhysRevSTAB.7.071001} {\bibfield  {journal} {\bibinfo  {journal}
  {Phys. Rev. ST Accel. Beams}\ }\textbf {\bibinfo {volume} {7}},\ \bibinfo
  {pages} {071001} (\bibinfo {year} {2004})}\BibitemShut {NoStop}%
\bibitem [{\citenamefont {Bagdasarian}\ \emph {et~al.}(2014)\citenamefont
  {Bagdasarian}, \citenamefont {Bertelli}, \citenamefont {Chiladze},
  \citenamefont {Ciullo}, \citenamefont {Dietrich}, \citenamefont {Dymov},
  \citenamefont {Eversmann}, \citenamefont {Fanourakis}, \citenamefont
  {Gaisser}, \citenamefont {Gebel}, \citenamefont {Gou}, \citenamefont
  {Guidoboni}, \citenamefont {Hejny}, \citenamefont {Kacharava}, \citenamefont
  {Kamerdzhiev}, \citenamefont {Lehrach}, \citenamefont {Lenisa}, \citenamefont
  {Lorentz}, \citenamefont {Magallanes}, \citenamefont {Maier}, \citenamefont
  {Mchedlishvili}, \citenamefont {Morse}, \citenamefont {Nass}, \citenamefont
  {Oellers}, \citenamefont {Pesce}, \citenamefont {Prasuhn}, \citenamefont
  {Pretz}, \citenamefont {Rathmann}, \citenamefont {Shmakova}, \citenamefont
  {Semertzidis}, \citenamefont {Stephenson}, \citenamefont {Stockhorst},
  \citenamefont {Str\"oher}, \citenamefont {Talman}, \citenamefont
  {Th\"orngren~Engblom}, \citenamefont {Valdau}, \citenamefont {Weidemann},\
  and\ \citenamefont {W\"ustner}}]{PhysRevSTAB.17.052803}%
  \BibitemOpen
  \bibfield  {author} {\bibinfo {author} {\bibfnamefont {Z.}~\bibnamefont
  {Bagdasarian}}, \bibinfo {author} {\bibfnamefont {S.}~\bibnamefont
  {Bertelli}}, \bibinfo {author} {\bibfnamefont {D.}~\bibnamefont {Chiladze}},
  \bibinfo {author} {\bibfnamefont {G.}~\bibnamefont {Ciullo}}, \bibinfo
  {author} {\bibfnamefont {J.}~\bibnamefont {Dietrich}}, \bibinfo {author}
  {\bibfnamefont {S.}~\bibnamefont {Dymov}}, \bibinfo {author} {\bibfnamefont
  {D.}~\bibnamefont {Eversmann}}, \bibinfo {author} {\bibfnamefont
  {G.}~\bibnamefont {Fanourakis}}, \bibinfo {author} {\bibfnamefont
  {M.}~\bibnamefont {Gaisser}}, \bibinfo {author} {\bibfnamefont
  {R.}~\bibnamefont {Gebel}}, \bibinfo {author} {\bibfnamefont
  {B.}~\bibnamefont {Gou}}, \bibinfo {author} {\bibfnamefont {G.}~\bibnamefont
  {Guidoboni}}, \bibinfo {author} {\bibfnamefont {V.}~\bibnamefont {Hejny}},
  \bibinfo {author} {\bibfnamefont {A.}~\bibnamefont {Kacharava}}, \bibinfo
  {author} {\bibfnamefont {V.}~\bibnamefont {Kamerdzhiev}}, \bibinfo {author}
  {\bibfnamefont {A.}~\bibnamefont {Lehrach}}, \bibinfo {author} {\bibfnamefont
  {P.}~\bibnamefont {Lenisa}}, \bibinfo {author} {\bibfnamefont
  {B.}~\bibnamefont {Lorentz}}, \bibinfo {author} {\bibfnamefont
  {L.}~\bibnamefont {Magallanes}}, \bibinfo {author} {\bibfnamefont
  {R.}~\bibnamefont {Maier}}, \bibinfo {author} {\bibfnamefont
  {D.}~\bibnamefont {Mchedlishvili}}, \bibinfo {author} {\bibfnamefont {W.~M.}\
  \bibnamefont {Morse}}, \bibinfo {author} {\bibfnamefont {A.}~\bibnamefont
  {Nass}}, \bibinfo {author} {\bibfnamefont {D.}~\bibnamefont {Oellers}},
  \bibinfo {author} {\bibfnamefont {A.}~\bibnamefont {Pesce}}, \bibinfo
  {author} {\bibfnamefont {D.}~\bibnamefont {Prasuhn}}, \bibinfo {author}
  {\bibfnamefont {J.}~\bibnamefont {Pretz}}, \bibinfo {author} {\bibfnamefont
  {F.}~\bibnamefont {Rathmann}}, \bibinfo {author} {\bibfnamefont
  {V.}~\bibnamefont {Shmakova}}, \bibinfo {author} {\bibfnamefont {Y.~K.}\
  \bibnamefont {Semertzidis}}, \bibinfo {author} {\bibfnamefont {E.~J.}\
  \bibnamefont {Stephenson}}, \bibinfo {author} {\bibfnamefont
  {H.}~\bibnamefont {Stockhorst}}, \bibinfo {author} {\bibfnamefont
  {H.}~\bibnamefont {Str\"oher}}, \bibinfo {author} {\bibfnamefont
  {R.}~\bibnamefont {Talman}}, \bibinfo {author} {\bibfnamefont
  {P.}~\bibnamefont {Th\"orngren~Engblom}}, \bibinfo {author} {\bibfnamefont
  {Yu.}\ \bibnamefont {Valdau}}, \bibinfo {author} {\bibfnamefont
  {C.}~\bibnamefont {Weidemann}}, \ and\ \bibinfo {author} {\bibfnamefont
  {P.}~\bibnamefont {W\"ustner}},\ }\bibfield  {title} {\enquote {\bibinfo
  {title} {Measuring the polarization of a rapidly precessing deuteron beam},}\
  }\href {\doibase 10.1103/PhysRevSTAB.17.052803} {\bibfield  {journal}
  {\bibinfo  {journal} {Phys. Rev. ST Accel. Beams}\ }\textbf {\bibinfo
  {volume} {17}},\ \bibinfo {pages} {052803} (\bibinfo {year}
  {2014})}\BibitemShut {NoStop}%
\bibitem [{\citenamefont {Aster}\ \emph {et~al.}(2018)\citenamefont {Aster},
  \citenamefont {Borchers},\ and\ \citenamefont
  {Thurber}}]{aster2018parameter}%
  \BibitemOpen
  \bibfield  {author} {\bibinfo {author} {\bibfnamefont {R.C.}\ \bibnamefont
  {Aster}}, \bibinfo {author} {\bibfnamefont {B.}~\bibnamefont {Borchers}}, \
  and\ \bibinfo {author} {\bibfnamefont {C.H.}\ \bibnamefont {Thurber}},\
  }\href@noop {} {\emph {\bibinfo {title} {Parameter Estimation and Inverse
  Problems}}}\ (\bibinfo  {publisher} {Elsevier Science},\ \bibinfo {year}
  {2018})\BibitemShut {NoStop}%
\bibitem [{\citenamefont {Wolski}(2014)}]{doi:10.1142/p899}%
  \BibitemOpen
  \bibfield  {author} {\bibinfo {author} {\bibfnamefont {Andrzej}\ \bibnamefont
  {Wolski}},\ }\href {\doibase 10.1142/p899} {\emph {\bibinfo {title} {Beam
  Dynamics in High Energy Particle Accelerators}}}\ (\bibinfo  {publisher}
  {IMPERIAL COLLEGE PRESS},\ \bibinfo {year} {2014})\ \Eprint
  {http://arxiv.org/abs/https://www.worldscientific.com/doi/pdf/10.1142/p899}
  {https://www.worldscientific.com/doi/pdf/10.1142/p899} \BibitemShut {NoStop}%
\bibitem [{\citenamefont {Sagan}(2006)}]{Sagan:Bmad2006}%
  \BibitemOpen
  \bibfield  {author} {\bibinfo {author} {\bibfnamefont {D.}~\bibnamefont
  {Sagan}},\ }\bibfield  {title} {\enquote {\bibinfo {title} {{Bmad: A
  relativistic charged particle simulation library}},}\ }\bibfield  {booktitle}
  {\emph {\bibinfo {booktitle} {{Computational accelerator physics.
  Proceedings, 8th International Conference, ICAP 2004, St. Petersburg, Russia,
  June 29-July 2, 2004}}},\ }\href {\doibase
  https://doi.org/10.1016/j.nima.2005.11.001} {\bibfield  {journal} {\bibinfo
  {journal} {Nucl. Instrum. Meth.}\ }\textbf {\bibinfo {volume} {A558}},\
  \bibinfo {pages} {356--359} (\bibinfo {year} {2006})},\ \bibinfo {note}
  {proceedings of the 8th International Computational Accelerator Physics
  Conference}\BibitemShut {NoStop}%
\bibitem [{\citenamefont {Smith}(2013)}]{10.5555/2568154}%
  \BibitemOpen
  \bibfield  {author} {\bibinfo {author} {\bibfnamefont {Ralph~C.}\
  \bibnamefont {Smith}},\ }\href@noop {} {\emph {\bibinfo {title} {Uncertainty
  Quantification: Theory, Implementation, and Applications}}}\ (\bibinfo
  {publisher} {Society for Industrial and Applied Mathematics},\ \bibinfo
  {address} {USA},\ \bibinfo {year} {2013})\BibitemShut {NoStop}%
\bibitem [{\citenamefont {Offermann}\ \emph {et~al.}(2015)\citenamefont
  {Offermann}, \citenamefont {Mac}, \citenamefont {Nguyen}, \citenamefont
  {Clénet}, \citenamefont {Gersem},\ and\ \citenamefont {Hameyer}}]{7093523}%
  \BibitemOpen
  \bibfield  {author} {\bibinfo {author} {\bibfnamefont {P.}~\bibnamefont
  {Offermann}}, \bibinfo {author} {\bibfnamefont {H.}~\bibnamefont {Mac}},
  \bibinfo {author} {\bibfnamefont {T.~T.}\ \bibnamefont {Nguyen}}, \bibinfo
  {author} {\bibfnamefont {S.}~\bibnamefont {Clénet}}, \bibinfo {author}
  {\bibfnamefont {H.~De}\ \bibnamefont {Gersem}}, \ and\ \bibinfo {author}
  {\bibfnamefont {K.}~\bibnamefont {Hameyer}},\ }\bibfield  {title} {\enquote
  {\bibinfo {title} {Uncertainty quantification and sensitivity analysis in
  electrical machines with stochastically varying machine parameters},}\ }\href
  {\doibase 10.1109/TMAG.2014.2354511} {\bibfield  {journal} {\bibinfo
  {journal} {IEEE Transactions on Magnetics}\ }\textbf {\bibinfo {volume}
  {51}},\ \bibinfo {pages} {1--4} (\bibinfo {year} {2015})}\BibitemShut
  {NoStop}%
\bibitem [{\citenamefont {Adelmann}(2018)}]{adelmann2018uncertainty}%
  \BibitemOpen
  \bibfield  {author} {\bibinfo {author} {\bibfnamefont {Andreas}\ \bibnamefont
  {Adelmann}},\ }\href@noop {} {\enquote {\bibinfo {title} {On uncertainty
  quantification in particle accelerators modelling},}\ } (\bibinfo {year}
  {2018}),\ \Eprint {http://arxiv.org/abs/1509.08130} {arXiv:1509.08130
  [physics.acc-ph]} \BibitemShut {NoStop}%
\bibitem [{\citenamefont {York}\ \emph {et~al.}(2004)\citenamefont {York},
  \citenamefont {Evensen}, \citenamefont {L\'{o}pez~Mart\'{i}nez},\ and\
  \citenamefont {De~Basabe~Delgado}}]{doi:10.1119/1.1632486}%
  \BibitemOpen
  \bibfield  {author} {\bibinfo {author} {\bibfnamefont {Derek}\ \bibnamefont
  {York}}, \bibinfo {author} {\bibfnamefont {Norman~M.}\ \bibnamefont
  {Evensen}}, \bibinfo {author} {\bibfnamefont {Margarita}\ \bibnamefont
  {L\'{o}pez~Mart\'{i}nez}}, \ and\ \bibinfo {author} {\bibfnamefont
  {Jon\'{a}s}\ \bibnamefont {De~Basabe~Delgado}},\ }\bibfield  {title}
  {\enquote {\bibinfo {title} {Unified equations for the slope, intercept, and
  standard errors of the best straight line},}\ }\href {\doibase
  10.1119/1.1632486} {\bibfield  {journal} {\bibinfo  {journal} {American
  Journal of Physics}\ }\textbf {\bibinfo {volume} {72}},\ \bibinfo {pages}
  {367--375} (\bibinfo {year} {2004})},\ \Eprint
  {http://arxiv.org/abs/https://doi.org/10.1119/1.1632486}
  {https://doi.org/10.1119/1.1632486} \BibitemShut {NoStop}%
\bibitem [{\citenamefont {Sudret}\ and\ \citenamefont
  {Der~Kiureghian}(2000)}]{sudret2000stochastic}%
  \BibitemOpen
  \bibfield  {author} {\bibinfo {author} {\bibfnamefont {Bruno}\ \bibnamefont
  {Sudret}}\ and\ \bibinfo {author} {\bibfnamefont {Armen}\ \bibnamefont
  {Der~Kiureghian}},\ }\href@noop {} {\emph {\bibinfo {title} {Stochastic
  finite element methods and reliability: a state-of-the-art report}}}\
  (\bibinfo  {publisher} {Department of Civil and Environmental Engineering,
  University of California},\ \bibinfo {year} {2000})\BibitemShut {NoStop}%
\bibitem [{\citenamefont {Sudret}(2008)}]{sudret2008global}%
  \BibitemOpen
  \bibfield  {author} {\bibinfo {author} {\bibfnamefont {Bruno}\ \bibnamefont
  {Sudret}},\ }\bibfield  {title} {\enquote {\bibinfo {title} {Global
  sensitivity analysis using polynomial chaos expansions},}\ }\href@noop {}
  {\bibfield  {journal} {\bibinfo  {journal} {Reliability Engineering \& System
  Safety}\ }\textbf {\bibinfo {volume} {93}},\ \bibinfo {pages} {964--979}
  (\bibinfo {year} {2008})}\BibitemShut {NoStop}%
\end{thebibliography}%
\end{document}